\documentclass{ejm}
\usepackage[utf8]{inputenc}
\usepackage{amsmath}
\usepackage{amsfonts}
\usepackage{amssymb}
\usepackage{graphicx}
\usepackage[colorlinks,citecolor = brown,linkcolor=violet,anchorcolor = blue]{hyperref}
\usepackage{tikz}
\usepackage[toc,page]{appendix}
\usepackage{wrapfig}
\usepackage{color}
\usepackage{graphicx}
\graphicspath{{figures/}} 
\usepackage{subcaption}
\usepackage{dirtytalk}
\numberwithin{equation}{subsection}
\usepackage[export]{adjustbox}

\usepackage[font={small,it}]{caption}

\usepackage{cleveref}
\crefname{section}{§}{§§}
\Crefname{section}{§}{§§}

\author{Sarafa A. Iyaniwura, Michael J. Ward} \title[Quorum Sensing
and Synchronization for Conditional Oscillators Coupled by Bulk
Diffusion]{ \textbf{Localized Signaling Compartments in 2-D Coupled by
    a Bulk Diffusion Field: Quorum Sensing and Synchronous
    Oscillations in the Well-Mixed Limit}}

\begin{document}

\maketitle

\begin{abstract}
  We analyze oscillatory instabilities for a coupled PDE-ODE system
  modeling the communication between localized spatially segregated
  dynamically active signaling compartments that are coupled through a
  passive extracellular bulk diffusion field in a bounded 2-D
  domain. Each signaling compartment is assumed to secrete a chemical
  into the extracellular medium (bulk region) and it can also sense
  the concentration of this chemical in the region around its
  boundary. This feedback from the bulk region, resulting from the
  entire collection of cells, in turn modifies the intracellular
  dynamics within each cell. In the limit where the signaling
  compartments are circular disks with a small common radius
  $\varepsilon \ll 1$ and where the bulk diffusivity is asymptotically
  large, a matched asymptotic analysis is used to reduce the
  dimensionless PDE-ODE system into a nonlinear ODE system with global
  coupling. For Sel'kov reaction kinetics, this ODE system for the
  intracellular dynamics and the spatial average of the bulk diffusion
  field is then used to investigate oscillatory instabilities in the
  dynamics of the cells that are triggered due to the global
  coupling. In particular, numerical bifurcation software on the ODEs
  is used to study the overall effect of coupling \textit{defective
    cells} (cells that behave differently from the remaining cells) to
  a group of identical cells. Moreover, when the number of cells is
  large, the Kuramoto order parameter is computed to predict the
  degree of phase synchronization of the intracellular
  dynamics. Quorum sensing behavior, characterized by the onset of
  collective behavior in the intracellular dynamics as the number of
  cells increases above a threshold, is also studied. Our analysis
  shows that the cell population density plays a dual role of
  triggering and then quenching synchronous oscillations in the
  intracellular dynamics.
\end{abstract}



\section{Introduction} 

 {Cell-cell communication is an important component of
  microbiological systems, which involves the sending and receiving of
  information from one cell to another. In a biological system
  consisting of unicellular organisms, cells coordinate and
  communicate with one another to accomplish tasks that cannot be
  achieved by a single cell. Those cells that are not in close
  proximity can only communicate through the extracellular space
  between them by both secreting and then sensing signaling chemicals
  in the extracellular medium, called autoinducers, and adjusting
  their intracellular activities accordingly. This bulk mediated form
  of cell-to-cell communication has been observed in many specific
  biological systems including, a collection of social amoebae
  \textit{Dictyostelium discoideum} where the secretion of cyclic
  adenosine monophosphate (cAMP) by the cells triggers an oscillatory
  response and helps guide the cells to aggregation
  (cf.~\cite{gregor2010}, \cite{nandy1998}), colonies of yeast cells
  in which the exchange of acetaldehyde (Aca) molecules leads to
  glycolytic oscillations (cf.~\cite{de2007dynamical}), and the
  emergence of bioluminescence for the marine bacterium {\em Vibrio
    fischeri} at large cell densities (cf.~\cite{qs_jump}).  In this
  context, the term quorum sensing refers to a process of cell-to-cell
  communication, mediated by a bulk diffusion field, that triggers
  some collective behavior of a group of cells as the cell density
  increases.

  Quorum sensing phenomenon where the collective dynamics involves a
  switch-like emergence of synchronous oscillations as the cell
  density increases past a threshold has been observed and studied in
  several specific biological systems (cf.~\cite{de2007dynamical},
  \cite{gregor2010}, \cite{review-yeast}, \cite{mina},
  \cite{leaman2018}) as well as in physicochemical systems involving
  small catalyst-loaded particles immersed in a chemical mixture
  (cf.~\cite{taylor1}, \cite{taylor2}, \cite{tinsley1},
  \cite{tinsley2}). An overview of some universal features for quorum
  sensing systems in biology are discussed in \cite{review-yeast}.  In
  other cell signalling problems, spatio-temporal oscillations
  associated with chemically active, but spatially localized, sites
  that are mediated by a bulk diffusion field also arise in models of
  certain gene regulatory networks \cite{hess}.}
 
\begin{figure}[htbp]
\begin{center}
\includegraphics[scale=0.9]{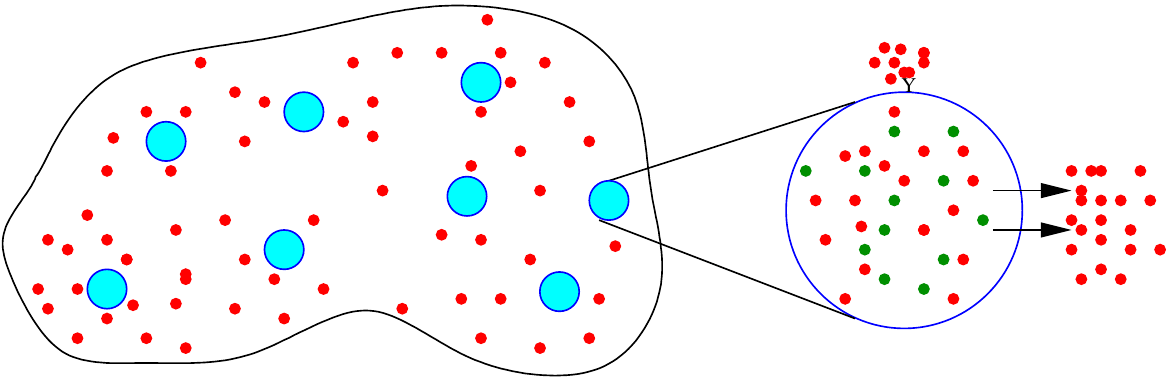}
\end{center}
\caption{A schematic diagram for \eqref{Dim_bulk} showing the
  dynamically active signaling compartments (in cyan) and the
  extracellular bulk region (unshaded) in a bounded 2-D domain. Each
  circular cell contains two signaling chemicals (red and green),
  but only the red chemical is secreted into the extracellular bulk
  region. A zoomed-in illustration of the secretion and feedback of
  chemicals into the cells is shown on the right.}
\label{onebulk_scheme}
\end{figure}

{Our goal herein is to use a theoretical framework based on a
  coupled PDE-ODE system to study the onset of synchronous
  intracellular dynamics in a collection of small signalling
  compartments or cells that are coupled through a 2-D passive bulk
  diffusion field. We begin by outlining the formulation of this
  coupled cell-bulk PDE-ODE model of \cite{jia2016}, which was
  inspired by the models introduced in \cite{muller2006} (see also
  \cite{muller2013} and \cite{Mueller2014uecke}). Let
  $\Omega \subset \mathbb{R}^2$ be a bounded domain that contains $m$
  disk-shaped dynamically active signalling compartments or ``cells''
  of a common radius $R$, denoted by $\Omega_j$ and centered at
  $\pmb{X}_j \in \Omega$ for $j=1,\dots,m$. We assume that the
  concentration $\mathcal{U}(\pmb{X},T)$ of the signaling molecule or
  autoinducer in the bulk or extracellular region
  $\Omega \setminus \cup_{j=1}^{m}\,\Omega_j$, which is confined
  within $\partial\Omega$, satisfies 
\begin{subequations}\label{Dim_bulk} 
\begin{align}
   \mathcal{U}_T  = & D_B \, \Delta \,
 \mathcal{U} - k_B \, \mathcal{U}\,, \quad T>0\,,\quad \pmb{X} \in
  \Omega \setminus \cup_{j=1}^{m}\,\Omega_j\,; \label{Dim_bulka}\\
  \, \partial_{n_{\pmb{X}}} \, \mathcal{U} = 0\,, \quad \pmb{X} \in
  \partial \Omega\,, & \qquad D_B \, \partial_{n_{\pmb{X}}} \, \mathcal{U}
  = \beta_{1,j} \, \mathcal{U} - \beta_{2,j} \, \mu_j^1\,,
   \qquad \pmb{X} \in \partial \Omega_j\,,\quad j = 1,2, \ldots, m\,.
                    \label{Dim_bulkb}
\end{align}
Here $D_B > 0$ and $k_B >0$ are the bulk diffusivity and the rate of
degradation of the signaling chemical in the bulk region,
respectively. In the Robin boundary condition \eqref{Dim_bulkb} on
each cell boundary, $\beta_{1,j}>0$ and $\beta_{2,j}>0$ are
dimensional constants modeling the permeability of the boundary for
the $j^{\text{th}}$ cell, while $\partial_{n_{\pmb{X}}}$ is the
outward normal derivative pointing into the bulk region. Inside each
dynamically active cell, which is assumed to be well-mixed, we suppose
that there are $n$ interacting species, represented by the
vector-field $ \pmb{\mu}_j \equiv (\mu_j^1,\ldots,\mu_j^n)^T$, that if
isolated from the bulk would interact according to the local
reaction kinetics $\pmb{F}_j \left( \pmb{\mu}_j/\mu_c \right)
$. However, owing to the transport across each cell boundary, the bulk
diffusion field of \eqref{Dim_bulk} is coupled to the local
kinetics within the $j^{\text{th}}$ cell by
\begin{equation}\label{Dim_Intra}
\begin{split}
  \frac{d \pmb{\mu}_j}{dT} & = k_R\, \mu_c \, \pmb{F}_j
  \left( \pmb{\mu}_j/\mu_c  \right)  + \pmb{e}_1 \int_{\partial \Omega_j}\,
  \left( \beta_{1,j} \, \mathcal{U} - \beta_{2,j} \, \mu_j^1 \right) \,
  \text{d}S_{\pmb{X}}\,,\qquad j = 1, \ldots, m\,.
\end{split}
\end{equation}
\end{subequations}
Here $\pmb{e}_1 \equiv (1,0,\ldots,0)^T$, $k_R > 0$ is the typical
reaction rate for the dimensionless local kinetics $\pmb{F}_j$, and
$\mu_c > 0$ is a typical scalar measure of $\pmb{\mu}_j$. In this
model, each cell secretes only one signaling chemical, labeled by
$\mu_j^1$, across its boundary into the bulk medium, and this
secretion is regulated by the efflux permeability parameter
$\beta_{2,j}$. However, each cell receives feedback across its
boundary, as controlled by the influx permeability parameter
$\beta_{1,j}$, from the entire collection of cells as mediated by the
bulk concentration field $\mathcal{U}(\pmb{X},T)$ for
$\pmb{X}\in \partial \Omega_j$. Figure \ref{onebulk_scheme} shows a
schematic diagram to illustrate the cell-bulk coupling in
\eqref{Dim_bulk} for the case where there are $n=2$ intracellular
species.

The coupled PDE-ODE model \eqref{Dim_bulk}, non-dimensionalized in \S
\ref{nondim_onebulk} and Appendix \ref{Append_A} and written in
dimensionless form below in \eqref{DimLess_bulk}, is analyzed in the
limit of a small common cell radius when the bulk diffusivity is
sufficiently large so that the bulk medium can be considered as
``well-mixed''. In this limit, where the leading-order bulk
concentration field is spatially homogeneous, a matched asymptotic
expansion analysis is used in \S \ref{Onebulk_WM_regime} to reduce the
dimensionless coupled PDE-ODE system \eqref{DimLess_bulk} into an
$nm+1$ component nonlinear ODE system, with global coupling, which
is given below in \eqref{WM_ODEsys_2D}. A similar reduction, but only
for the case of identical cells, was given in \S 5 of
\cite{jia2016}. We remark that our asymptotic reduction, starting from
a governing PDE-ODE coupled system, provides a systematic method for
deriving limiting globally coupled ODE systems characterizing
cell-cell communication through a well-mixed bulk medium. This is in
contrast to the more phenemologically-based ODE models used in
previous studies (cf. \cite{de2007dynamical}, \cite{spiral},
\cite{schwab}, \cite{mina}, \cite{review-design}). Our limiting ODE
system is related to those used in \cite{Rossler} and \cite{RosslerHeter}
to model communication between nonlinear oscillators coupled indirectly
through an external medium.

For the specific case of Sel'kov reaction kinetics \cite{Selkov}, the
ODE system \eqref{WM_ODEsys_2D} for the intracellular dynamics and the
spatial average of the bulk diffusion field is then used as a
conceptual model to investigate the emergence of oscillatory
instabilities in the intracellular dynamics that are triggered due to
the global coupling. In the absence of coupling to the bulk medium,
the Sel'kov kinetic parameters are chosen so that each cell is a {\em
  conditional oscillator}, in that its dynamics has only a stable
steady-state when uncoupled from the bulk. As a result, in our study
the emergence of intracellular oscillations is inherently due to the
cell-cell interaction, as mediated by the global coupling through the
well-mixed bulk region. However, in qualitative agreement with
experimental findings in \cite{de2007dynamical} and \cite{olson}
involving yeast cells that are typically non-oscillatory when
isolated, the Sel'kov parameters are chosen relatively close to the
threshold for the onset of limit-cycle oscillations for an isolated
cell.

With Sel'kov reaction kinetics, in \S \ref{GlobalBifur} we compute
global branches of steady-states and periodic solutions for the ODEs
\eqref{WM_ODEsys_2D} using the numerical bifurcation software XPPAUT
\cite{xpp2002} for a small collection of cells. Our goal is to show
how the permeability parameters, the Sel'kov kinetic parameters, and
the number of cells influence the emergence of intracellular
oscillations from a quiescent linearly stable steady-state. In our
numerical bifurcation study in \S \ref{GlobalBifur} we consider two
main scenarios: one where all the cells are identical with common
parameters, and the other  a single \textit{defective} cell,
which has either different permeabilities or kinetic parameters, 
is coupled to a group of identical cells. Our numerical results,
presented in the form of one- and two-parameter global bifurcation
diagrams, show that for a collection of identical cells synchronous
intracellular oscillations can occur in various parameter regimes as a
result of the global coupling, even when their individual dynamics are
linearly stable. Moreover, we show that small changes in the
permeabilities or the reaction kinetics from a single defective cell
can either extinguish or trigger intracellular oscillations of the
entire group of cells.  However, most typically, the range of these
parameters that yields stable periodic solutions are smaller for the
cases involving a defective cell, and this range decreases as the
number of identical cells increases. Our theoretical finding that a
single defective cell can trigger oscillations in the entire group is
qualitatively similar to the observations in \cite{gregor2010}, based
on live-cell imaging of the social bacteria {\em Dictyostelium
  discoideum}, that stochastic pulses originating from a discrete
location play a critical role in the onset of collective oscillations.
Our study of the effect of a defective cell on the entire group of
cells is also related to game-theoretic concepts, which have recently
been applied to quorum sensing behavior in microbial communities \cite{game}.

In \S \ref{Section_QS} we recast the ODE system \eqref{WM_ODEsys_2D}
in terms of a cell density parameter, defined by
$\rho\equiv {m/|\Omega|}$, in order to study the quorum sensing
behavior associated with a large collection of either identical cells
or a mixture of identical and defective cells. We show that the quorum
sensing transition is characterized by the switch-like emergence of
synchronous intracellular oscillations as $\rho$ crosses through a
critical threshold. This is followed by a switch-like quenching of
these oscillations as the cell density parameter crosses through a
second, larger, threshold value. When the cells are identical, we show
that these two threshold values are Hopf bifurcation points for a
cubic polynomial. For a mixture of identical and defective cells,
where a Sel'kov kinetic parameter is randomly chosen from some range, the
quorum sensing and quenching thresholds are identified from a
numerical computation of the Kuramoto order parameter, defined in
\eqref{QS_Kura_Par} (see \cite{RosslerHeter}, \cite{Rossler}). The
interval of the cell density parameter where the synchronization of
intracellular dynamics occurs was found, as expected, to decrease as
the number of defective cells increases.

In \S \ref{Sec:Discussion} we summarize some of our main results, and
we discuss possible extensions of this study for investigating some
specific biologically-based cell-cell communication problems that are
mediated by a bulk diffusion field. Related problems that fit within
the theoretical framework offered by the coupled PDE-ODE system
\eqref{Dim_bulk} are also discussed.}

\section{The coupled PDE-ODE model: Non-dimensionalization and the well-mixed limit}\label{One_bulk_species}

In this section we non-dimensionalize the coupled PDE-ODE model
\eqref{Dim_bulk} and study the dimensionless model in the large bulk
diffusivity limit in a bounded 2-D domain. In this limit, the bulk
region becomes well-mixed and the bulk concentration field becomes
spatially homogeneous. In this well-mixed limit, a matched asymptotic
analysis is used to reduce the dimensionless coupled PDE-ODE model
into a system of ODEs with global coupling.  This limiting system is
used to investigate oscillatory instabilities in the intracellular
dynamics and quorum sensing behavior.

\subsection{Non-dimensionalization}\label{nondim_onebulk}

We assume that the signaling compartments are circular with equal
radius $R_0$, which is small relative to the length-scale $L$ of the
domain. As such, we introduce a small scaling parameter
$\varepsilon \equiv R_0/L \ll 1$. With this assumption, in Appendix
\ref{Append_A} the coupled PDE-ODE model \eqref{Dim_bulk} is
non-dimensionalized such that the dimensionless spatio-temporal bulk
concentration field $U(\pmb{x},t)$ satisfies
\begin{subequations}\label{DimLess_bulk}
\begin{align}
  \tau \frac{\partial U}{\partial t} =&\,  D \, \Delta U -  \, U\,,
\quad t > 0\,,\quad \pmb{x} \in \Omega \setminus \cup_{j=1}^{m}\,
    \Omega_{\varepsilon_j}\,;\label{DimLess_bulka} \\
  \partial_n \, U =\, 0, \quad \pmb{x} \in \partial \Omega\,, & \qquad
  \varepsilon  D\, \partial_{n_j} U  = d_{1,j} \, U - d_{2,j} \, u_j^1\,,
 \quad \pmb{x} \in \partial \Omega_{\varepsilon_j}\,,\quad j = 1, \ldots, m\,,
    \label{DimLess_bulkb}
\end{align}
where $\Omega_{\varepsilon_j}$ is a circular region of radius
$\varepsilon \ll 1$ centered at $\pmb{x}_j \in \Omega$, representing
the $j^{\text{th}}$ cell, $D$ is the effective diffusion coefficient
in the bulk region, $\partial_n$ is the normal derivative pointing
outward on the boundary of the domain $\Omega$, and $\partial_{n_j}$
is the normal derivative on the boundary of the $j^{\text{th}}$ cell
pointing out of the cell into the bulk medium.  It is assumed that the
cells are well-separated from each other in the sense that
dist$(\pmb{x}_i,\pmb{x}_j) = \mathcal{O}(1)$ for $i \neq j$, and that
the center of each cell is at an $\mathcal{O}(1)$ distance from the
boundary of the domain $\Omega$,
i.e.~dist$(\pmb{x}_j,\partial \Omega) = \mathcal{O}(1)$ for
$j=1,\dots,m$ as $ \varepsilon \to 0$. The dimensionless bulk
concentration $U(\pmb{x},t)$ is coupled to the local kinetics within
the $j^{\text{th}}$ cell by
\begin{equation}\label{DimLess_Intra}
\begin{split}
  \frac{d \pmb{u}_j}{dt} & =  \, \pmb{F}_j \left( \pmb{u}_j  \right)  +
  \frac{\pmb{e}_1}{\varepsilon \tau} \int_{\partial \Omega_{\varepsilon_j}}\,
  ( d_{1,j} \, U - d_{2,j} \, u_j^1 ) \,\,\text{d}s\,,\quad j = 1, \ldots, m
  \,,
\end{split}
\end{equation}
\end{subequations}
where $\pmb{u}_j\equiv (u_j^1,\dots,u_j^n)^T$ is a dimensionless
vector representing the $n$ chemical species in the $j^{\text{th}}$
cell, $\pmb{e}_1\equiv (1,0,\dots,0)^T$, while $d_{1,j} >0$ and
$d_{2,j} >0$ are dimensionless permeability parameters modelling the
influx and efflux on the $j^{\text{th}}$ cell boundary,
respectively. The intracellular, or local, kinetics within the
$j^{\text{th}}$ cell is specified by the vector-field
$\pmb{F}_j \left( \pmb{u}_j \right)$. In \eqref{DimLess_bulk}, the
dimensionless parameters are defined in terms of the
parameters of \eqref{Dim_bulk} by
\begin{equation}\label{Variables}
\begin{split}
  \tau \equiv \frac{k_R}{k_B}\,,\qquad D \equiv \frac{D_B}{k_B L^2}\,,
  \qquad d_{1,j} \equiv \frac{\varepsilon \beta_{1,j}}{k_B L}\,,
  \qquad d_{2,j} \equiv \frac{\varepsilon \beta_{2,j} L}{k_B} \,.
\end{split}
\end{equation}
With this non-dimensionalization, $\tau$ is large if the intracellular
reactions occur quickly with respect to the time it takes the
autoinducer signal to decay in the bulk medium.  The dimensionless
permeabilities $d_{1,j}$ and $d_{2,j}$ are taken to be
$\mathcal{O}(1)$, which implies that $\beta_{1,j}$ and $\beta_{2,j}$
are $\mathcal{O}(\varepsilon^{-1})$ since the length-scale $L$ of the
domain and the rate $k_B$ of chemical decay in the bulk region are
$\mathcal{O}(1)$. In our analysis below, this scaling of the
permeabilities is necessary in order to have significant secretion and
feedback of chemicals into each signaling compartment in the limit
$\varepsilon \to 0$.

\subsection{Asymptotic analysis in the well-mixed limit} \label{Onebulk_WM_regime}

In this subsection, we study the dimensionless coupled PDE-ODE model
\eqref{DimLess_bulk} in the well-mixed limit
$D \gg \mathcal{O}(\nu^{-1}) \gg \mathcal{O}(1)$, where
$\nu \equiv -1/\log\varepsilon$. In this limit, the method of matched
asymptotic expansions is used to reduce the coupled model
\eqref{DimLess_bulk} into a system of ODEs with global coupling.

For $D \gg \mathcal{O}(\nu^{-1})$, we expand
$U(\pmb{x},t)$ in the (outer) bulk region at ${\mathcal O}(1)$ distances
from the cells as
\begin{equation}\label{WM_Uexpnd}
\begin{split}
U(\pmb{x},t) = U_0(\pmb{x},t) + \frac{1}{D}\, U_1(\pmb{x},t) + \cdots
\end{split}
\end{equation}
Upon substituting \eqref{WM_Uexpnd} into \eqref{DimLess_bulk}, and
collecting terms in powers of $D$, we obtain the leading-order problem 
\begin{equation}\label{WM_leadOrd}
\begin{split}
  \Delta U_0 = 0\,, \quad & \pmb{x} \in \Omega \setminus
  \{\pmb{x}_1, \ldots,\pmb{x}_m \}\,;\qquad
\partial_n U_0 = 0, \quad \pmb{x} \in \partial \Omega \,,
\end{split}
\end{equation}
which has the solution $U_0 \equiv U_0(t)$. The next order problem for
$U_1$ in the outer bulk region is 
\begin{equation}\label{WM_orderOne}
\begin{split}
  \Delta U_1 = U_0 +\tau  U_{0t} \,, \quad & \pmb{x} \in \Omega \setminus
  \{\pmb{x}_1, \ldots,\pmb{x}_m \}\,;\qquad
\partial_n U_1 = 0\,, \quad  \pmb{x} \in \partial \Omega\,,
\end{split}
\end{equation}
which must be supplemented by singularity conditions for $U_1$ as
$\pmb{x} \to \pmb{x}_j$, for each $j=1,\ldots,m$.

To determine these conditions, we consider the inner region defined
within an $\mathcal{O}(\varepsilon)$ neighborhood of the
$j^{\text{th}}$ cell. In this region, we introduce the inner variables
$\pmb{y} = \varepsilon^{-1} (\pmb{x} - \pmb{x}_j)$ and
$U(\pmb{x}, t ) = \mathcal{U} (\pmb{x}_j + \varepsilon \pmb{y}, t)$,
with $\rho = |\pmb{y}|$.  Upon writing \eqref{DimLess_bulka} and
\eqref{DimLess_bulkb} in terms of the inner variables, we obtain
\begin{align}\label{WM_dimlessInner}
\tau \, \mathcal{U}_{t} & = \frac{D}{\varepsilon^2}\, \Delta_{
  \pmb{y}}\, \mathcal{U} -  \, \mathcal{U}\,,  \quad 1 <  \rho  < \infty\,;
                          \qquad
 D\, \partial_{\rho} \, \mathcal{U}  = d_{1,j} \, \mathcal{U} - d_{2,j} \,
   u_j^1\,, \,\,\, \text{on} \,\,\, \rho = 1\,, \qquad j = 1, \ldots, m\,.
\end{align}
Since $D \gg \mathcal{O}(1)$, we expand the inner solution
$\mathcal{U}(\rho,t) $ in powers of $D$ as 
\begin{equation}\label{WM_UexpndInner}
\begin{split}
  \mathcal{U}(\rho,t) = \mathcal{U}_0(\rho,t) + \frac{1}{D}\,
  \mathcal{U}_1(\rho,t) + \cdots\,.
\end{split}
\end{equation}
Upon substituting \eqref{WM_UexpndInner} into \eqref{WM_dimlessInner}, and
collecting terms in powers of $D$, we derive the leading-order inner
problem 
\begin{equation}\label{WM_leadOrdInner}
\begin{split}
  \Delta_{\rho} \, \mathcal{U}_0 \equiv \frac{1}{\rho} \frac{\partial
    }{\partial \rho} \left( \rho \frac{\partial {\mathcal U}_0}{\partial\rho}
  \right) = 0\,, & \quad 1< \rho <\infty\,;\qquad \partial_\rho \,
  \mathcal{U}_0 = 0\,, \quad \text{on} \quad \rho = 1\,,
\end{split}
\end{equation}
near the $j^{\text{th}}$ cell. The solution to this problem which
matches the outer solution as $\rho\to\infty$ is simply
$\mathcal{U} _0 \equiv U_0(t)$.

We proceed to consider the next order problem in the inner region
given by
\begin{equation}\label{WM_NextOrd_Inner2}
\begin{split}
\Delta_{\rho}\, \mathcal{U}_1 &= 0\,,  \quad 1< \rho  < \infty\,;\qquad
D\, \partial_{\rho} \, \mathcal{U}_1   =  \,d_{1,j} \, \mathcal{U}_0 -
d_{2,j} \, u_j^1\,, \quad \text{on} \quad \rho = 1\,,\qquad j = 1, \ldots, m\,.
\end{split}
\end{equation}
Allowing logarithmic growth as $\rho \to \infty$, the radially
symmetric solution to this problem is 
\begin{equation}\label{WM_UniqSol}
\begin{split}
  \mathcal{U}_{1,j} = (d_{1,j}\, \mathcal{U}_0 - d_{2,j} \,u_j^1) \log\rho + c_{j}
  \,,\qquad j = 1, \ldots, m\,,
\end{split}
\end{equation}
where $c_j$ for $j = 1, \ldots, m$ are unknown constants to be
determined.  We write this solution in terms of the outer
variables and substitute the resulting expression into the
inner expansion \eqref{WM_UexpndInner}. For $\rho \to \infty$, this
yields that
\begin{equation}\label{WM_UniqExp}
\begin{split}
  \mathcal{U}  \sim U_0 + \frac{1}{D} \left[ (d_{1,j}\, U_0 - d_{2,j} u_j^1) \,
    \log|\pmb{x} - \pmb{x}_j |  + \frac{1}{\nu} (d_{1,j}\, U_0 - d_{2,j} u_j^1)
    \,  + c_{j} \right] + \cdots,\quad j = 1, \ldots, m\,.
\end{split}
\end{equation}
Since $D \gg \mathcal{O}(\nu^{-1})$, where
$\nu = {-1/\log\varepsilon}$, we have $1/{D\nu} \ll 1$, which implies
that $U_0$ is the largest term in the far-field behavior
\eqref{WM_UniqExp} of the inner solution. This estimate establishes
the range of $D$ for which our analysis is valid. Upon matching the
far-field behaviour of the inner solution given in \eqref{WM_UniqExp}
to the outer solution, we obtain the singularity behaviour of the
outer solution $U_1$ as $\pmb{x} \to \pmb{x}_j$. In this way, the
complete outer problem for $U_1$ is given by
\begin{subequations}\label{WM_ComplorderOne}
\begin{align}
  \Delta U_1 &= U_0 +\tau  U_{0t} \,, \quad  \pmb{x} \in
        \Omega \setminus \{\pmb{x}_1, \dots,\pmb{x}_m \}\,;\qquad
               \partial_n U_1 = 0\,, \quad  \pmb{x} \in \partial \Omega\,;
               \label{WM_PDE} \\ \vspace{0.5cm}
  U_1  & \sim (d_{1,j}\,  U_0 - d_{2,j}\, u_j^1) \, \log |\pmb{x} - \pmb{x}_j |\,,
         \quad \text{as} \quad \pmb{x} \to \pmb{x}_j\,, \quad
         j = 1, \ldots, m\,. \label{WM_Match1}
\end{align}
\end{subequations}
By representing the singularity behaviour for $U_1$ given in
\eqref{WM_Match1} in terms of a delta function for the PDE 
\eqref{WM_PDE}, the problem for $U_1$ is equivalently written as
\begin{equation}\label{WM_PDEdelta}
\begin{split}
  \Delta U_1  &= (\tau  U_{0t} + U_0) + 2\pi \sum_{j=1}^{m} (d_{1,j}\, U_0 -
  d_{2,j}\, u_j^1) \,\delta(\pmb{x} - \pmb{x}_j)\,,  \quad
  \pmb{x} \in \Omega\,;\qquad
\partial_n U_1 = 0\,, \quad  \pmb{x} \in \partial \Omega\,.
\end{split}
\end{equation}
Upon integrating \eqref{WM_PDEdelta} over the domain $\Omega$ and then
applying the divergence theorem, we obtain
\begin{equation}\label{OuterSolution}
\begin{split}
  U_{0}^{\prime} = - \frac{1}{\tau} U_0  -
  \frac{ 2\pi}{\tau|\Omega|}   \sum_{j=1}^{m}  (d_{1,j}\, U_0 - d_{2,j} u_j^1)\,,
\end{split}
\end{equation}
where $|\Omega|$ is the area of $\Omega$. The chemical concentration
in the bulk region is described at leading-order by this ODE, where
the first term on the right hand side of the equation represents the
decay of chemical in the bulk region, and the summation averages the
net flux of chemical through the boundaries of all the signaling
compartments.

To represent the solution to \eqref{WM_PDEdelta} we introduce the
unique Neumann Green's function $G(\pmb{x};\pmb{x}_j)$, which satisfies
\begin{subequations}\label{WM_NUEGREEN}
\begin{align}
  \Delta G &= \frac{1}{|\Omega|} - \delta(\pmb{x} - \pmb{x}_j)\,, \quad
             \pmb{x} \in \Omega\,;\qquad \partial_n G  = 0\,, \quad  \pmb{x}
             \in \partial \Omega\,;\label{WM_NUEGREENA}\\
    G(\pmb{x};\pmb{x}_j) \sim - &\frac{1}{2\pi} \log|\pmb{x} - \pmb{x}_j| +
 R_j\,, \quad \text{as} \quad \pmb{x} \to \pmb{x}_j\,, \quad \text{and}
  \quad \int_{\Omega} \, G \, \text{d}\pmb{x} =0 \,, \label{WM_NUEGREENB}
\end{align}
\end{subequations}
where $R_j \equiv R(\pmb{x}_j)$ is its regular part at
$\pmb{x} = \pmb{x}_j$ for $j = 1, \ldots, m$. Without loss of
generality, we impose $\int_{\Omega} U_1 \, \text{d}\pmb{x}= 0$, so
that the spatial average of $U$ in the bulk region to order
${\mathcal O}(\nu)$ is simply $U_0$. Then, the solution to
\eqref{WM_PDEdelta} is 
\begin{equation}\label{WM_U1}
  U_1 = -2\pi \sum_{i=1}^{m} \, (d_{1,i}\, U_0 - d_{2,i} u_i^1)\,
  G(\pmb{x};\pmb{x}_i)\,.
\end{equation}
Upon substituting \eqref{WM_U1} into the outer expansion
\eqref{WM_Uexpnd}, we obtain the two-term asymptotic expansion 
\begin{equation}\label{WM_UOuter}
\begin{split}
  U \sim U_0 - \frac{2\pi}{D}   \sum_{i=1}^{m} \,
  (d_{1,i}\, U_0 - d_{2,i}\,u_i^1)\, G(\pmb{x};\pmb{x}_i) + \cdots \,,
\end{split}
\end{equation}
for the outer solution in the bulk region.  Then, we expand
\eqref{WM_U1} as $\pmb{x} \to \pmb{x}_j$ and use the singularity
behaviour of the Neumann Green's function given in
\eqref{WM_NUEGREENB} to determine the constants $c_j$ for
$j=1,\ldots,m$ in \eqref{WM_UniqSol}. Upon matching the local behavior
of \eqref{WM_UOuter} to \eqref{WM_UniqExp}, we obtain that
\begin{equation}\label{WM_C}
\begin{split}
  c_j = - \left( \frac{1}{\nu} + 2\pi\, R_j \right) (d_{1,j}\,  U_0 -
  d_{2,j} u_j^1)\,  - 2\pi \sum_{i \neq j}^{m} \,
  (d_{1,i}\, U_0 - d_{2,i} u_i^1)\, G(\pmb{x}_j,\pmb{x}_i)\,, \qquad j=1,\ldots,m
  \,.
\end{split}
\end{equation}
In this way, \eqref{WM_UniqExp} and \eqref{WM_C} provide a two-term
asymptotic expansion of the inner solution $\mathcal{U}(\rho,t)$, which
is valid within an $\mathcal{O} (\varepsilon)$ neigbourhood of the
$j^{\text{th}}$ cell. This inner solution is given by
\begin{equation}\label{WM_UniqExp2}
\begin{split}
  \mathcal{U}  = U_0 + \frac{1}{D} \left[ (d_{1,j}\, U_0 - d_{2,j} u_j^1) \,
    \log \rho  + \frac{1}{\nu} (d_{1,j}\, U_0 - d_2 u_j^1) \,  + c_{j} \right]
  + \cdots \,,\qquad j = 1, \ldots, m\,.
\end{split}
\end{equation}
Lastly, we substitute the leading-order inner solution $U_0(t)$ into
\eqref{DimLess_Intra} and evaluate the integral over the boundary of
the $j^{\text{th}}$ cell. This provides the global coupling of the
intracellular dynamics to the bulk solution.

In summary, in the well-mixed limit $D\gg {\mathcal O}(\nu^{-1})$, where
$\nu\equiv {-1/\log\varepsilon}$ and $\varepsilon\ll 1 $, the coupled
PDE-ODE model \eqref{DimLess_bulk} reduces to the following $nm+1$
dimensional nonlinear ODE system where the intracellular dynamics for
$\pmb{u}_j$ are globally coupled through the spatially uniform bulk
diffusion field $U_0(t)$:
\begin{equation}\label{WM_ODEsys_2D}
\frac{dU_0}{dt}  = -  \frac{U_0}{\tau}   -\frac{ 2\pi}{\tau |\Omega|}
\sum_{j=1}^{m}  (d_{1,j}\, U_0 - d_{2,j} u_j^1)\,; \qquad
\frac{d \pmb{u}_j}{dt}   =  \, \pmb{F}_j  \left( \pmb{u}_j  \right)  +
\frac{2\pi \pmb{e}_1}{\tau}\, ( d_{1,j} \, U_0 - d_{2,j} \, u_j^1 )\,,
\quad j = 1, \ldots, m\,.
\end{equation}
Here $\pmb{e}_1 = (1,0,..., 0)^T$, $m$ is the total number of
signaling compartments/cells, $|\Omega|$ is the area of $\Omega$,
$\pmb{u}_j = (u_j^1, \dots, u_j^n)^T$ is an $n$-dimensional vector
whose entries are the $n$ chemical species in the $j^{\text{th}}$
cell, and $\pmb{F} _j(\pmb{u}_j)$ is the vector-field describing the
local reaction kinetics within the $j^{\text{th}}$ cell.

\section{Global bifurcation diagrams} \label{GlobalBifur}

In this section, we compute global bifurcation diagrams for
\eqref{WM_ODEsys_2D} using the numerical bifurcation software XPPAUT
\cite{xpp2002} for the case where the intracellular species interact
according to Sel'kov reaction kinetics. Sel'kov kinetics provides a
two-component system of ODEs, which was originally used to qualitatively
model oscillations in the glycolytic pathway (cf.~\cite{Selkov}). For
convenience of notation, we write the intracellular variable as
$\mathbf{u} = (v , w)^T$, and the local Sel'kov reaction kinetics
$\pmb{F}(\mathbf{u})$ as $\pmb{F} (v, w) = (F (v, w), G(v, w))^T$,
where
\begin{equation}\label{Selkov}
\begin{split}
  F (v, w)\equiv \alpha w + wv^2 -v\,, \qquad G(v, w) \equiv
  \epsilon_0( \mu - (\alpha w + w v^2))\,.
\end{split}
\end{equation}
Here $\alpha > 0$, $\mu > 0$ and $\epsilon_0>0$ are the Sel'kov reaction
parameters. The steady-state for Sel'kov dynamics ${dv/dt}=F(v,w)$ and
${dw/dt}=G(v,w)$ is $v_e=\mu$ and $w_e={\mu/(\alpha+\mu^2)}$, while the
trace and determinant of the Jacobian matrix $J_e$ for this steady-state
is readily calculated as
\begin{equation}\label{selkov:trace}
  \det(J_e)=\epsilon_0 (\alpha + \mu^2) \,, \qquad
  \text{tr}(J_e)=2w_ev_e -1 -\det(J_e) = \frac{2\mu^2}{\alpha+\mu^2}
  - 1 - \epsilon_0(\alpha+\mu^2) \,.
\end{equation}
Since $\det(J_e)>0$, we conclude for the Sel'kov reaction kinetics
that a single isolated cell, decoupled from the bulk medium, has a
steady-state that is linearly stable if and only if
$\text{tr}(J_e)<0$. The Hopf bifurcation boundary in the $\alpha$
versus $\mu$ parameter plane, as obtained by setting
$\text{tr}(J_e)=0$ is given by
\begin{equation}\label{selkov:boundary}
  \alpha = -\mu^2 -\frac{1}{2\epsilon_0} + \frac{\sqrt{1+8\epsilon_0\mu^2}}
  {2\epsilon_0} \,.
\end{equation}
By constructing a trapping region and appealing to the
Poincare-Bendixson theorem, it is well-known that when the
steady-state is unstable there is a time-periodic solution,
characterized by a limit cycle, for the isolated cell.  In contrast,
when the steady-state is linearly stable there are no time-periodic
oscillations with Sel'kov kinetics.

\begin{figure}[htbp]
\begin{center}
\includegraphics[height=4.3cm,width=0.40 \textwidth]{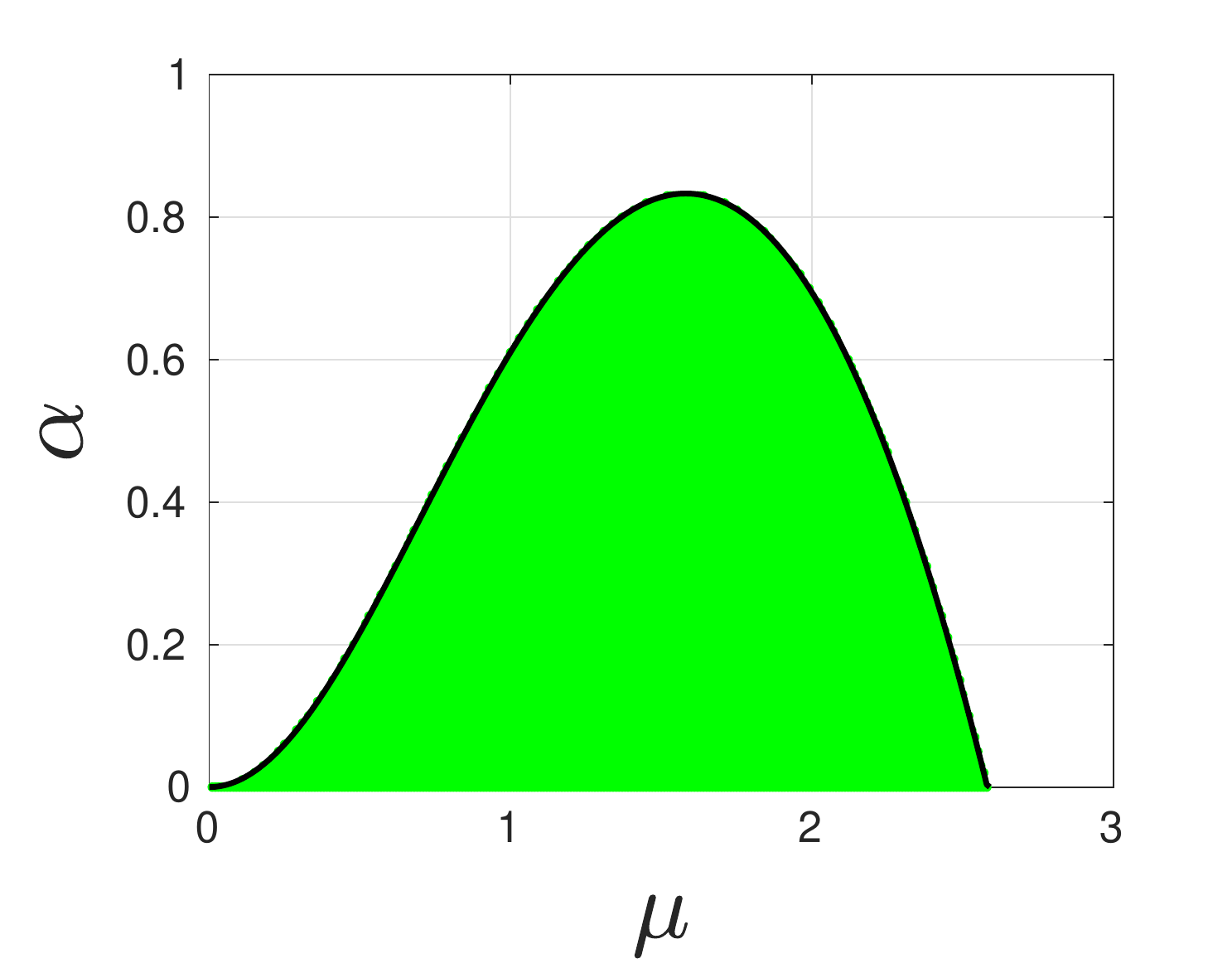}
\end{center}
\caption{The region of instability (green-shaded) in the $\alpha$
  versus $\mu$ parameter plane for the steady-state of a single
  isolated cell, decoupled from the bulk region, that undergoes
  Sel'kov kinetics \eqref{Selkov} with $\epsilon_0=0.15$. Within this
  region, where $\text{tr}(J_e)>0$ in \eqref{selkov:trace}, there is a
  time-periodic solution (limit cycle) for an isolated cell. On the
  boundary of this region, given by \eqref{selkov:boundary}, is where
  Hopf bifurcations occur.  In the unshaded region the steady-state is
  linearly stable.}
\label{fig:selkov}
\end{figure}

In our study of the ODE system \eqref{WM_ODEsys_2D}, the Sel'kov
kinetic parameters are chosen so that the steady-state of the
intracellular dynamics within each cell is linearly stable, and hence
no intracellular oscillations occur, when there is no coupling to the
bulk diffusion field. When $\epsilon_0=0.15$, this corresponds to
choosing a pair $(\mu,\alpha)$ not within the green-shaded region in
Figure \ref{fig:selkov}. Our goal is to investigate intracellular
oscillations that are triggered as a result of the global coupling
through the bulk medium. Our results below are presented in the form
of one- and two-parameter global bifurcation diagrams computed from
\eqref{WM_ODEsys_2D} using XPPAUT \cite{xpp2002}. When the cells are
all identical, the bifurcation parameters are common to all of
them. However, for the cases involving a defective cell, the
bifurcation parameter is that of the defective cell only, which
is identified as the cell with index $j=1$.
 
\subsubsection{Instability triggered by the rate of influx and efflux of the bulk chemical}\label{Perm_2D}

We begin by studying the effect of the permeability parameters $d_{1}$
and $d_{2}$ on the intracellular dynamics. Recall that $d_{1,j}$ is
the rate of chemical feedback (influx) from the bulk into the
$j^{\text{th}}$ cell, while $d_{2,j}$ is the rate of secretion of
chemical (efflux) by the cell into the bulk region. Our goal below is
to investigate synchronous oscillations in the cells that are
triggered by these parameters. One particular focus is to study how
the intracellular dynamics of a collection of identical cells is
altered due to changes in either the efflux or influx rate of {\em one
  specific cell}. The numerical bifurcation diagrams shown below
exhibit how the permeabilities of such a single ``defective'' cell can
significantly alter the parameter region where collective
intracellular oscillations of the entire group of cells can occur.

\begin{figure}[htbp]
    \centering
    \begin{subfigure}[t]{0.22\textwidth}
        \includegraphics[width=\textwidth,valign=t]{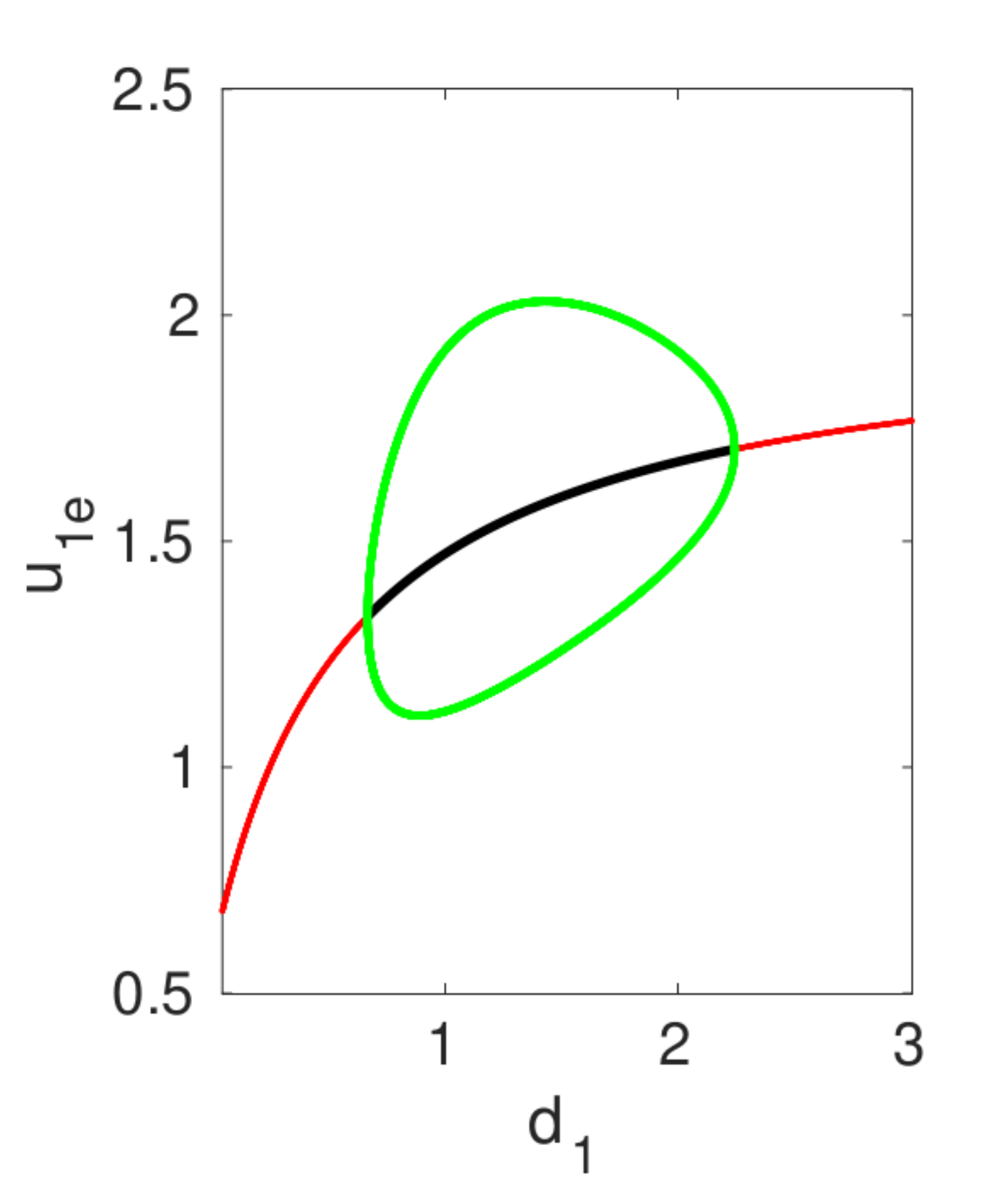}
        \caption{Three identical cells.}
        \label{Bifur_d1A}
    \end{subfigure}
    \quad
    \begin{subfigure}[t]{0.22\textwidth}
        \includegraphics[width=\textwidth,valign=t]{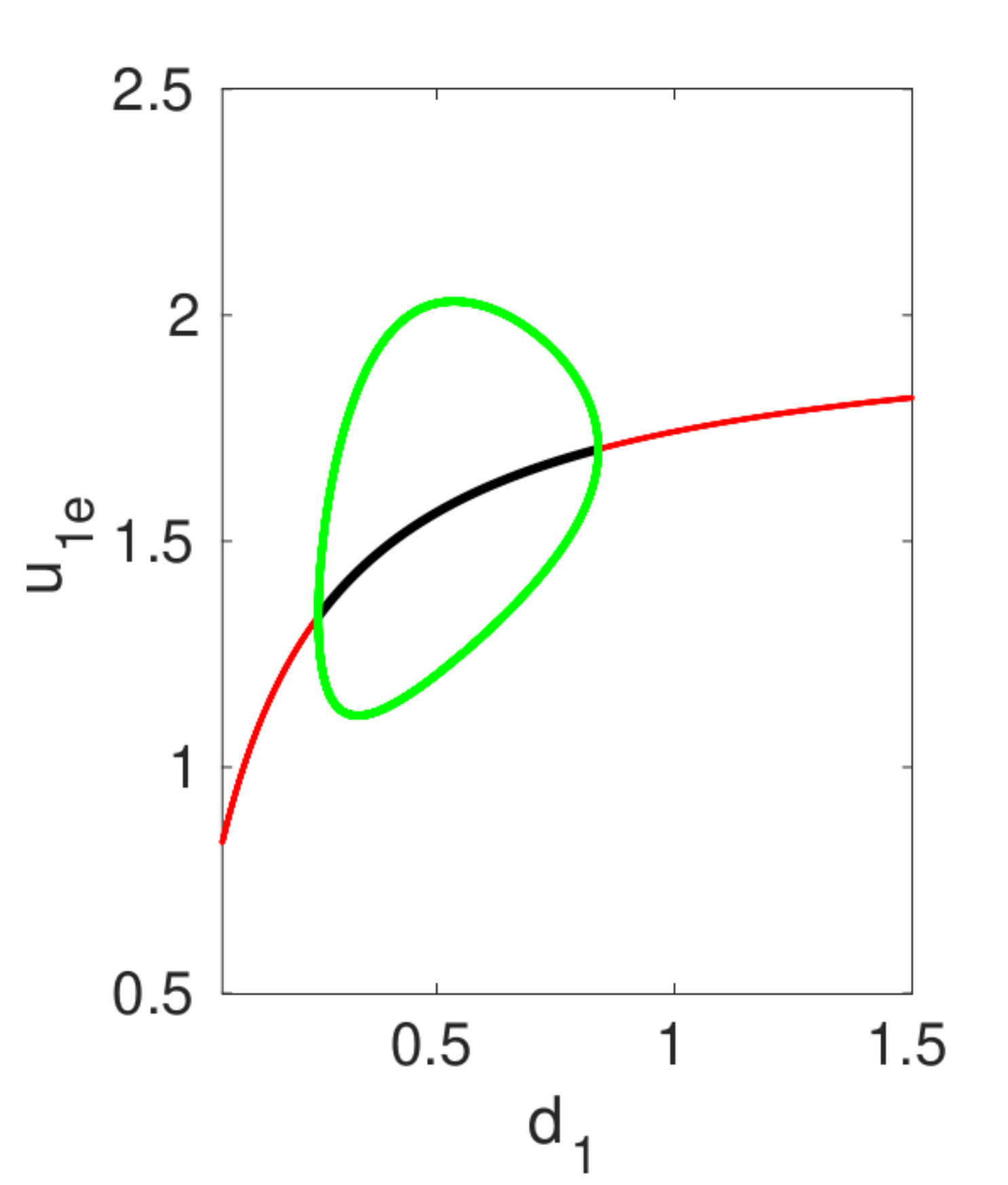}
        \caption{Eight identical cells}
        \label{Bifur_d1B}
    \end{subfigure}
    \quad
    \begin{subfigure}[t]{0.22\textwidth}
        \includegraphics[width=\textwidth,valign=t]{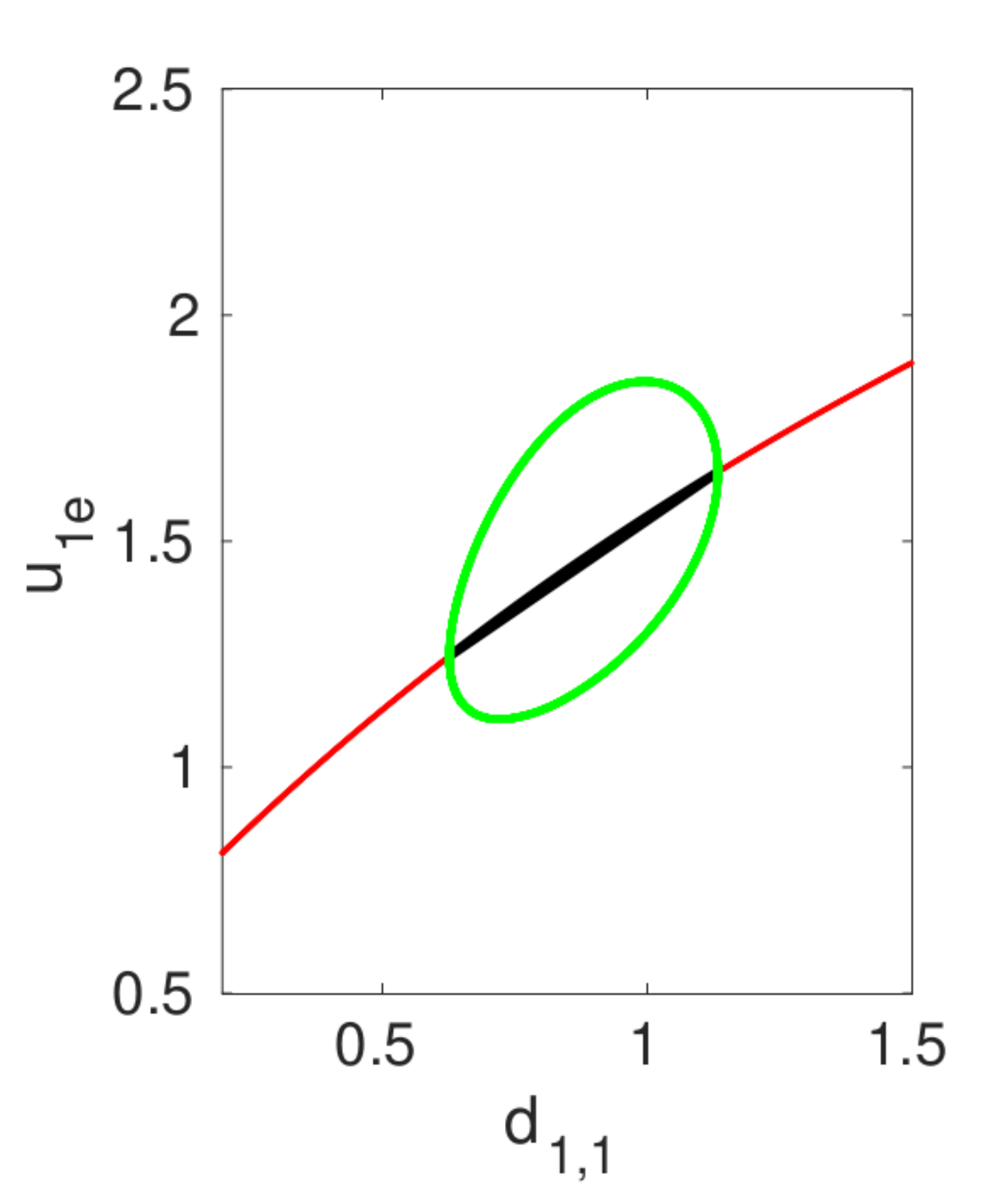}
        \caption{Two identical cells  and a defective cell.}
        \label{Bifur_d1_3cells}
    \end{subfigure}
    \quad
    \begin{subfigure}[t]{0.22\textwidth}
        \includegraphics[width=\textwidth,valign=t]{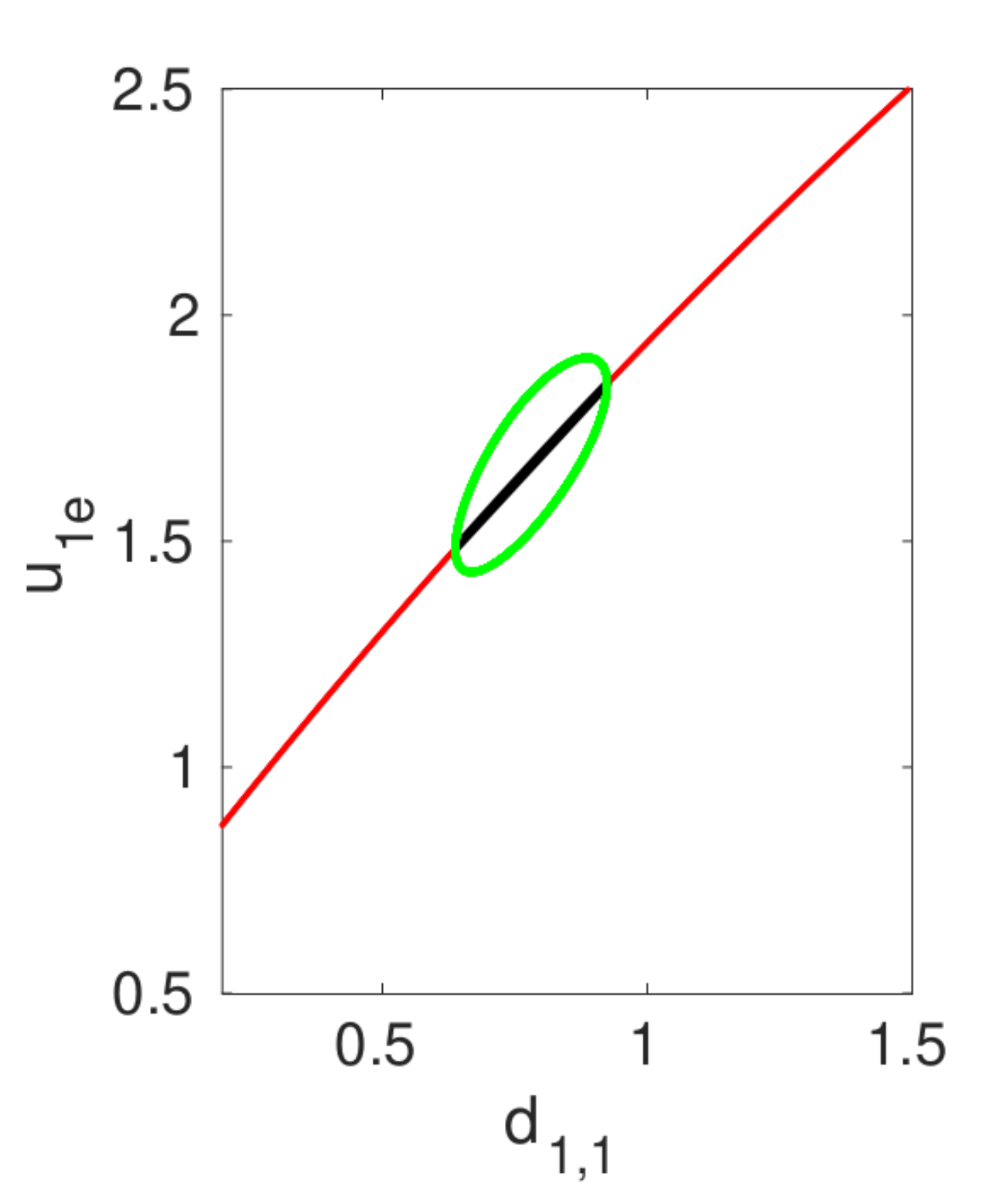}
        \caption{Seven identical cells and a defective cell}
        \label{Bifur_d1_8cells}
    \end{subfigure}
    \caption{Global bifurcation diagrams for $u_{1e}$ versus $d_{1}$,
      as computed from the ODEs \eqref{WM_ODEsys_2D}, showing
      steady-states and global branches of periodic solutions. The
      thin red lines and bold black lines correspond to linearly
      stable and unstable steady-state solution branches,
      respectively. The green loops indicate linearly stable branches
      of periodic solutions. The parameters are $\tau=0.5$,
      $ \alpha =0.9$, $\epsilon_0 = 0.15$, and $\mu = 2$, with
      $d_{2,j} \equiv d_{2}= 0.2$ for $j=1,\ldots,m$ and
      $d_{1,j} = 0.8$ for $j=2,\dots,m$. (a) and (b) are for $m=3$ and
      $m=8$ identical cells, with the range of $d_{1}$ yielding
      synchronous oscillations is computed as
      $ 0.7613 \leq d_{1} \leq 2.241$ and
      $ 0.2517 \leq d_{1} \leq 0.8403$, respectively. (c) is for $m=3$
      cells, where two of cells are identical with the remaining one
      defective. The Hopf bifurcation points are $d_{1,1} = 0.6277$
      and $d_{1,1} = 1.134$. (d) is for $m=8$ cells (seven identical
      and a defective), with Hopf bifurcations occurring at
      $d_{1,1} = 0.639$ and $d_{1,1} = 0.9248$. In (c) and (d), the
      bifurcation parameter is that of the defective cell only.
    }\label{Bifur_d1}
  \end{figure}

In Figures~\ref{Bifur_d1A} and \ref{Bifur_d1B} global branches of
steady-state and periodic solutions for \eqref{WM_ODEsys_2D} are shown
as the permeability parameter $d_{1}$ for $m=3$ and $m=8$ identical
cells, respectively, is varied. In contrast, in
Figures~\ref{Bifur_d1_3cells} and \ref{Bifur_d1_8cells} we plot
similar bifurcation diagrams but for the case where a single defective
cell, having a modified permeability parameter $d_{1,1}$, is coupled
to the remaining identical cells of Figures~\ref{Bifur_d1A} and
\ref{Bifur_d1B}. In both sets of figures the parameters used are
$\alpha =0.9$, $ \epsilon_0 = 0.15$, $\mu = 2$, $\tau=0.5$ and
$d_{2} = 0.2$. In Figures~\ref{Bifur_d1_3cells} and
\ref{Bifur_d1_8cells}, $d_{1} = 0.8$ is used for the identical cells.
We observe from these figures that the range of $d_{1}$ for which
periodic solutions occur for \eqref{WM_ODEsys_2D} is bounded for this
parameter set. Since $d_{1}$ is the rate of chemical feedback into the
cells, a small value of $d_{1}$ implies little feedback of the bulk
chemical into the cells and, hence, a weak coupling between the
cells. As a result, since each cell has a linearly stable steady-state
and no time-periodic oscillations when uncoupled from the bulk, one
would expect no synchronous oscillations when $d_1$ is too small. This
intuition is confirmed in
Figures~\ref{Bifur_d1A}--\ref{Bifur_d1B}. Alternatively, when $d_{1}$
is large relative to $d_{2}$ (rate of secretion of chemical into the
extracellular medium), the secreted chemical is almost immediately fed
back into the cells, which leads to a lower chemical concentration in
the bulk region. As a result, the coupling between the cells is again
weak, and one would expect that the steady-state is linearly stable
and that no intracellular oscillations occur. This intuition is again
confirmed in Figures~\ref{Bifur_d1A}--\ref{Bifur_d1B}. Later in our
numerical study (see Figure \ref{Bifur_2ParA} below), we use
two-parameter bifurcation diagrams to show that some specific ratios
of $d_1$ and $d_2$ are required in order to trigger oscillatory
instabilities of the steady-state. For $m=3$ identical cells (see
Figure~\ref{Bifur_d1A}), the range of $d_{1}$ that yields synchronous
oscillations is $ 0.7613 \leq d_{1} \leq 2.241$, while for $m=8$ (see
Figure~\ref{Bifur_d1B}) that range is
$ 0.2517 \leq d_{1} \leq 0.8403$.  This indicates that the range of
$d_{1}$ for which stable periodic solutions occur shrinks as the
number of cells increases, and as a result it becomes increasingly
difficult to synchronize the dynamics of the cells using the
permeability parameter $d_{1}$ when $d_2$ is
fixed. Figure~\ref{Bifur_d1_3cells} shows the result for two identical
cells coupled to a defective cell, with intracellular oscillations
occurring for $ 0.6277 \leq d_{1,1} \leq 1.134$. In
Figure~\ref{Bifur_d1_8cells}, where seven identical cells are coupled
to a defective cell, intracellular oscillations occur only when
$0.639 \leq d_{1,1} \leq 0.9248$. Similar to the case of identical
cells, the range of $d_{1,1}$ for which linearly stable periodic
solutions exist is larger when $m=3$ as compared to when $m=8$.
{Most importantly, however, Figures~\ref{Bifur_d1_3cells} and
  \ref{Bifur_d1_8cells} clearly show how rather small changes in the
  influx rate of a single defective cell, relative to the common value
  of the group, can act as a switch that quenches, or extinguishes,
  intracellular oscillations for the {\em entire collection} of
  cells.}

\begin{figure}[htbp]
    \centering
    \begin{subfigure}[t]{0.22\textwidth}
        \includegraphics[width=\textwidth,valign=t]{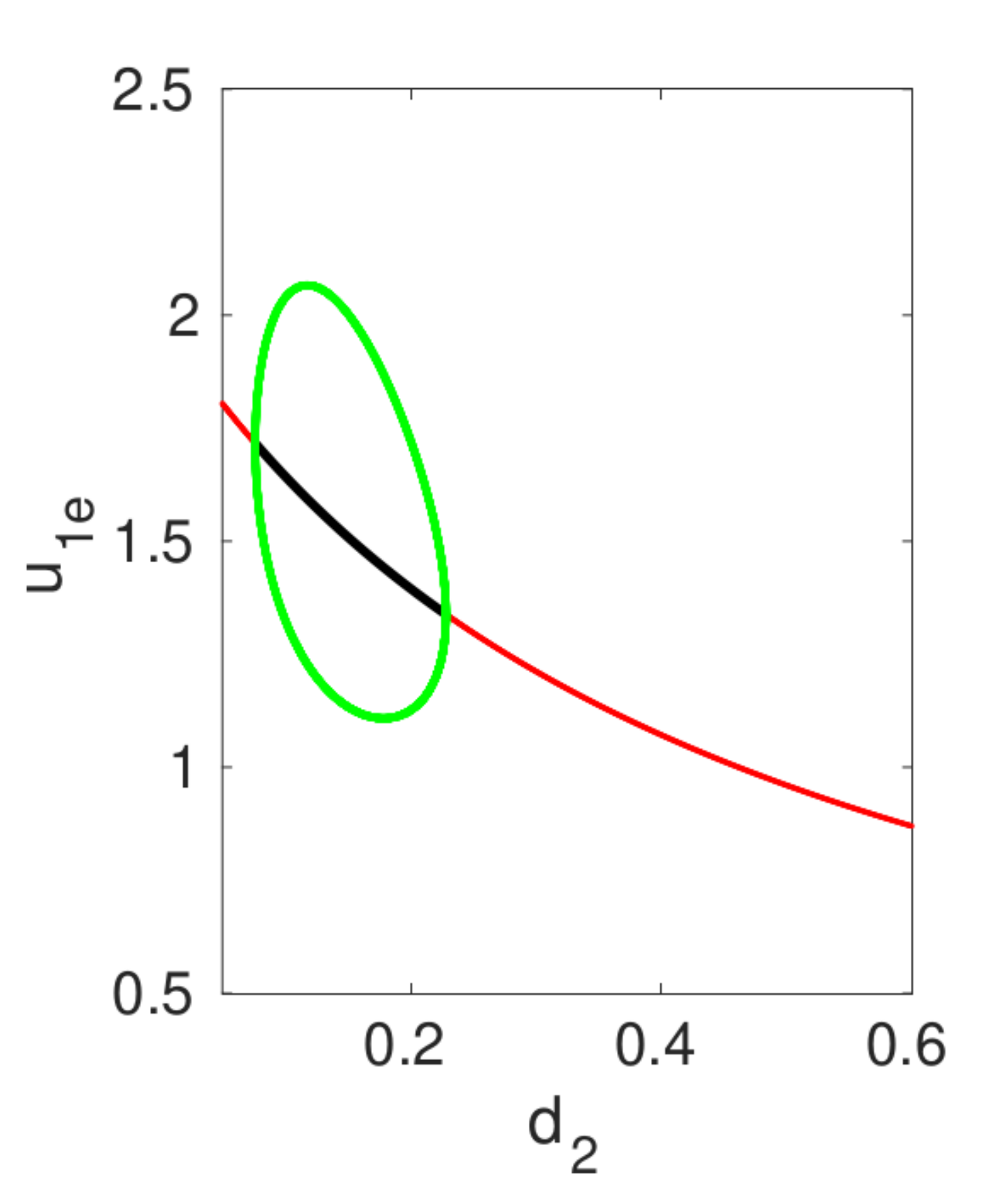}
        \caption{Three identical cells}
        \label{Bifur_d2_3cellsID}
    \end{subfigure}
    \quad 
    \begin{subfigure}[t]{0.23\textwidth}
        \includegraphics[width=\textwidth,valign=t]{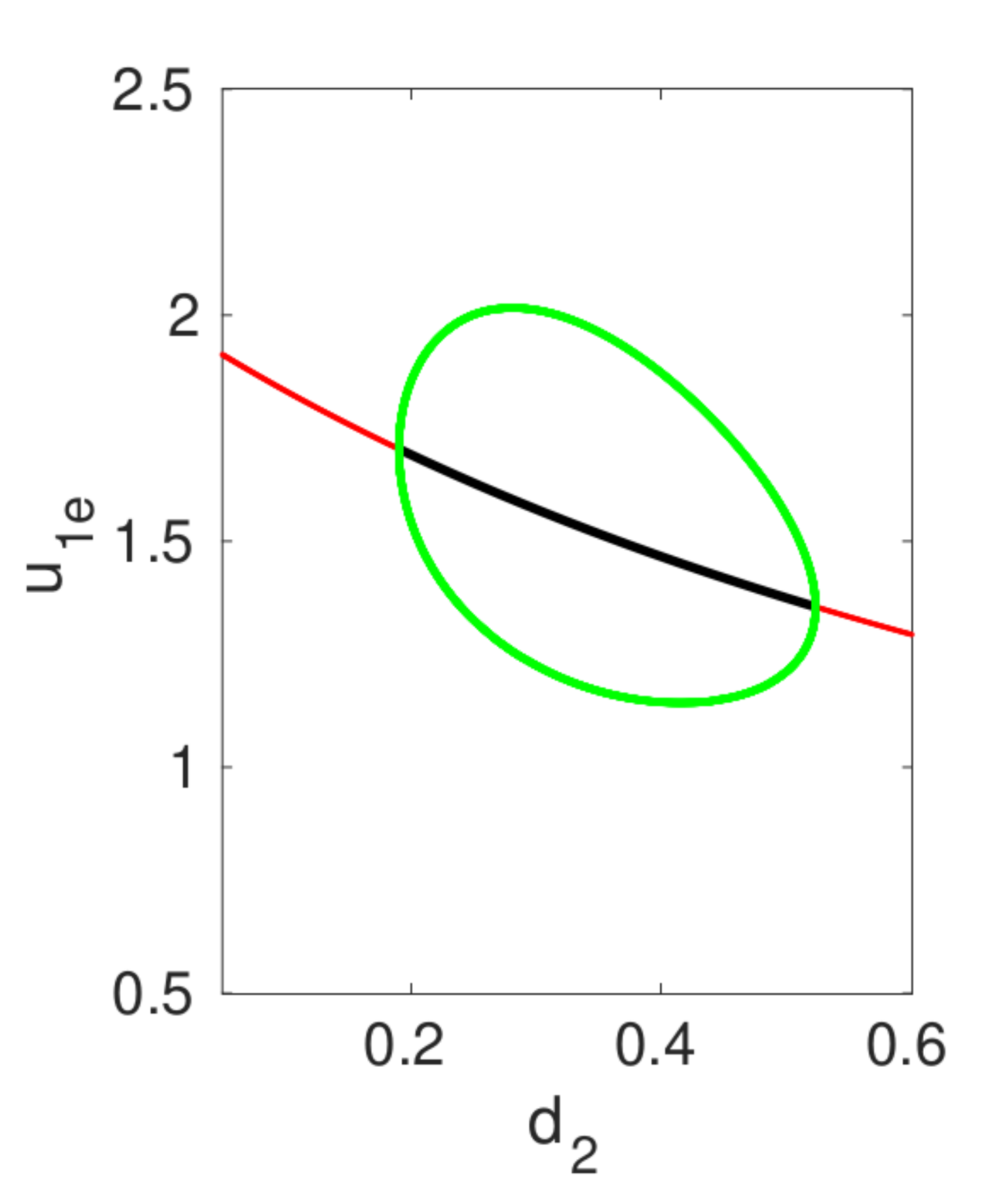}
        \caption{Eight identical cells}
        \label{Bifur_d2_8cellsID}
    \end{subfigure}
     \quad 
	\begin{subfigure}[t]{0.22\textwidth}
        \includegraphics[width=\textwidth,valign=t]{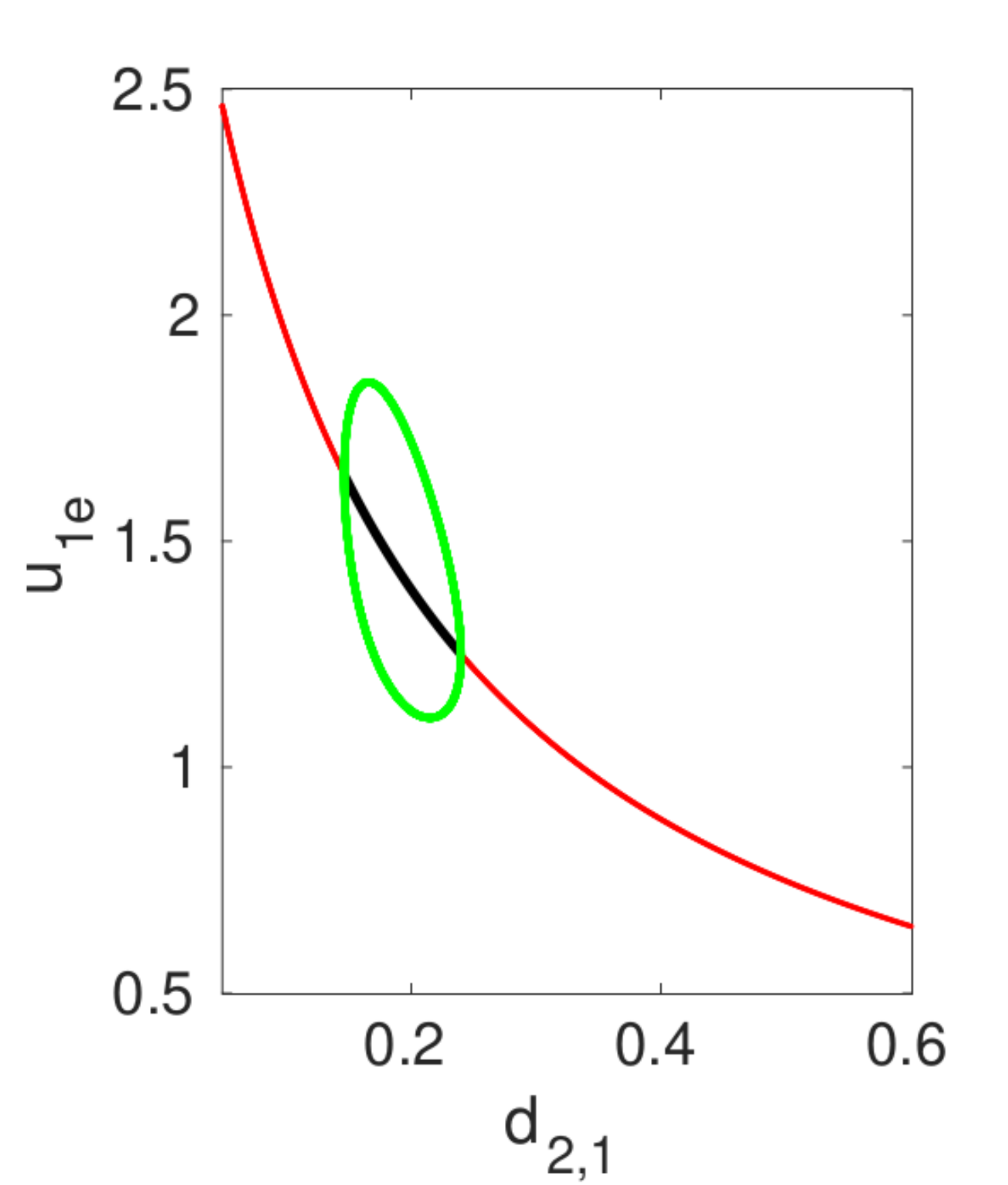}
        \caption{Two identical cells  and a defector cell.}
        \label{Bifur_d2_3cells}
    \end{subfigure}
    \quad 
    \begin{subfigure}[t]{0.22\textwidth}
        \includegraphics[width=\textwidth,valign=t]{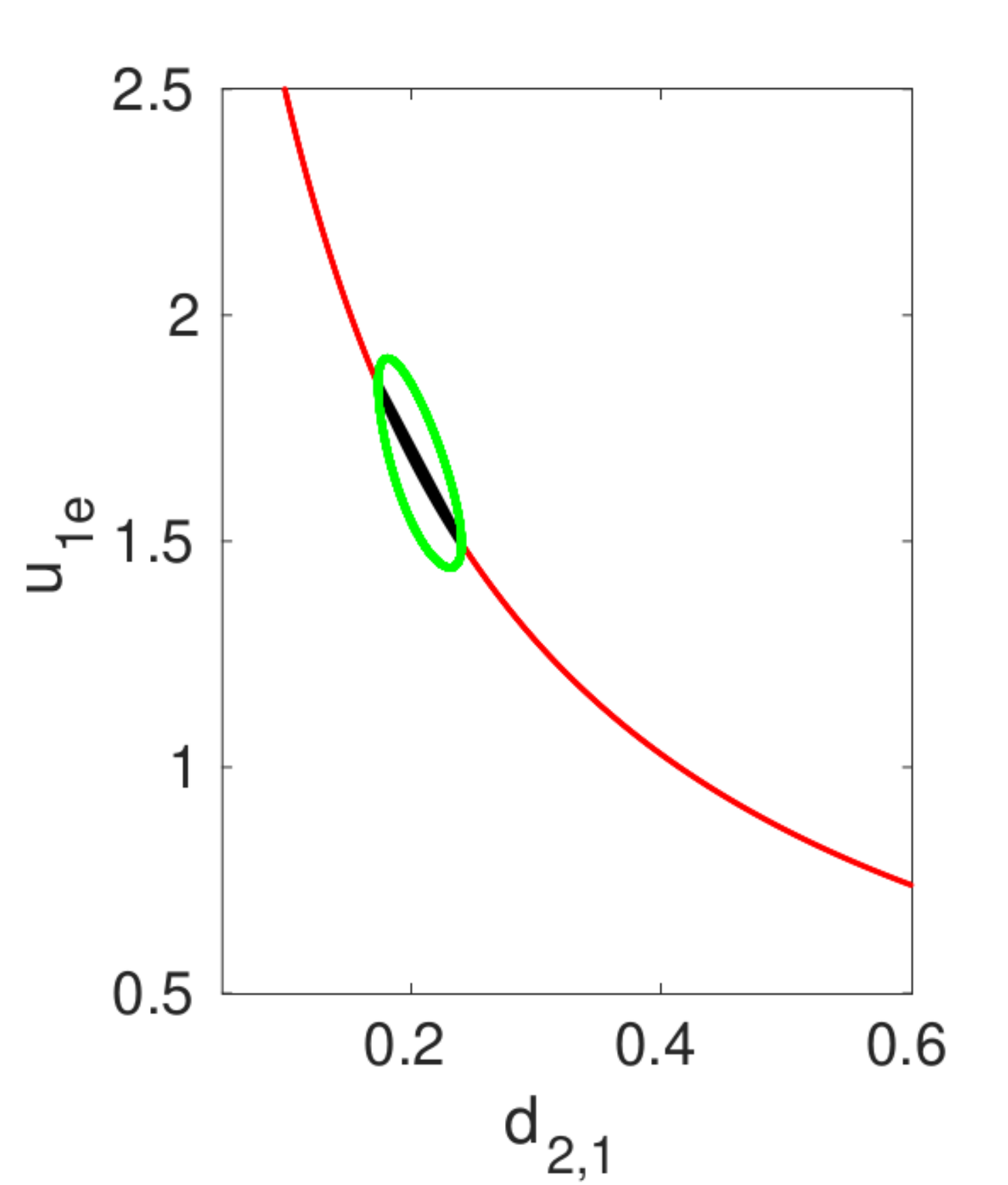}
        \caption{Seven identical cells  and a defector cell.}
        \label{Bifur_d2_8cells}
    \end{subfigure}
    \caption{Global bifurcation diagrams for $u_{1e}$ versus $d_{2}$,
      as computed from the ODEs \eqref{WM_ODEsys_2D}, showing
      steady-states and global branches of periodic solutions. The
      labeling of the curves is the same as in Figure \ref{Bifur_d1}.
      The parameters are $\tau=0.5$, $ \alpha =0.9$,
      $\epsilon_0 = 0.15$, and $\mu = 2$, with
      $d_{1,j} \equiv d_{1}= 0.8$ for $j=1,\ldots,m$ and
      $d_{2,j} = 0.2$ for $j=2,\dots,m$. (a) and (b) are for $m=3$ and
      $m=8$ identical cells, and the range of $d_{2}$ yielding
      synchronous oscillations is computed as
      $0.07624 \leq d_2 \leq 0.2283$ and
      $0.1908 \leq d_2 \leq 0.5229$, respectively. (c) is for $m=3$
      cells, where two of cells are identical, and the remaining one
      is defective. The Hopf bifurcation points are $d_{2,1} = 0.1471$
      and $d_{2,1} = 0.2402$. (d) is for $m=8$ cells (seven identical
      and one defective), with Hopf bifurcations occurring at
      $d_{2,1} = 0.1743$ and $d_{2,1} = 0.2407$. In (c) and (d), the
      bifurcation parameter is that of the defective cell
      only.} \label{Bifur_d2}
\end{figure}

  The global bifurcation diagrams where the permeability parameter
  $d_{2}$ (the efflux rate of secretion of chemicals into the bulk
  region by the cells) is the bifurcation parameter are shown in
  Figure~\ref{Bifur_d2}. The other parameter values are the same as
  those in Figure~\ref{Bifur_d1}.  Similar to the results for $d_1$ in
  Figure~\ref{Bifur_d1}, the range of $d_{2}$ for which stable
  periodic solutions occur is bounded above and below for all the
  scenarios considered. Linearly stable steady-state solutions exist
  for small values of $d_{2}$ because the cells do not secrete enough
  chemical required to trigger oscillations. On the other hand, when
  the rate of secretion of chemicals is large relative to the rate of
  feedback of chemicals into the cells, there is insufficient chemical
  concentration within the cells to trigger oscillations. For $m=3$
  and $m=8$ identical cells, we find that linearly stable synchronous
  periodic solutions occur on the ranges
  $0.07624 \leq d_2 \leq 0.2283$ and $0.1908 \leq d_2 \leq 0.5229$,
  respectively. However, unlike the results in Figures \ref{Bifur_d1A}
  and \ref{Bifur_d1B}, the range of $d_2$ that predicts synchronous
  oscillations becomes larger as the number of identical cells
  increases. For the case involving a defective cell, when $m=3$ (two
  identical cells and one defective cell) the Hopf bifurcation points
  are $d_{2,1} = 0.1471$ and $d_{2,1} = 0.2402$, while for $m=8$
  (seven identical and one defective cell) the Hopf bifurcation points
  are $d_{2,1} = 0.1743$ and $d_{2,1} = 0.2407$. Since the range of
  $d_{2,1}$ that predicts stable periodic solutions is smaller when
  $m=8$ relative to when $m=3$, this shows that when there is a
  defector cell it is more difficult to achieve synchronous
  oscillations as the overall population of cells increases. Comparing
  the identical cells cases (Figures \ref{Bifur_d2_3cellsID} and
  \ref{Bifur_d2_8cellsID}) with the cases involving a defective cell
  (Figures \ref{Bifur_d2_3cells} and \ref{Bifur_d2_8cells}), we
  observe that the range of $d_{2}$ that predicts linearly stable
  periodic solutions is larger, as expected, when the cells are
  identical. {This indicates that cell-cell communication is
    more difficult to achieve when a single cell is secreting chemical
    at a different rate as compared to the other cells. More
    specifically, we observe that intracellular oscillations of the
    group can be extinguished by a defective cell that reduces its
    efflux rate.}  From comparing the results in Figures \ref{Bifur_d1}
  and \ref{Bifur_d2}, we notice that the range of $d_{1}$ that
  synchronizes the intracellular dynamics of the cells is larger than
  that of $d_{2}$ for all the cases considered. This shows, for this
  parameter set, that it is easier to trigger communication among the
  cells using the rate of chemical feedback or influx from the bulk
  into the cells, rather than the rate of secretion of chemicals into
  the bulk.
  
\begin{figure}[htbp]
    \centering
    \begin{subfigure}[t]{0.22\textwidth}
        \includegraphics[width=\textwidth,valign=t]{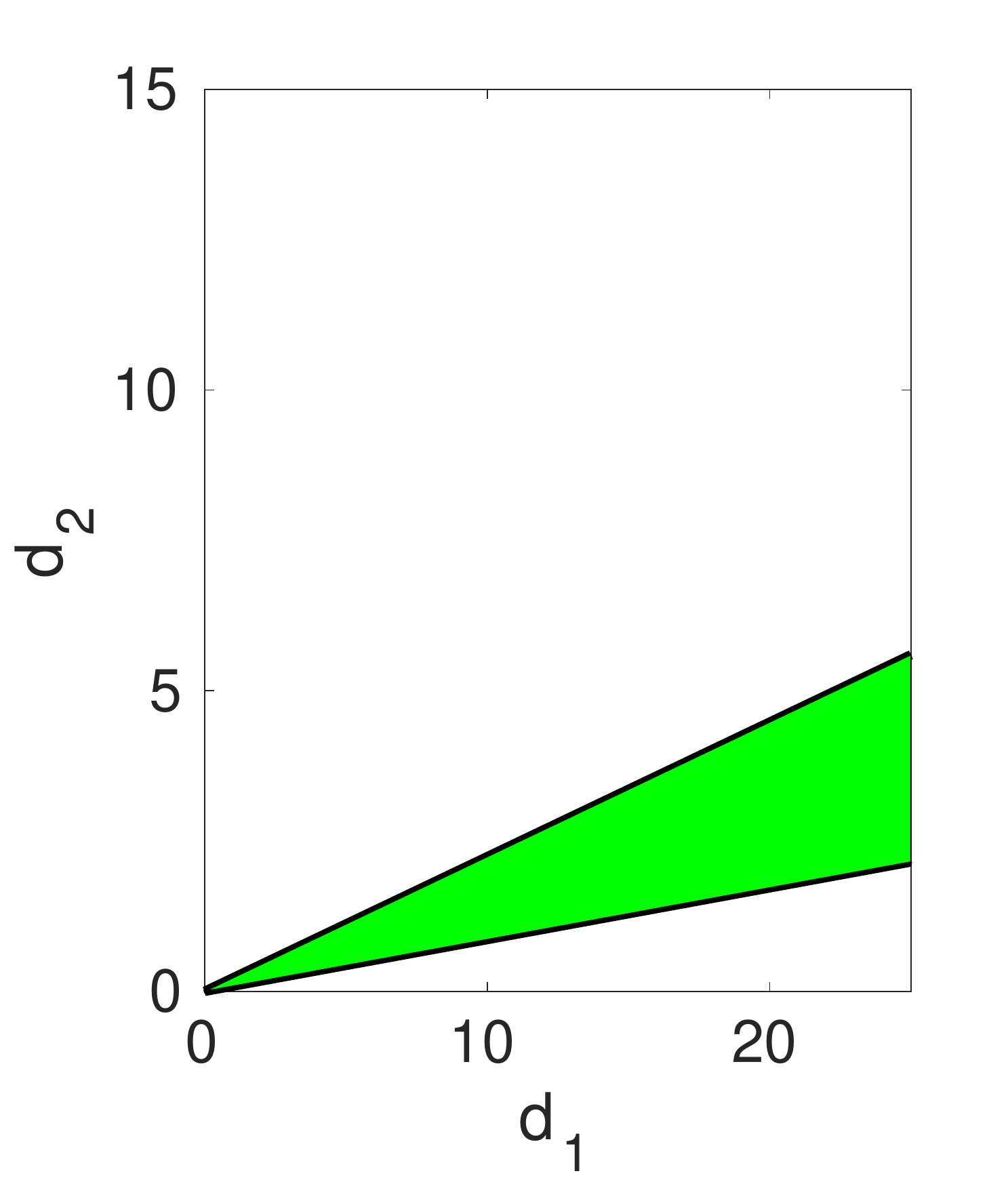}
        \caption{Three identical cells}
        \label{Bifur_2Par_Id_3cellsA}
    \end{subfigure}
    \quad 
    \begin{subfigure}[t]{0.22\textwidth}
        \includegraphics[width=\textwidth,valign=t]{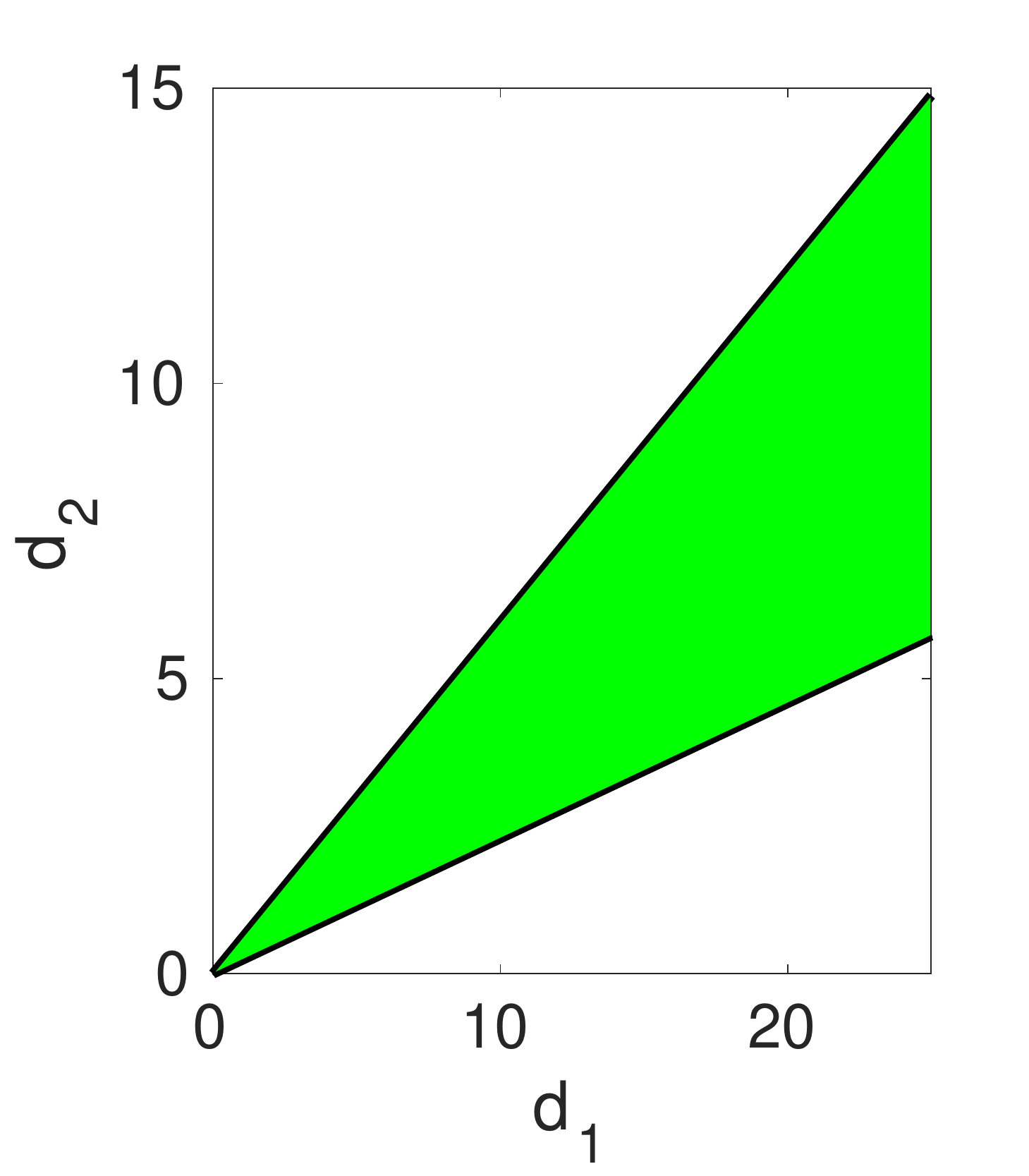}
        \caption{Eight identical cells}
        \label{Bifur_2Par_Id_8cellsA}
    \end{subfigure} 
    \quad
\begin{subfigure}[t]{0.22\textwidth}
        \includegraphics[width=\textwidth,valign=t]{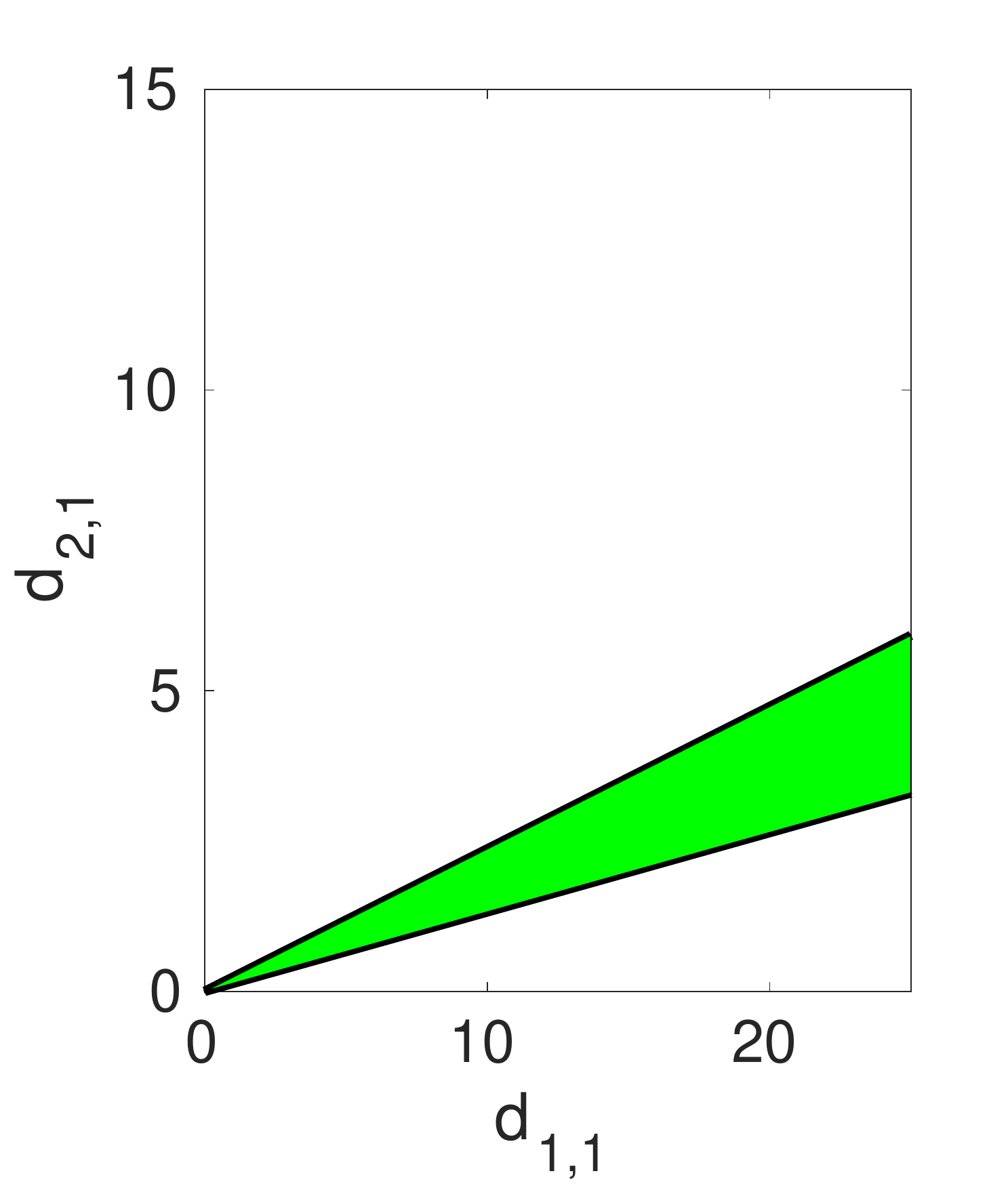}
        \caption{Two identical cells  and a defective cell}
        \label{Bifur_2Par_NonId_3cellsA}
    \end{subfigure}
    \quad
    \begin{subfigure}[t]{0.22\textwidth}
        \includegraphics[width=\textwidth,valign=t]{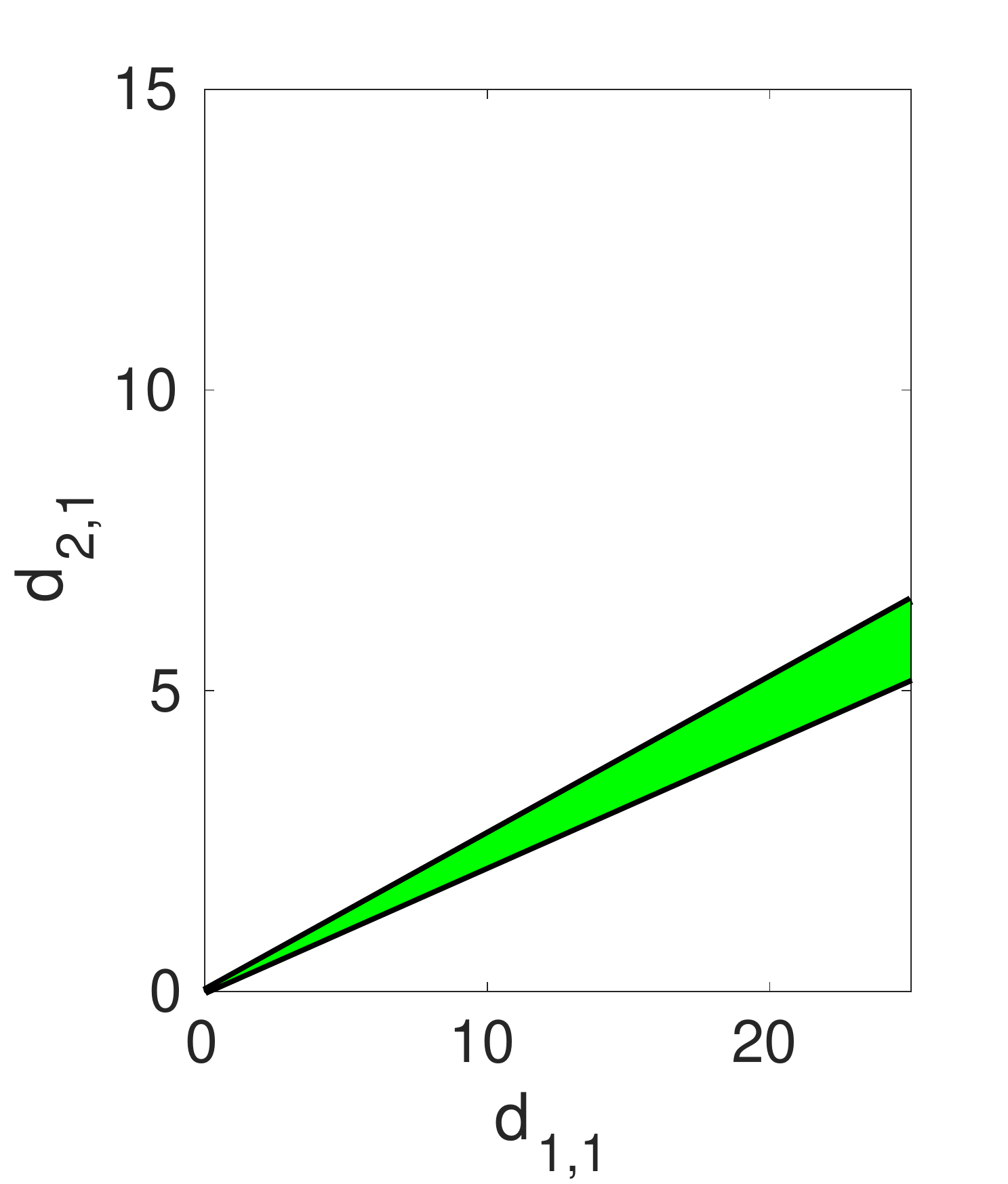}
        \caption{Seven identical cells and a defective cell}
        \label{Bifur_2Par_NonId_8cellsA}
    \end{subfigure}    
    \caption{Two-parameter Hopf bifurcation diagrams in the plane of
      permeabilities, as computed from the ODEs \eqref{WM_ODEsys_2D},
      showing regions of stable periodic solutions (green) and regions
      of linearly stable steady-state solutions (unshaded). The
      Sel'kov parameters are $\alpha = 0.9$, $\mu = 2$,
      $\epsilon_0=0.15$, and we fix $\tau = 0.5$. The results in (a)
      and (b) are for $m=3$ and $m=8$ identical cells, respectively,
      having common permeability parameters $d_1$ and $d_2$. (c) shows
      the result for $m=3$ cells (two identical and a defective cell),
      while the result in (d) is for $m=8$ cells (seven identical and
      a defective cell). For (c) and (d), the permeability parameters
      $d_{1,1}$ and $d_{2,1}$ are for the defective cell only, while
      they are fixed as $d_{1,j}=0.8$ and $d_{2,j}=0.2$ for
      $j=2, \dots,m$, for the remaining identical
      cells.}\label{Bifur_2ParA}
\end{figure}

Next, we present two-parameter global bifurcation diagrams in terms of
the permeability parameters $d_{1}$ and
$d_{2}$. Figures~\ref{Bifur_2Par_Id_3cellsA} and
\ref{Bifur_2Par_Id_8cellsA} are for $m=3$ and $m=8$ identical cells,
respectively, while Figures~\ref{Bifur_2Par_NonId_3cellsA} and
\ref{Bifur_2Par_NonId_8cellsA} are for the cases involving a defective
cell.  The remaining parameters are the same as those used in Figures
\ref{Bifur_d1} and \ref{Bifur_d2}. In Figures
~\ref{Bifur_2Par_Id_3cellsA}--\ref{Bifur_2Par_NonId_8cellsA}, stable
periodic solutions are predicted in the green wedge-shaped regions,
while linearly stable steady-state solutions occur in the unshaded
regions.  From Figure \ref{Bifur_2ParA} the region of intracellular
oscillations are bounded by two straight black lines, consisting of
Hopf bifurcation values for the steady-state solution that correspond
to the one-parameter bifurcation diagrams in Figures \ref{Bifur_d1}
and \ref{Bifur_d2}. In other words, the one-parameter bifurcation
diagrams in Figures \ref{Bifur_d1} and \ref{Bifur_d2} represent to
taking vertical and horizontal slices through the parameter plane
shown in Figures \ref{Bifur_2ParA}, respectively. {We observe that as
the number of identical cells increases there is a larger region in
the permeability parameter space that lead to linearly stable
synchronous intracellular oscillations (see Figure
\ref{Bifur_2Par_Id_8cellsA}). Moreover, the permeability parameter
space that yields oscillations is relatively larger when
the cells are all identical (Figure \ref{Bifur_2Par_Id_3cellsA} and
\ref{Bifur_2Par_Id_8cellsA}), as compared to when a single defective
cell has permeability parameters that are different from those of the
group (Figure \ref{Bifur_2Par_NonId_3cellsA} and
\ref{Bifur_2Par_NonId_8cellsA}).  However, this is not the case for
the scenario involving a defective cell. For this case, as the number
of remaining identical cells increases, it becomes increasingly
difficult to trigger intracellular oscillations of the group through
changes in the permeability parameters of a single defective cell (see
Figures \ref{Bifur_2Par_NonId_3cellsA} and
\ref{Bifur_2Par_NonId_8cellsA}).}

In Figure \ref{GL_1} of Appendix \ref{Append_B} we show a gallery of
time-dependent solutions to the ODE system \eqref{WM_ODEsys_2D} for
permeability parameter values sampled from the bifurcation diagrams in
Figures \ref{Bifur_d1}--\ref{Bifur_2ParA}.

\subsubsection{Instability triggered by local reaction kinetics of the cells}

In the previous subsection, we studied how intracellular oscillations
are triggered by modifying the rate of influx ($d_{1}$) and efflux
($d_{2}$) of signaling chemicals across the cell boundaries. In
this subsection, we use the ODE system \eqref{WM_ODEsys_2D} to study
intracellular oscillations that are triggered by changes in the
parameters of the local reaction kinetics. More specifically, in
\eqref{Selkov}, the Sel'kov parameters $\mu > 0$ and $\alpha > 0$ are
used as bifurcation parameters, while $\epsilon_0 = 0.15$ is fixed.
In the results below, the permeabilities were fixed at
$d_{1,j}=0.8$ and $d_{2,j}=0.2$ for $j=1, \dots,m$.

\begin{figure}[htbp]
    \centering
    \begin{subfigure}[t]{0.22\textwidth}
        \includegraphics[width=\textwidth]{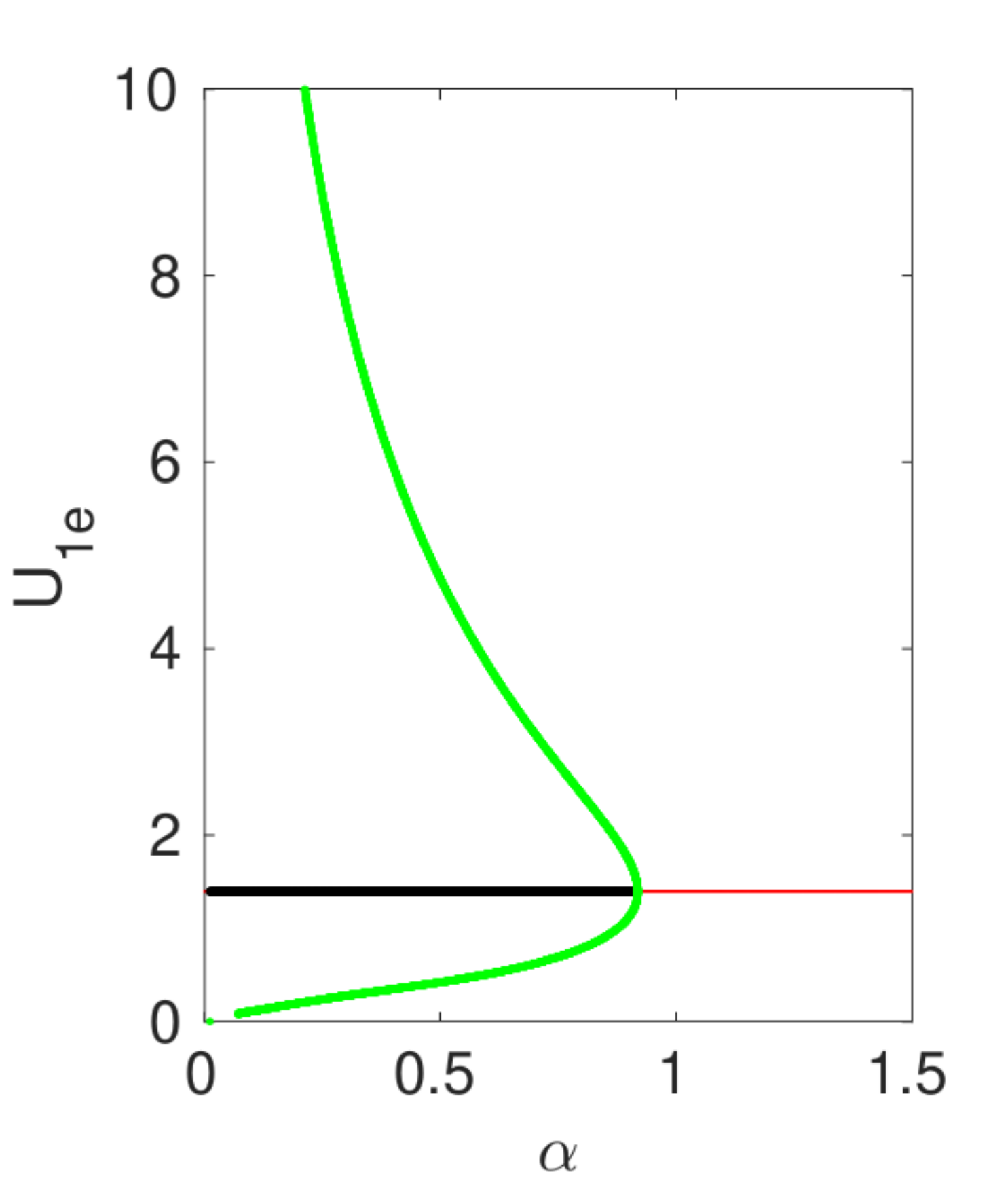}
        \caption{Three identical cells}
        \label{Bf_alpha3ID}
    \end{subfigure}
    \quad 
    \begin{subfigure}[t]{0.22\textwidth}
        \includegraphics[width=\textwidth]{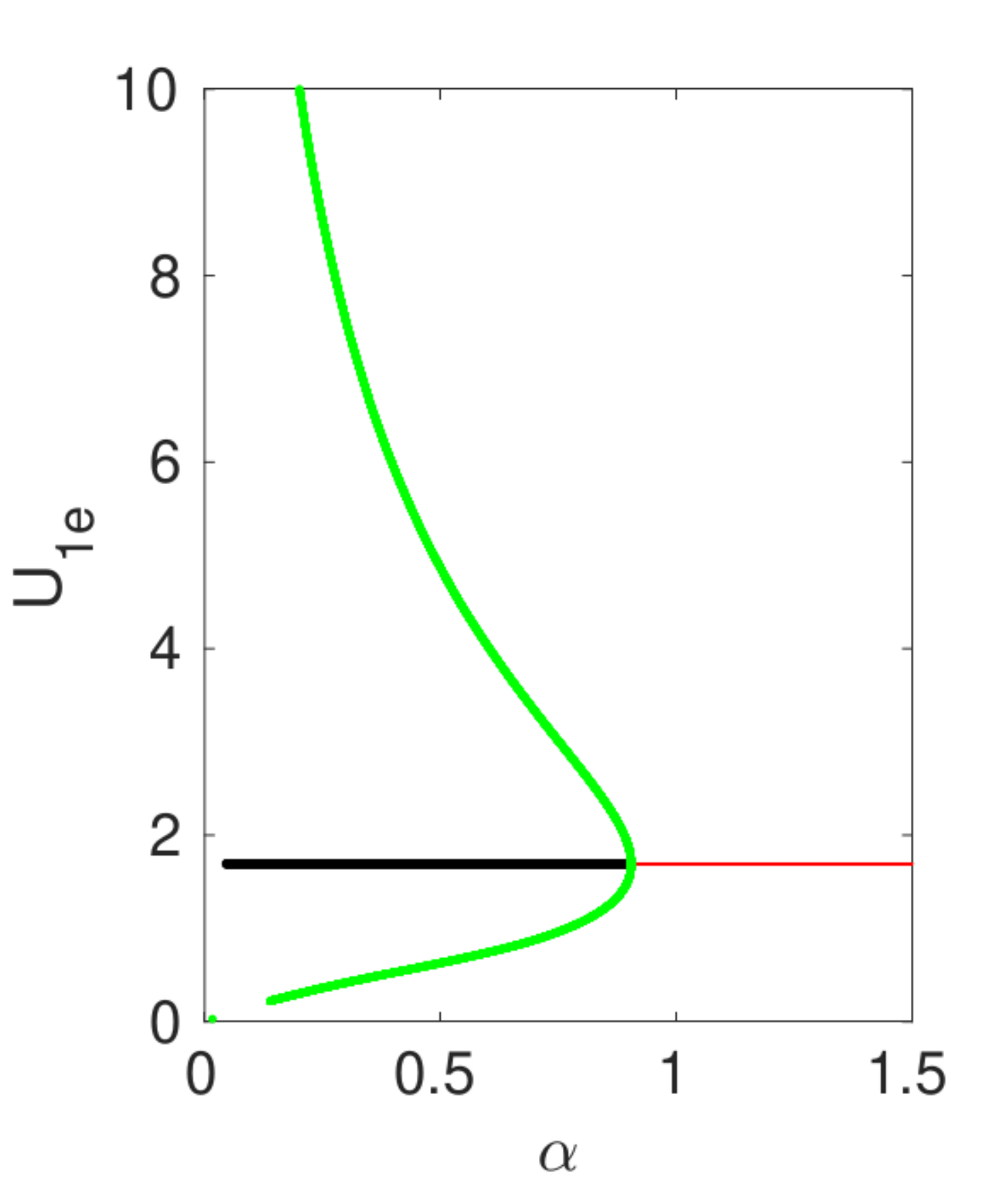}
        \caption{Eight identical cells}
        \label{Bf_alpha8ID}
    \end{subfigure}
    \quad
    \begin{subfigure}[t]{0.22\textwidth}
        \includegraphics[width=\textwidth]{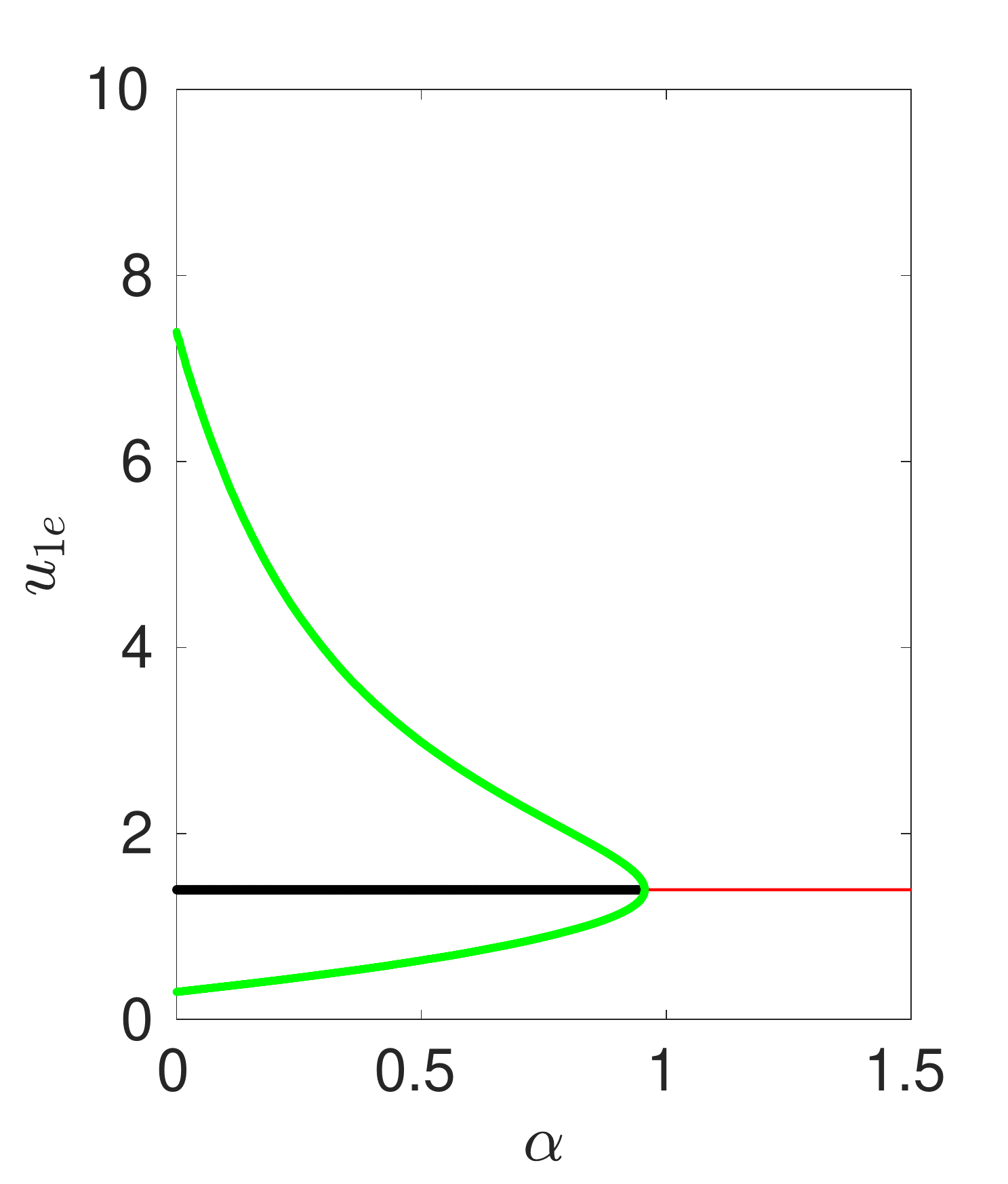}
        \caption{Two identical cells and a defective cell}
        \label{Bf_alpha3}
    \end{subfigure}
    \quad 
    \begin{subfigure}[t]{0.22\textwidth}
        \includegraphics[width=\textwidth]{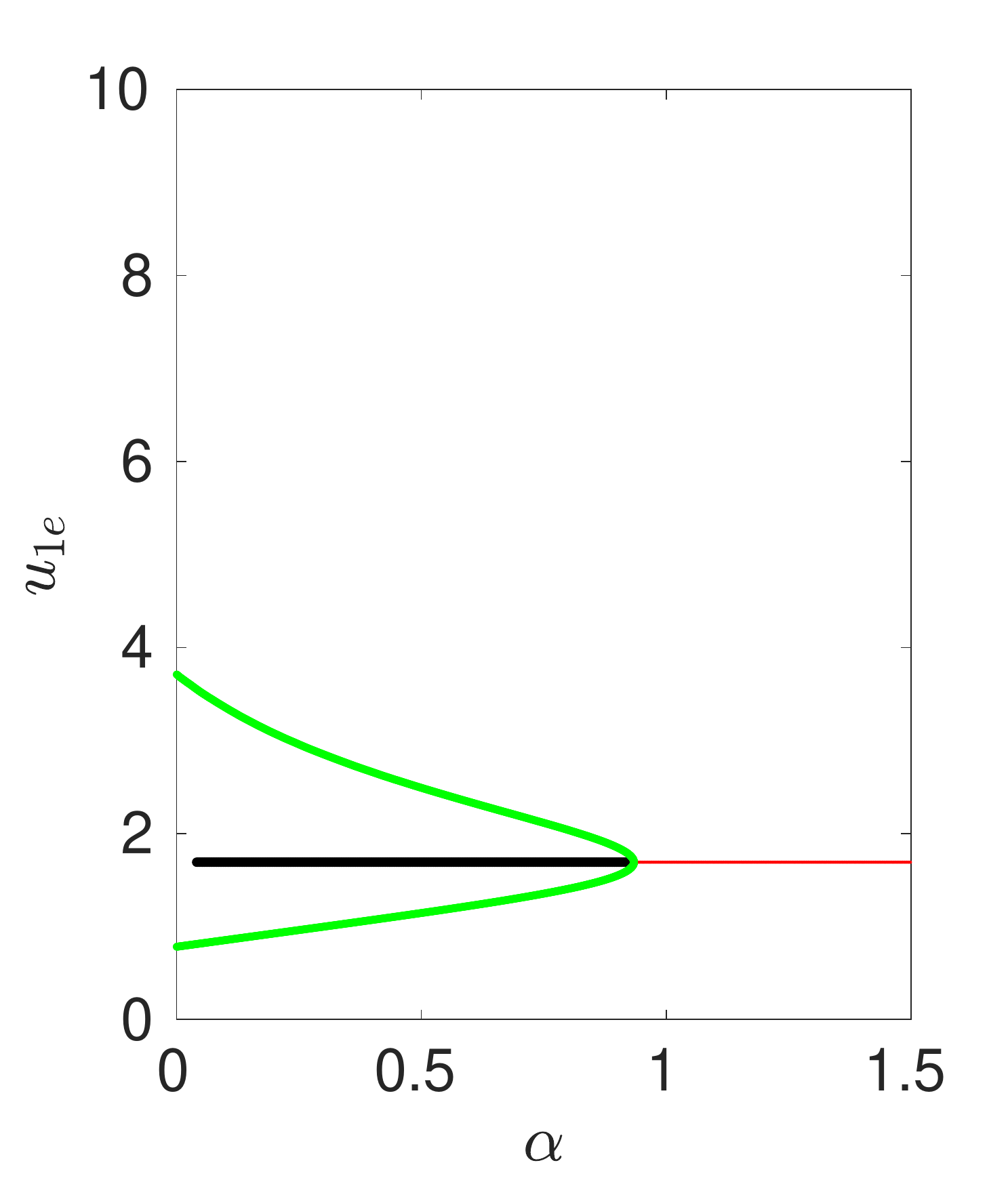}
        \caption{Seven identical cells and a defective cell}
        \label{Bf_alpha8}
    \end{subfigure}
    \caption{Global bifurcation diagrams for $u_{1e}$ versus $\alpha$,
      as computed from the ODEs \eqref{WM_ODEsys_2D}, showing
      steady-states and global branches of periodic solutions.  The
      thin red lines and bold black lines correspond to linearly
      stable and unstable steady-state solution branches,
      respectively, while the green curves are linearly stable
      branches of periodic solutions. The parameters are $\tau=0.5$,
      $\epsilon_0 = 0.15$, $\mu = 2$, with $d_{1,j} =0.8$ and
      $d_{2,j}=0.2$ for $j=1,\ldots,m$. (a) and (b) are for $m=3$ and
      $m=8$ identical cells, respectively. When $m=3$, the bifurcation
      point is $\alpha_c = 0.9182$, while $\alpha_c = 0.9042$ for
      $m=8$. (c) is for two identical cells and a defective cell, with
      a Hopf bifurcation occurring at $\alpha_c = 0.9562$, and (d) is
      for $m=8$, where seven of the cells are identical and the
      remaining one is defective. The Hopf bifurcation occurs at
      $\alpha_c =0.9342$.  The bifurcation parameter in (c) and (d) is
      that of the defective cell only, and $\alpha = 0.9$ is fixed for
      the identical cells.}\label{Bf_alpha}
\end{figure}

\begin{figure}[htbp]
    \centering
    \begin{subfigure}[t]{0.33\textwidth}
        \includegraphics[width=\textwidth, height=4.2cm]{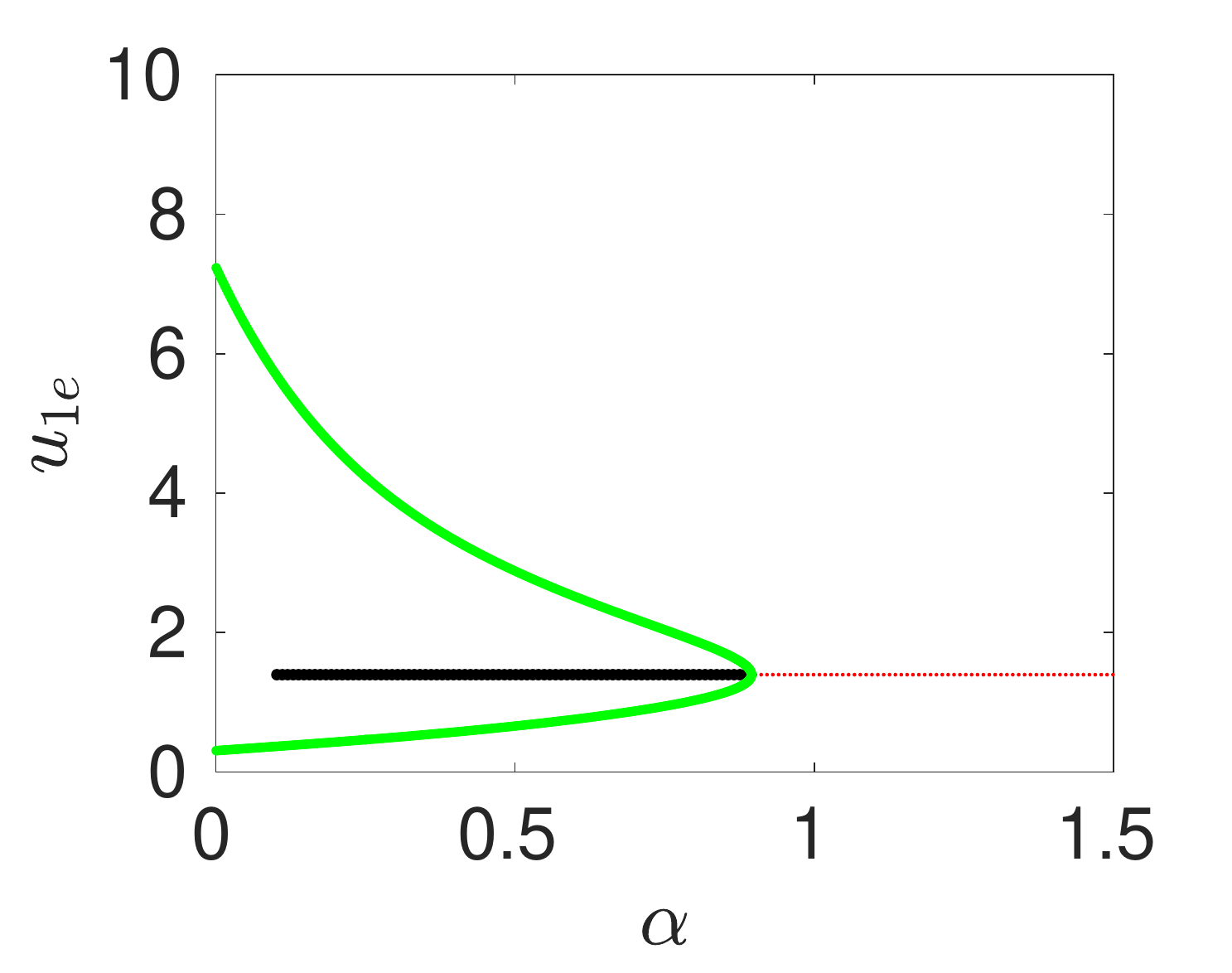}
        \caption{Two identical cells and a defective cell}
        \label{nBf_alpha3}
    \end{subfigure}
    \quad 
    \begin{subfigure}[t]{0.33\textwidth}
        \includegraphics[width=\textwidth,height=4.2cm]{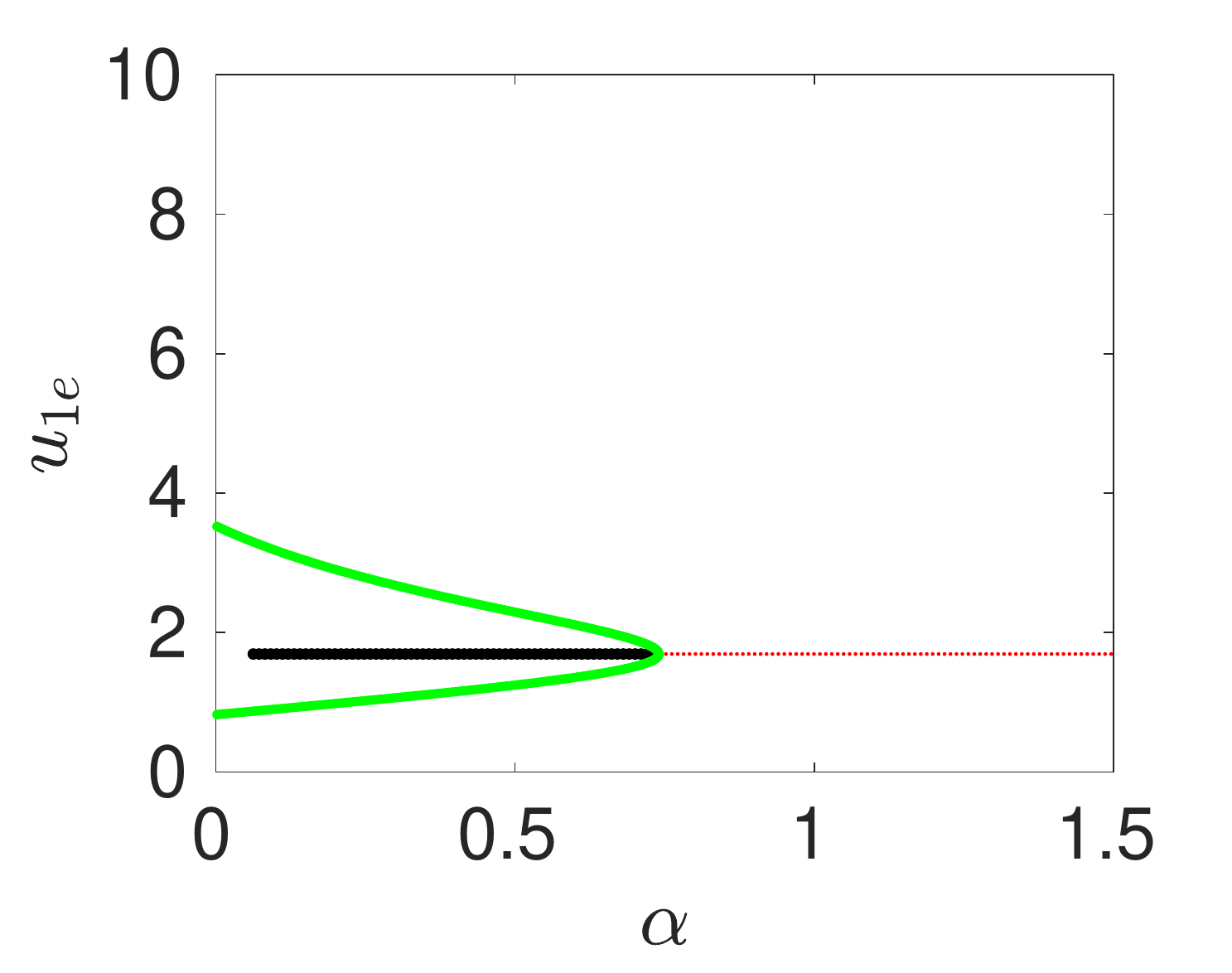}
        \caption{Seven identical cells and a defective cell}
        \label{nBf_alpha8}
    \end{subfigure}
    \caption{Global bifurcation diagram for $u_{1e}$ versus the
      Sel'kov parameter $\alpha$ for the defective cell.  Same caption
      and parameter values as in Figure \ref{Bf_alpha} except that now
      $\alpha=0.93$ for the identical cells.  When $\alpha=0.93$,
      there are no intracellular oscillations for $m=3$ or $m=8$ (see
      Figures \ref{Bf_alpha3ID} and \ref{Bf_alpha8ID}).  As $\alpha$
      decreases below the thresholds $\alpha_c\approx 0.8953$ for
      $m=3$ and $\alpha_c\approx 0.7404$ for $m=8$, the defector cell
      triggers intracellular oscillations for the entire group of
      cells. On the range $0.694<\alpha<\alpha_c$, the defector cell
      still has a linearly stable steady-state if uncoupled from the
      bulk.}\label{nBf_alpha}
\end{figure}

Figure \ref{Bf_alpha} shows the stability of the steady-state
solutions of the ODEs \eqref{WM_ODEsys_2D} as $\alpha > 0$ is
varied. We observe from this figure that there is only one Hopf
bifurcation point, $ \alpha_c$ for $\alpha > 0$. For any value of
$\alpha \leq \alpha_c$, the ODE system has linearly stable periodic
solutions with increasing amplitude as $\alpha$ decreases. Linearly
stable steady-state solutions exist for $\alpha > \alpha_c$. Figures~
\ref{Bf_alpha3ID} and \ref{Bf_alpha8ID} are for $m=3$ and $m=8$
identical cells, respectively. The amplitudes of the predicted
oscillations for $m=3$ and $m=8$ are roughly similar, and the Hopf
bifurcation threshold for $m=3$ at $\alpha_c = 0.9182$ is a little
larger than for $m=8$ where $\alpha_c = 0.9042$.  Observe from Figure
\ref{fig:selkov} that for a single isolated cell, uncoupled from the
bulk, a Hopf bifurcation occurs when $\alpha\approx 0.694$ as computed
from \eqref{selkov:boundary}. For the defective cell case (Figures
\ref{Bf_alpha3} and \ref{Bf_alpha8}), when $m=3$ (two identical cells
with $\alpha=0.9$ and a defective cell), the Hopf bifurcation occurs
at $\alpha_c = 0.9562$, while $\alpha_c= 0.9342$ when $m=8$ cells
(seven identical cells with $\alpha=0.9$ and a defective
cell). {As expected, similar to the case of identical cells,
  the Hopf bifurcation value of $\alpha$ for the defective cell is
  smaller than, and therefore closer to the stability boundary shown
  in Figure \ref{fig:selkov} (see equation \eqref{selkov:boundary}),
  as the number of cells increases. Most importantly, since for
  $\alpha=0.9$ there would be intracellular oscillations in the
  identical cells when the defective cell is absent, we observe from
  Figures \ref{Bf_alpha3} and \ref{Bf_alpha8} that when there is a
  defective cell then an increase in its kinetic parameter $\alpha$
  above $\alpha_c$ will extinguish the intracellular oscillations of
  the entire group of cells. In contrast, from Figures
  \ref{nBf_alpha3} and Figures \ref{nBf_alpha8} we observe that a
  defective cell can trigger collective oscillations in a group of
  identical cells when the kinetic parameter of the defective cell
  decreases below a threshold. In Figures \ref{nBf_alpha3} and Figures
  \ref{nBf_alpha8} we fix $\alpha=0.93$ so that that no oscillations
  are possible for this group of identical cells. However, by adding a
  defective cell, we observe that intracellular oscillations for the
  entire group are triggered when the kinetic parameter for the
  defector satisfies $\alpha<\alpha_c\approx 0.8953$ for $m=3$ and
  $\alpha<\alpha_c\approx 0.7404$ for $m=8$. Since
  $\alpha_c>0.694$, the defective cell is still a conditional
  oscillator, and it would have a linearly stable steady-state when
  uncoupled from the bulk.}

\begin{figure}[htbp]
    \centering
    \begin{subfigure}[t]{0.22\textwidth}
        \includegraphics[width=\textwidth]{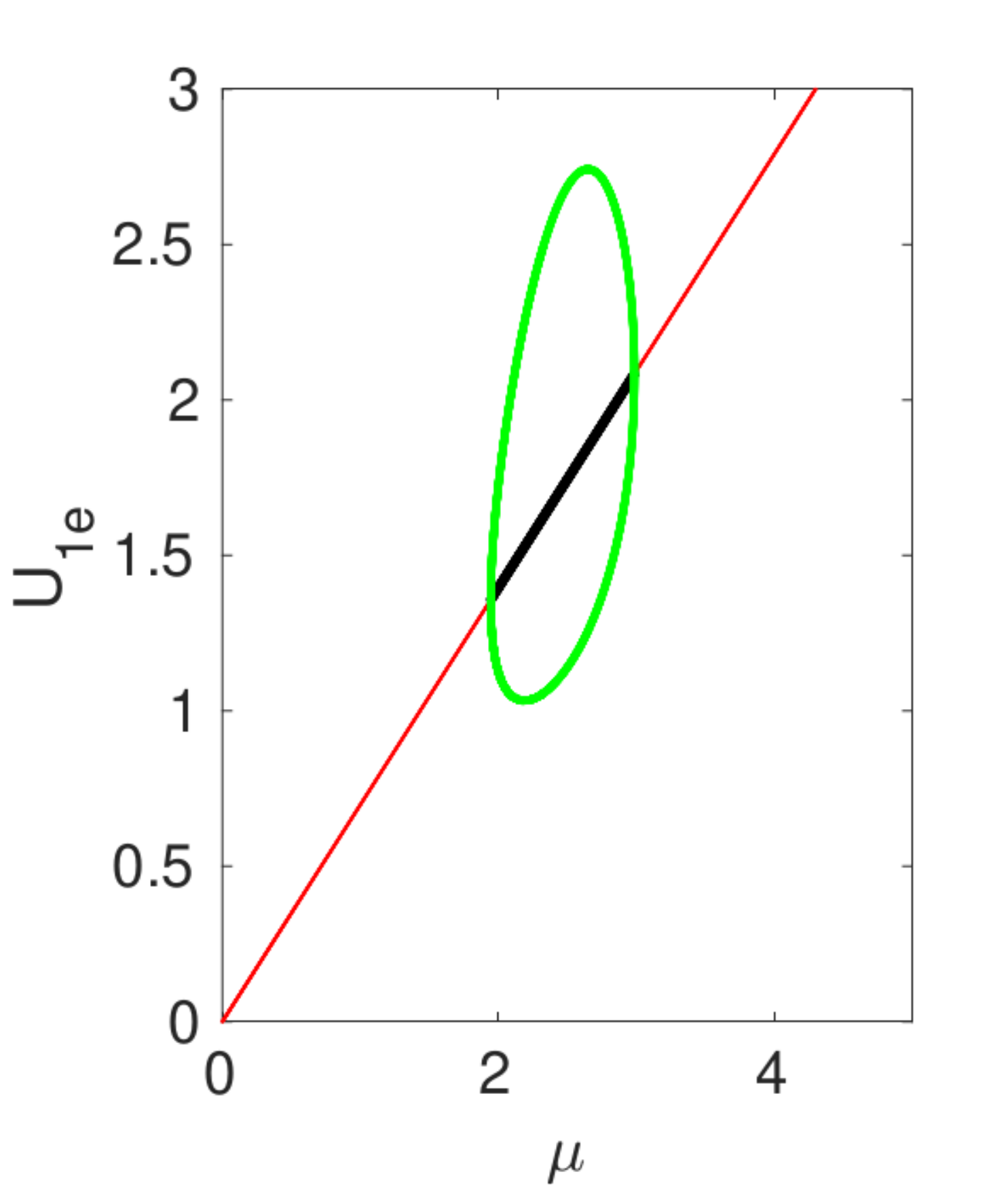}
        \caption{Three identical cells}
        \label{Bf_mu3ID}
    \end{subfigure}
    \quad 
    \begin{subfigure}[t]{0.22\textwidth}
        \includegraphics[width=\textwidth]{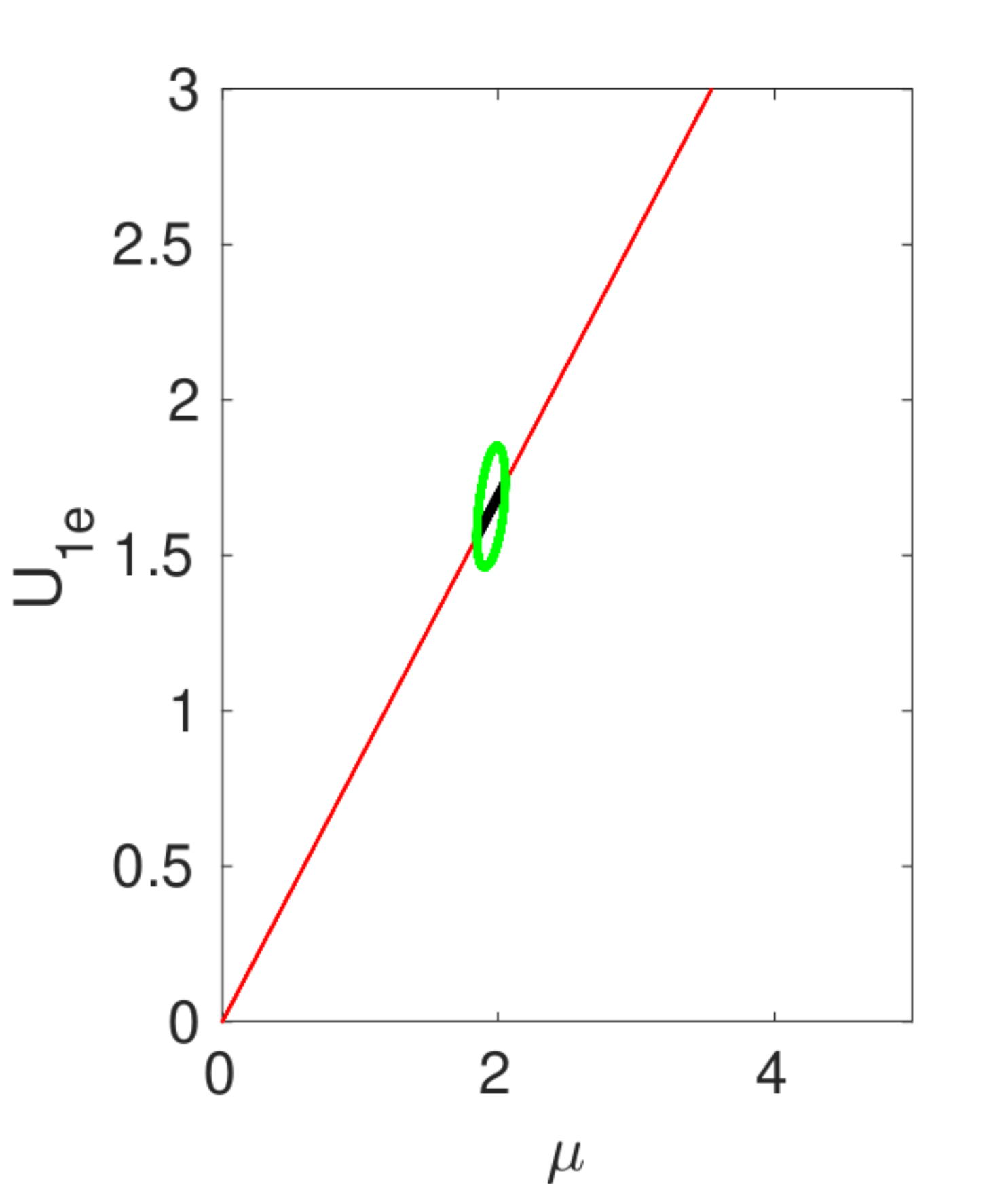}
        \caption{Eight identical cells}
        \label{Bf_mu8ID}
    \end{subfigure}
	\quad
	\begin{subfigure}[t]{0.22\textwidth}
        \includegraphics[width=\textwidth]{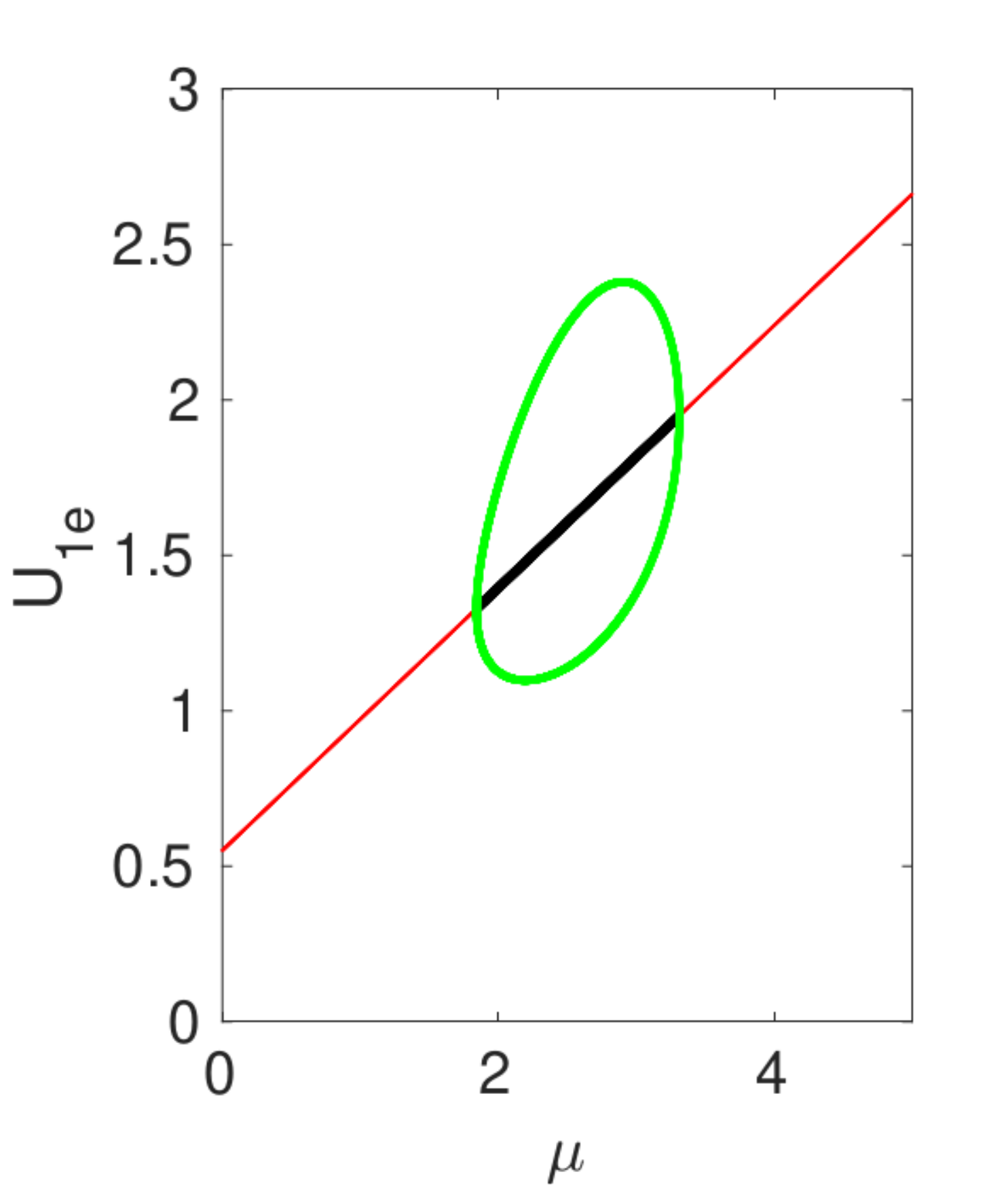}
        \caption{Two identical cells and a defective cell}
        \label{Bf_mu3}
    \end{subfigure}
    \quad 
    \begin{subfigure}[t]{0.22\textwidth}
        \includegraphics[width=\textwidth]{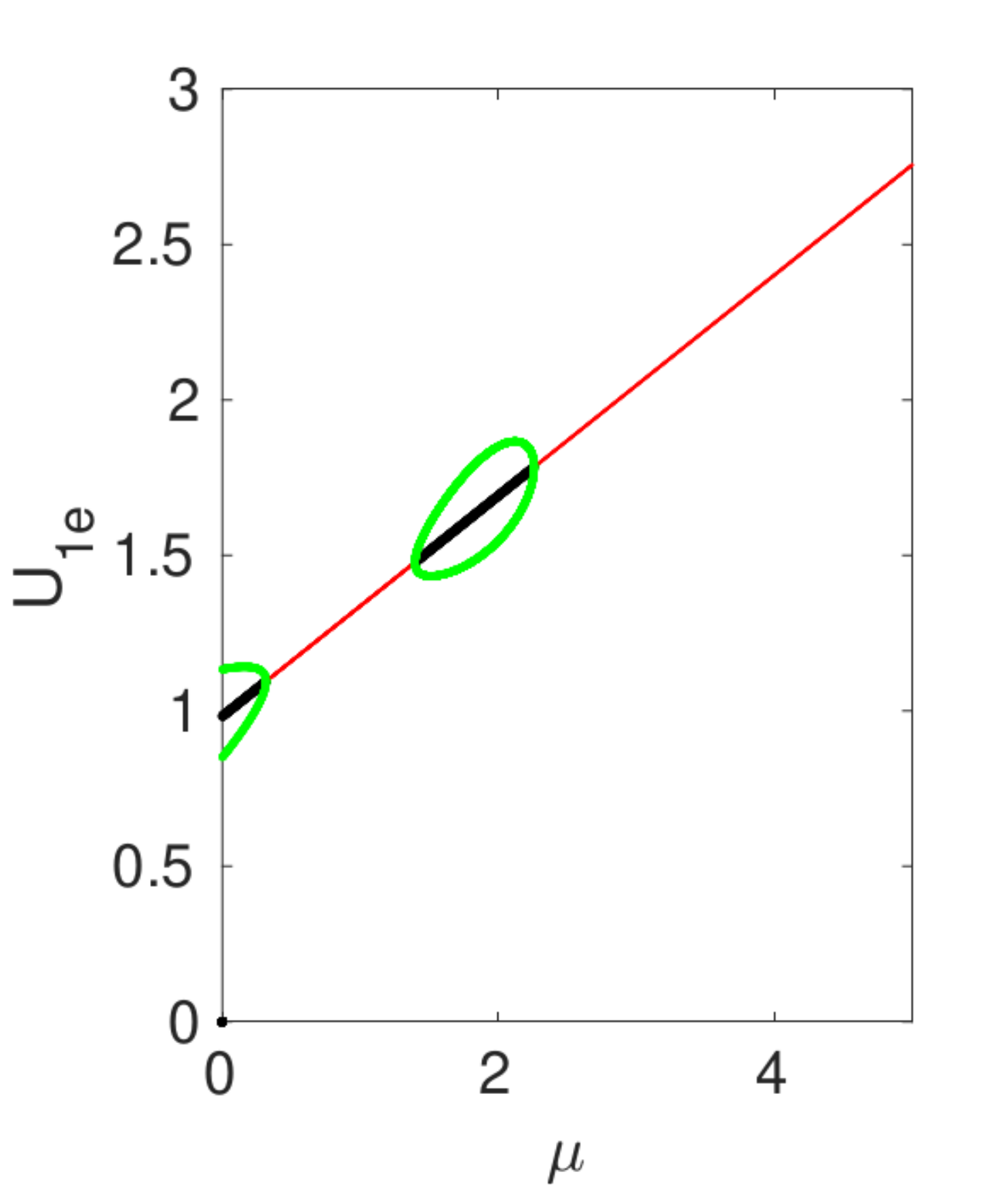}
        \caption{Seven identical cells and a defective cell}
        \label{Bf_mu8}
    \end{subfigure}
    \caption{Global bifurcation diagrams for $u_{1e}$ versus $\mu$,
      computed from the ODEs \eqref{WM_ODEsys_2D}, showing
      steady-states and global branches of periodic solutions. The
      labeling of the curves is the same as in Figure \ref{Bf_alpha}.
      The parameters are $\tau=0.5$, $\epsilon_0 = 0.15$,
      $\alpha=0.9$, with $d_{1,j} =0.8$ and $d_{2,j}=0.2$ for
      $j=1,\ldots,m$. (a) and (b) are for $m=3$ and $m=8$ identical
      cells respectively.  The Hopf bifurcation points are
      $\mu_1 = 1.946$ and $\mu_2 = 2.987$ for $m=3$, and
      $\mu_1 = 1.845$ and $\mu_2 = 2.05$ for $m=8$. (c) is for two
      identical cells (with $\mu=2$) and a defective cell, with Hopf
      bifurcation points at $\mu_1 = 1.842$ and $\mu_2=3.312$. (d) is
      for $m=8$ cells, where seven of the cells are identical (with
      $\mu=2$) and the remaining one is defective. There are three
      Hopf bifurcation points for this case:
      $\mu_1 = 0.3164, \mu_2 = 1.396$ and $ \mu_3 = 2.26$. The
      bifurcation parameter in (c) and (d) is that of the defective
      cell only.} \label{Bf_mu}
\end{figure}

The global bifurcation diagrams with respect to the Sel'kov parameter
$\mu>0$ are shown in Figure \ref{Bf_mu}. For an isolated cell,
uncoupled from the bulk, a horizontal slice at a sufficiently small
value of $\alpha$ through the instability region in
Figure \ref{fig:selkov} shows that there are two Hopf bifurcation values
of $\mu$ for an isolated cell.  The results in Figure \ref{Bf_mu} show
how this instability region of an isolated cell is modified due to
coupling with the bulk and in the presence of a defective
cell. Similar to the results shown above when $\alpha$ was the bifurcation
parameter, the range of $\mu$ that yields synchronous oscillations
shrinks as the number of identical cells increases (see
Figures~\ref{Bf_mu3ID} and \ref{Bf_mu8ID}). For $m=3$ identical cells,
Hopf bifurcations occur at $\mu_1 = 1.946$ and $\mu_2 = 2.987$, and
when $m=8$ they are at $\mu_1 = 1.845$ and $\mu_2 = 2.05$. For the
case of two identical cells (with $\mu=2$) and a defective cell, the
Hopf bifurcation points are $\mu_1 = 1.842$ and $ \mu_2=3.312$ (Figure
\ref{Bf_mu3}). Figure \ref{Bf_mu8} shows the result for $m=8$ (seven
identical cells with $\mu=2$ and a defective cell), which has three
Hopf bifurcation points. This result shows that linearly stable
periodic solutions exist for $0<\mu < 0.3164$ and for
$1.396 < \mu < 2.26$. By comparing the results in Figures \ref{Bf_mu3}
and \ref{Bf_mu8} we notice that the range of $\mu$ that predicts
linearly stable periodic solutions is smaller when $m=8$ than when
$m=3$, even though synchronous oscillations are predicted in two
different intervals when $m=8$.  This agrees with our earlier
observation that an increase in the population of identical cells
makes synchronization more difficult when they are coupled to a
defective cell.

\begin{figure}[htbp]
    \centering
    \begin{subfigure}[t]{0.22\textwidth}
        \includegraphics[width=\textwidth,valign=t]{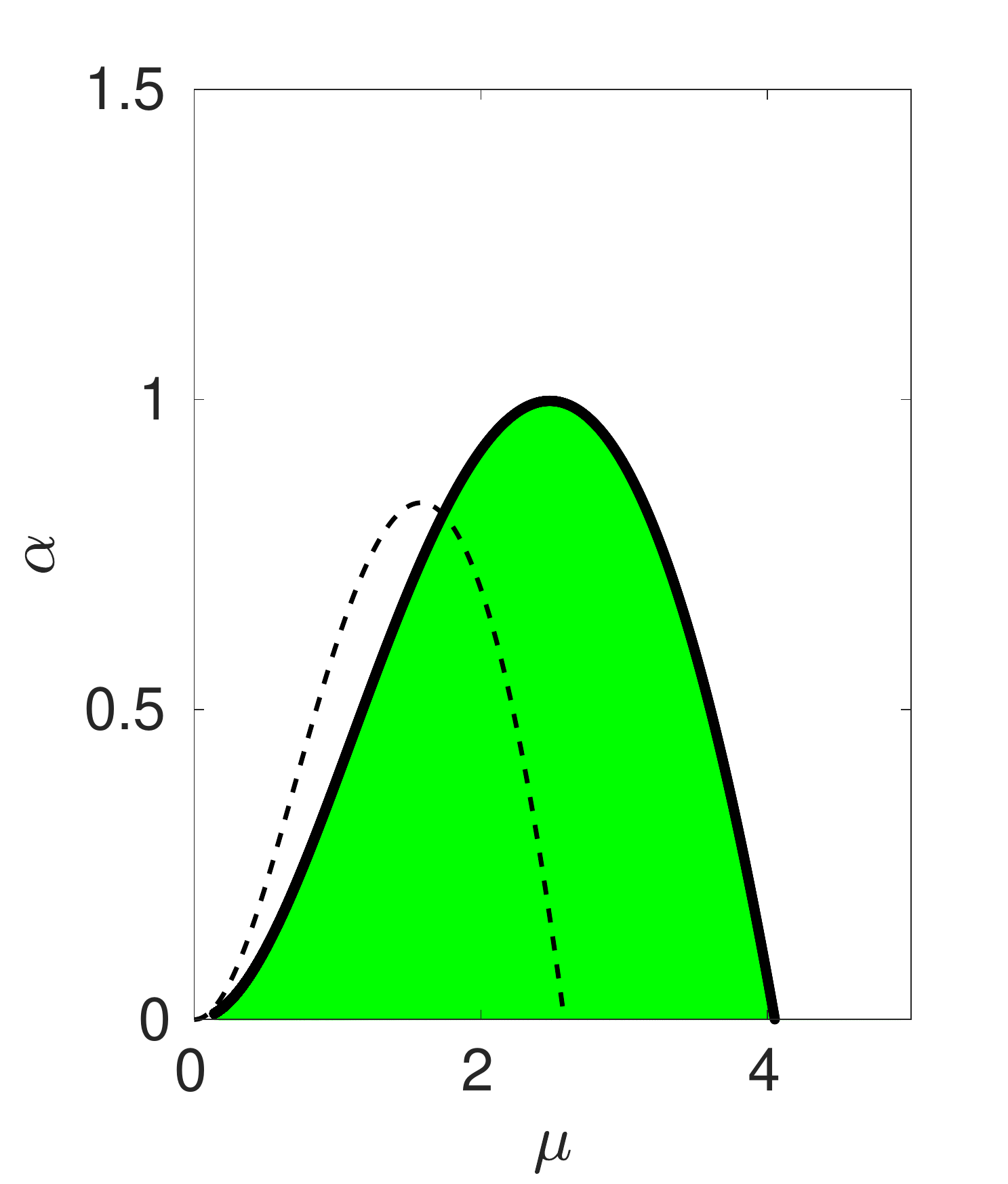}
        \caption{Three identical cells.}
        \label{Bifur_2Par_Id_3cells}
    \end{subfigure}
    \quad 
    \begin{subfigure}[t]{0.22\textwidth}
        \includegraphics[width=\textwidth,valign=t]{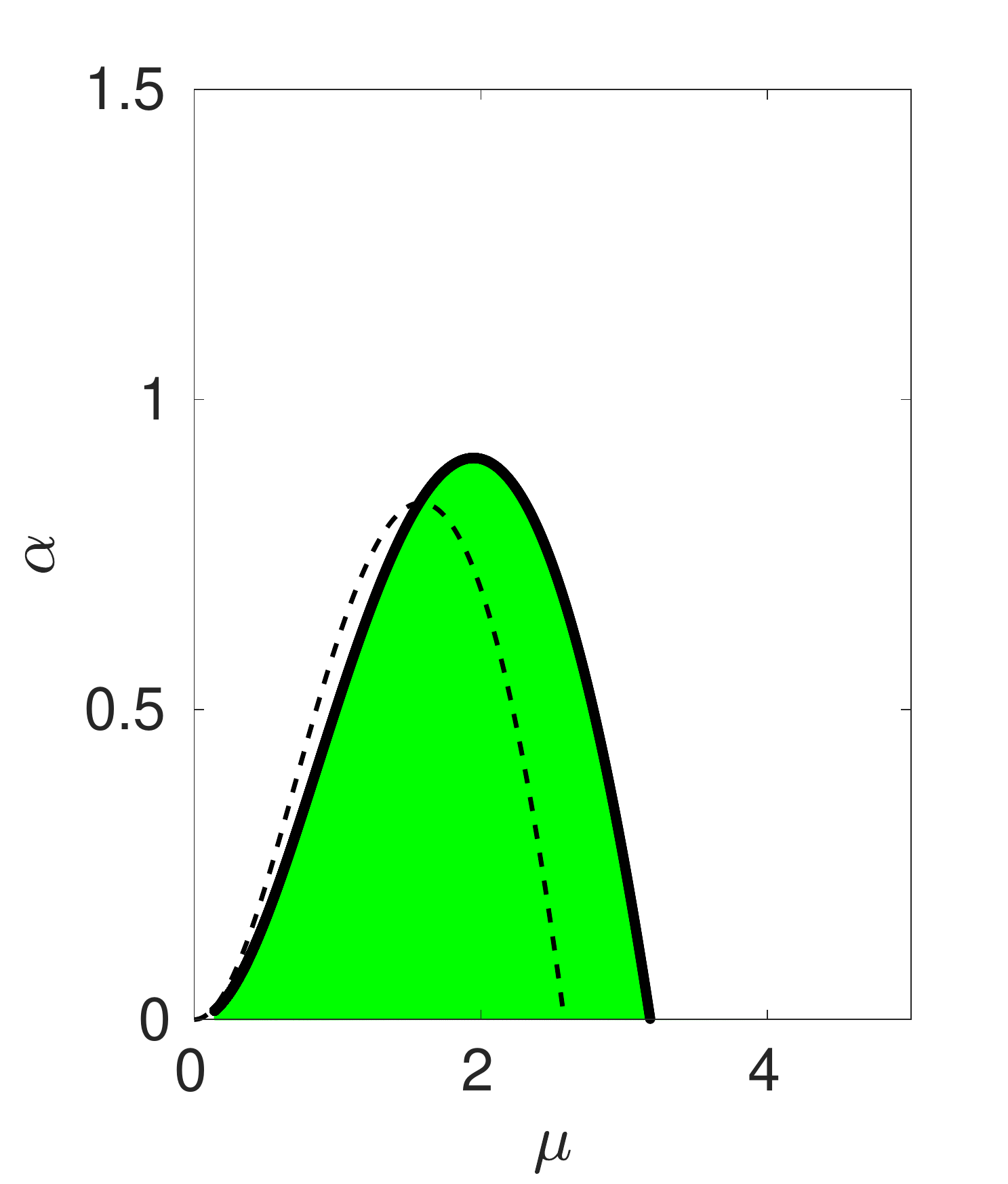}
        \caption{Eight identical cells.}
        \label{Bifur_2Par_Id_8cells}
    \end{subfigure} 
    \quad 
\begin{subfigure}[t]{0.22\textwidth}
        \includegraphics[width=\textwidth,valign=t]{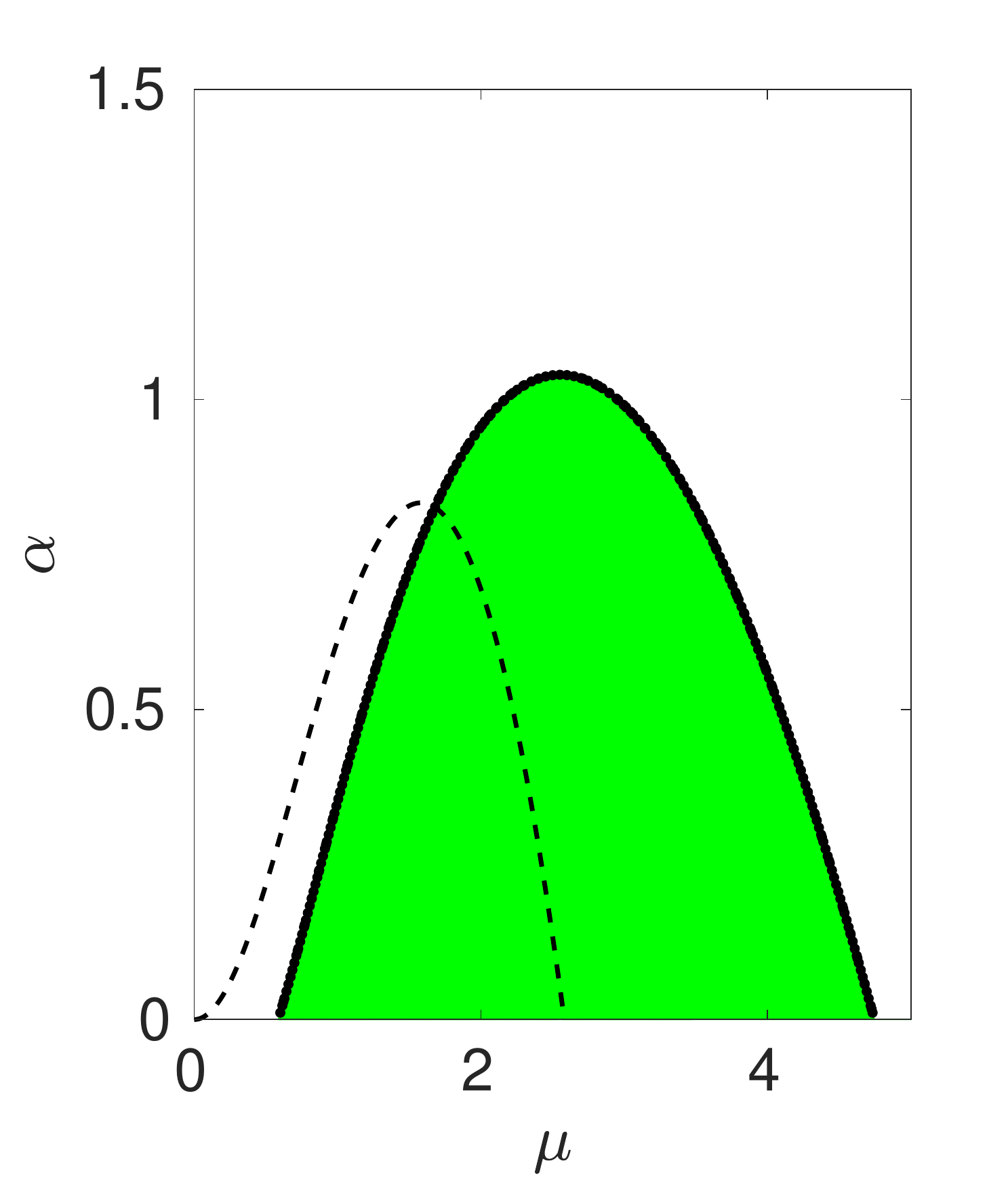}
        \caption{Two identical cells and a defective cell.}
        \label{Bifur_2Par_NonId_3cells}
    \end{subfigure}
    \quad 
    \begin{subfigure}[t]{0.22\textwidth}
        \includegraphics[width=\textwidth,valign=t]{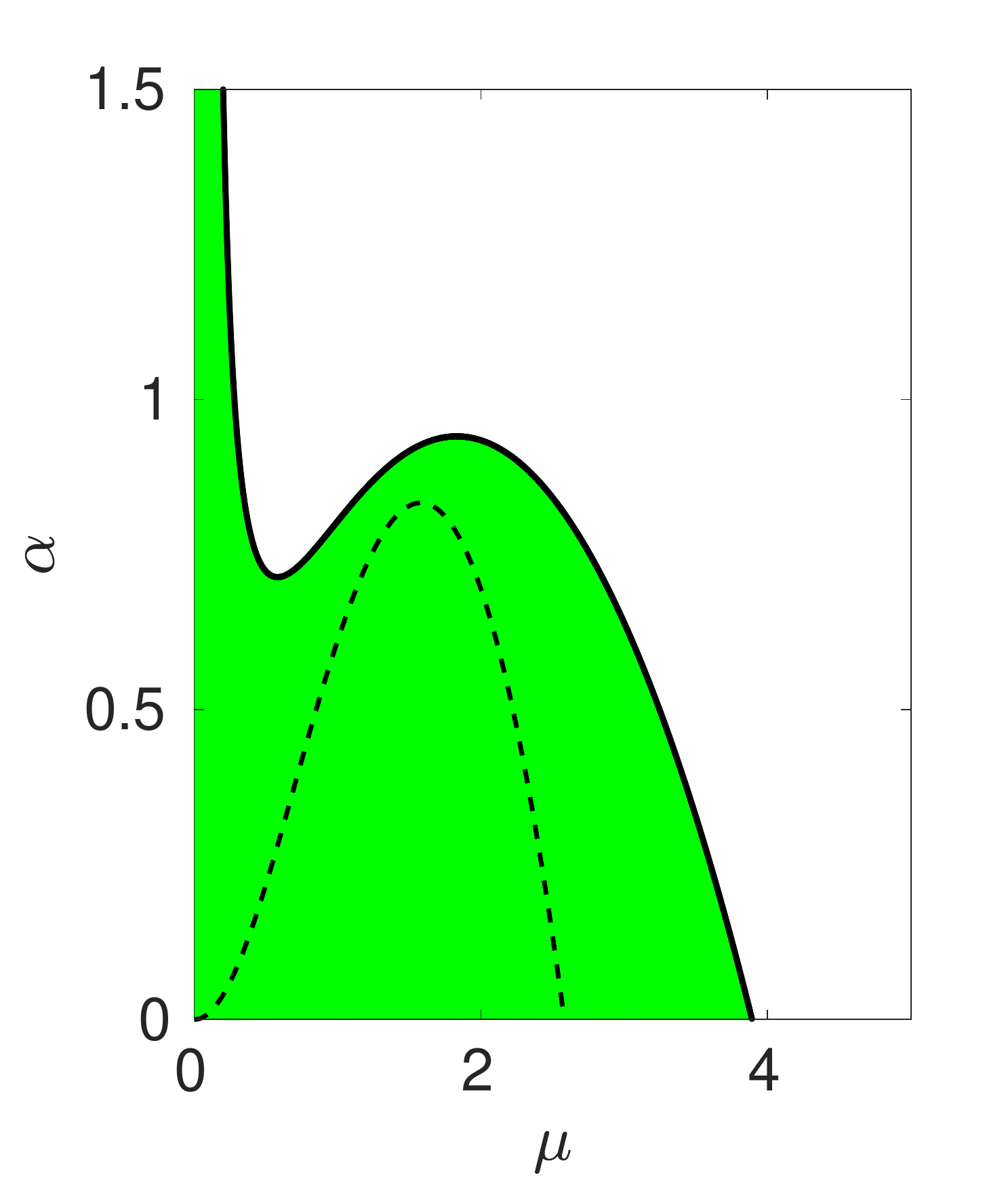}
        \caption{Seven identical cells and a defective cell.}
        \label{Bifur_2Par_NonId_8cells}
    \end{subfigure}    
    \caption{Two-parameter bifurcation diagrams in the $\alpha$ versus
      $\mu$ parameter plane, computed from the ODEs
      \eqref{WM_ODEsys_2D}, showing regions of linearly stable
      periodic solutions (shaded in green) and linearly stable
      steady-state solutions (unshaded). {The black dashed
        curve in each figure is the Hopf bifurcation boundary shown in
        Figure~\ref{fig:selkov} for a single isolated cell, as given
        by \eqref{selkov:boundary}.} The parameters are
      $\epsilon_0=0.15$, $\tau=0.15$ and with $d_{1,j}=0.8$ and
      $d_{2,j}=0.2$ for $j=1,\dots,m$. (a) and (b) are for $m=3$ and
      $m=8$ identical cells, respectively, while (c) and (d) are for
      $m=3$ (two identical cells and a defective cell) and $m=8$
      (seven identical cells and a defective cell), respectively.  The
      bifurcation parameter in (c) and (d) is for the defective cell
      only, with $\mu =2$ and $\alpha=0.9$ fixed for the identical
      cells. } \label{Bifur_2Par}
\end{figure}

In Figure \ref{Bifur_2Par}, we present two-parameter bifurcation
diagrams for the Sel'kov parameters $\alpha$ and $\mu$. In this
figure, the green-shaded region is where linearly stable periodic
solutions occur, while the unshaded region predicts linearly stable
steady-state solutions. The results in Figures \ref{Bf_alpha} and
\ref{Bf_mu} correspond to vertical and horizontal slices through the
two-parameter plane in Figure \ref{Bifur_2Par}. {The black
  dashed curve in Figure \ref{Bifur_2Par}(a-d) is the Hopf bifurcation
  boundary of a single isolated cell, uncoupled from the bulk, as
  given in \eqref{selkov:boundary}. In each subfigure, for any pair
  $(\mu,\alpha)$ in a green-shaded region that lies above this dashed
  curve, we conclude that the cell-cell coupling from the bulk medium
  is the trigger for intracellular oscillations that would otherwise
  not occur when the cells are decoupled from the bulk. In contrast,
  and most notably for three cells, we observe from Figures
  \ref{Bf_mu3ID} and \ref{Bf_mu3} that there are narrow parameter
  regimes where the black dashed curve is above the green
  region. These thin regimes correspond to where an oscillatory
  instability of an isolated cell is quenched by cell-cell coupling
  through the bulk medium.

  When the cells are all identical, we observe by comparing Figure
  \ref{Bf_mu3ID} and Figure \ref{Bf_mu8ID} that the region where
  synchronous oscillations occur for $m=3$ cells is slightly larger
  than that for $m=8$ cells.} This feature is in contrast to our
observation from Figures~ \ref{Bifur_2Par_Id_3cellsA} and
\ref{Bifur_2Par_Id_8cellsA} when the permeabilities were used as
bifurcation parameters, where the parameter region of intracellular
oscillations get larger as the number of cells increases. For the
defective cell case, the existence of an additional regime of periodic
solutions, as observed in Figure \ref{Bf_mu8}, is apparent in Figure
\ref{Bifur_2Par_NonId_8cells}.  This new regime predicts linearly
stable periodic solutions for large values of $\alpha$, provided that
$\mu > 0$ is small enough, which is not apparent in the one-parameter
bifurcation diagram for $\alpha$ presented in Figure
\ref{Bf_alpha8}. We notice from the result in Figure
\ref{Bifur_2Par_Id_8cells} that synchronous oscillations are not
predicted as $\mu$ approaches zero when the cells are all
identical. This indicates that the stable periodic solutions observed
in Figure~\ref{Bifur_2Par_NonId_8cells} as $\mu \to 0$ are triggered
by the existence of a defective cell.

In Figure \ref{GL1_1} of Appendix \ref{Append_B} we show a gallery of
time-dependent solutions to the ODE system \eqref{WM_ODEsys_2D} for
different Sel'kov kinetic parameters $\alpha$ and $\mu$ sampled from
the bifurcation diagrams in Figures \ref{Bf_alpha}, \ref{Bf_mu} and
\ref{Bifur_2Par}.

\subsubsection{Instabilities due to increasing the number of cells}

\begin{figure}[htbp]
    \centering
    \begin{subfigure}[t]{0.22\textwidth}
        \includegraphics[width=\textwidth,valign=t]{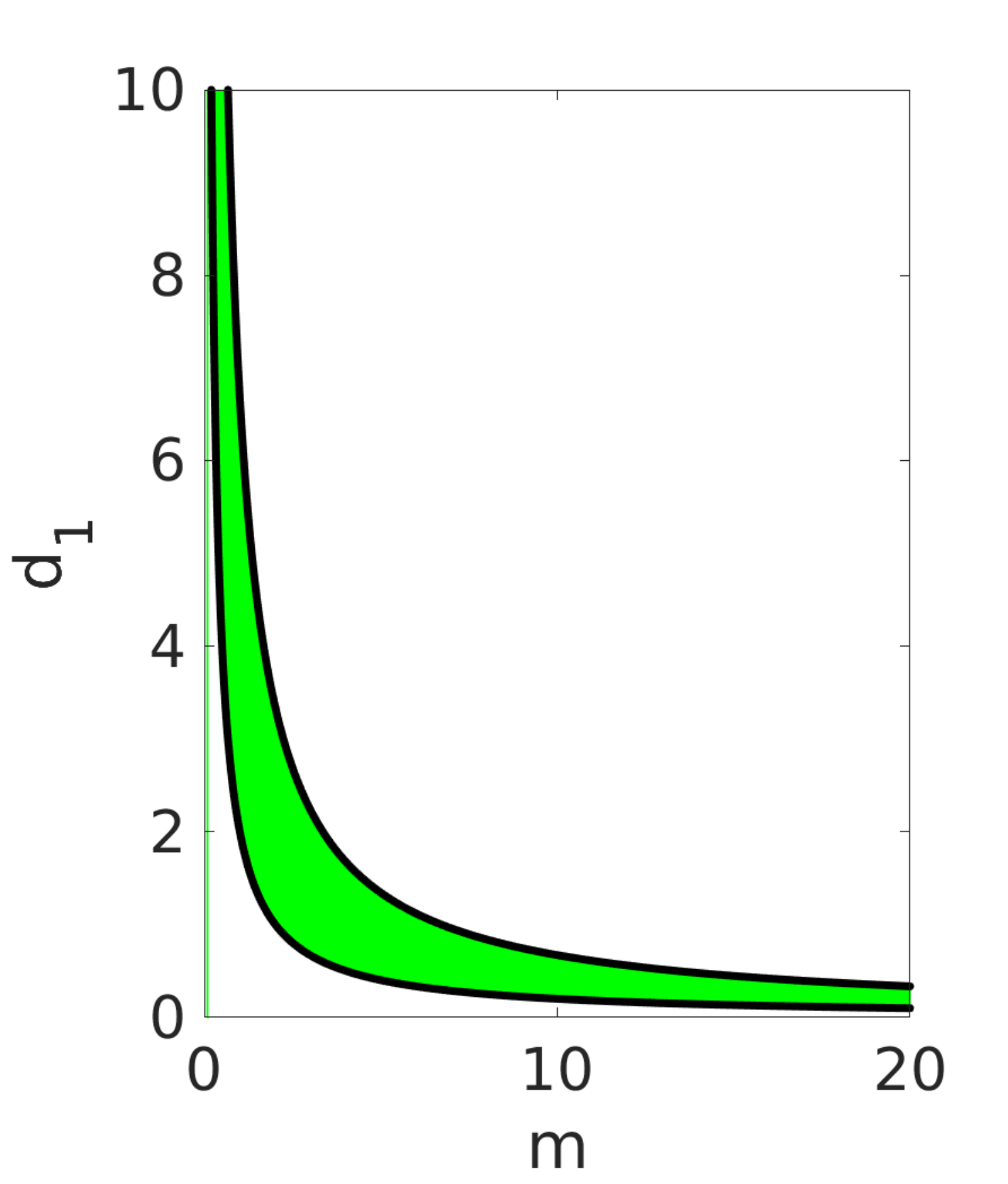}
        \caption{$d_{1}$ versus $m$.}
        \label{ID_d1_m}
    \end{subfigure}
    \quad 
    \begin{subfigure}[t]{0.22\textwidth}
        \includegraphics[width=\textwidth,valign=t]{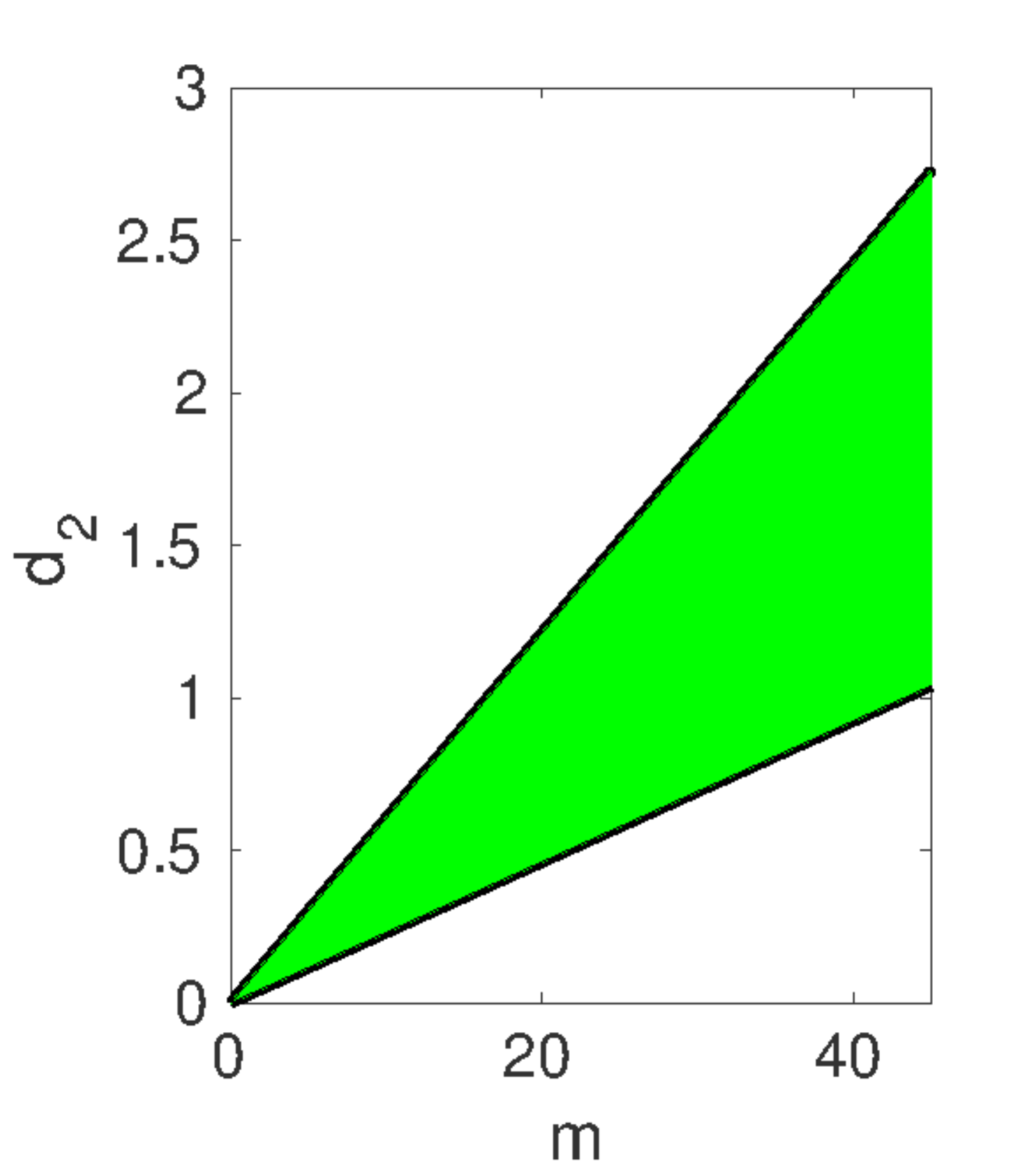}
        \caption{$d_{2}$ versus $m$.}
        \label{ID_d2_m}
    \end{subfigure} 
    \quad 
\begin{subfigure}[t]{0.22\textwidth}
        \includegraphics[width=\textwidth,valign=t]{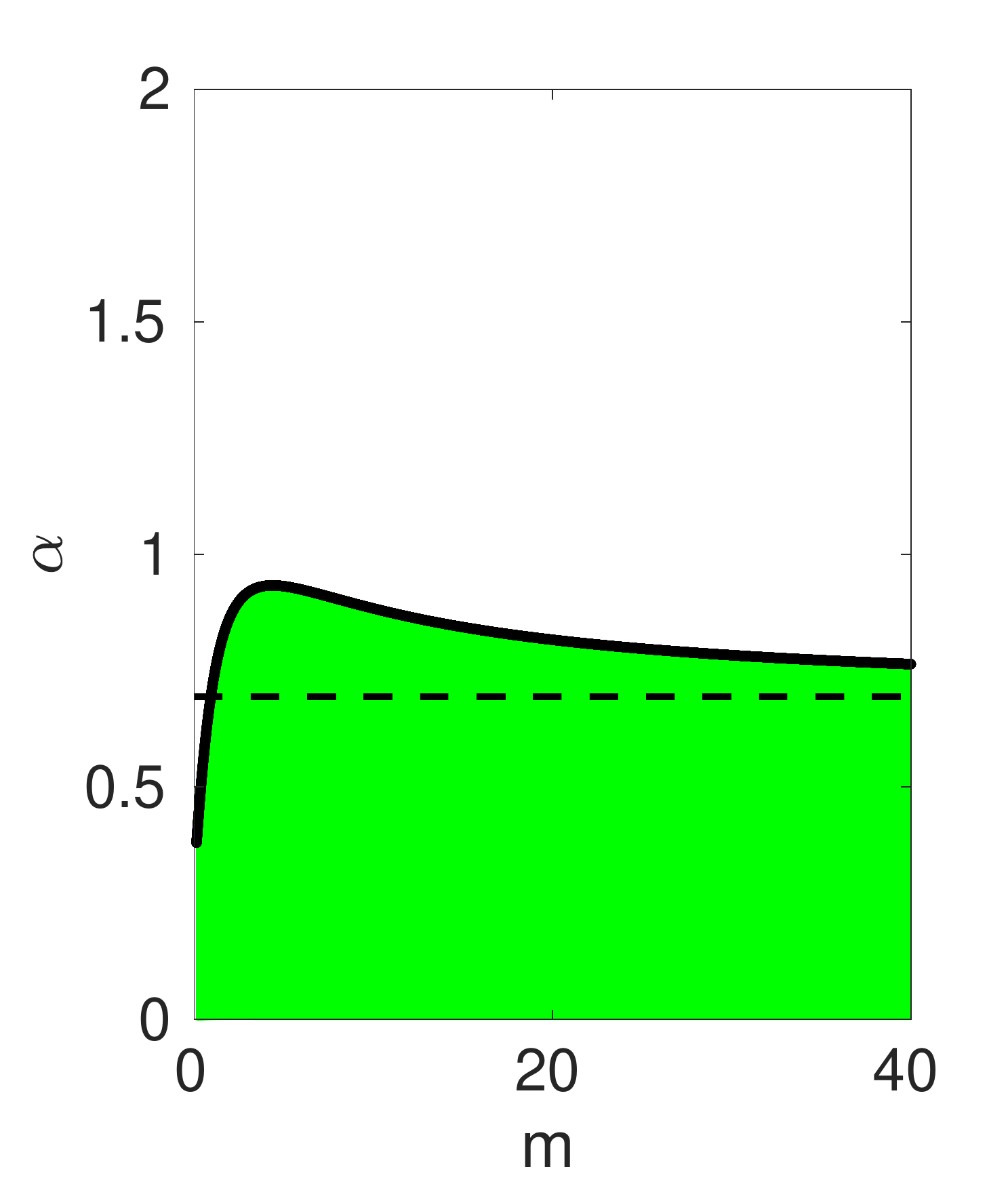}
        \caption{$\alpha$ versus $m$.}
        \label{ID_alpha_m}
    \end{subfigure}
    \quad 
    \begin{subfigure}[t]{0.22\textwidth}
        \includegraphics[width=\textwidth,valign=t]{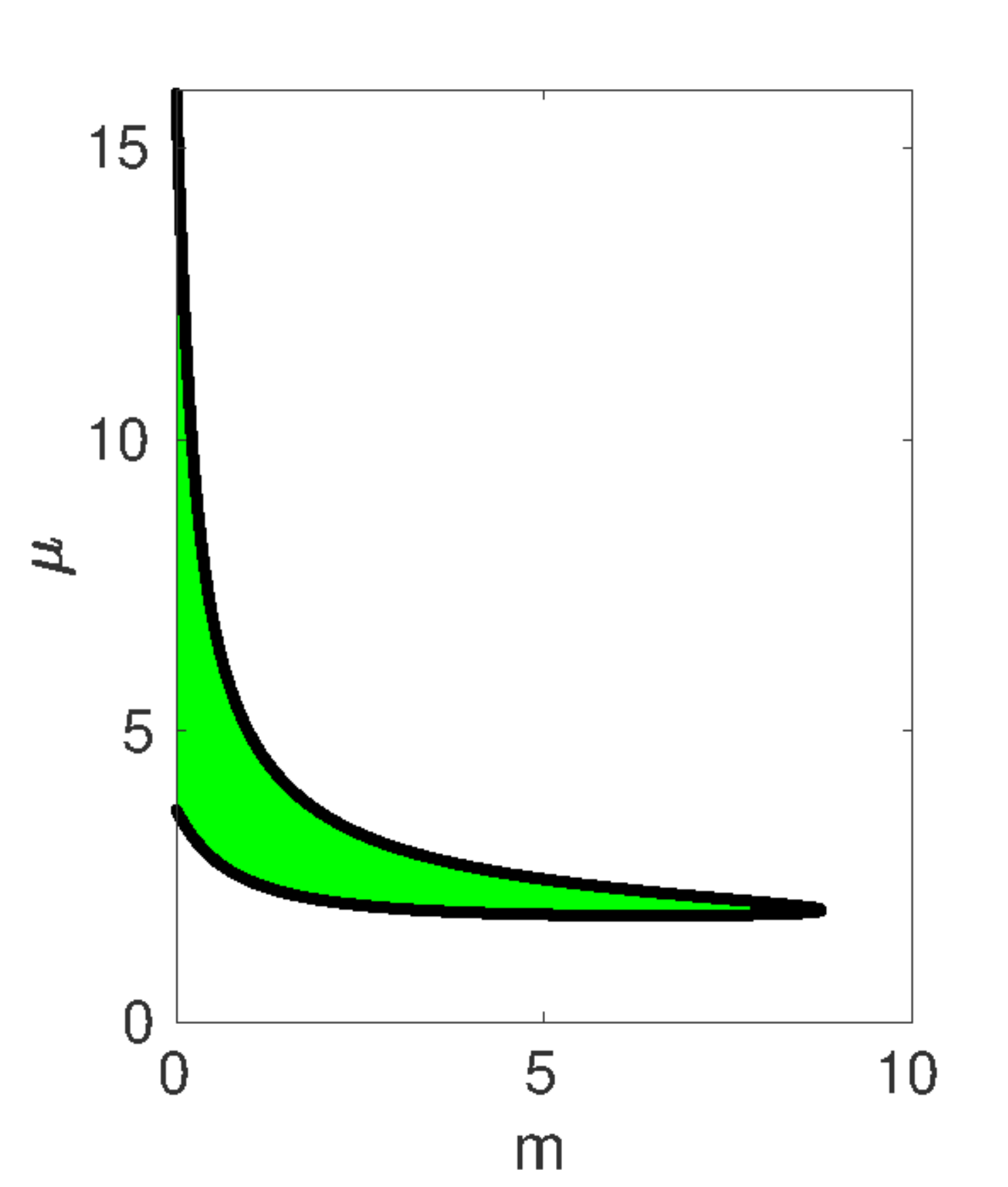}
        \caption{$\mu$ versus $m$.}
        \label{ID_Mu_m}
    \end{subfigure}      	    
    \caption{Two-parameter bifurcation diagrams computed from the ODEs
      \eqref{WM_ODEsys_2D} showing regions of instability with respect
      to certain parameters and the number of identical
      cells. Linearly stable periodic solutions are predicted in the
      green-shaded regions, while the unshaded regions correspond to
      linearly stable steady-state solutions. The parameters are
      $d_{1,j}=0.8$ and $d_{2,j}=0.2$ for $j=1,\dots,m$, with
      $\mu = 2$, $\alpha = 0.9$, $\epsilon_0=0.15$ and $\tau = 0.5$
      (except when used as a bifurcation parameter). (a) and (b) are for
      the permeability parameters $d_{1}$ and $d_{2} $, respectively,
      (c) shows the result for $\alpha$. {The black dashed
        horizontal line at $\alpha\approx 0.694$ is the Hopf
        bifurcation threshold of a single isolated cell when uncoupled
        from the bulk.} (d) is for $\mu$.}
     \label{Bifur_M}
\end{figure}

\begin{figure}[htbp]
    \centering
    \begin{subfigure}[t]{0.22\textwidth}
      \includegraphics[width=\textwidth,valign=t]{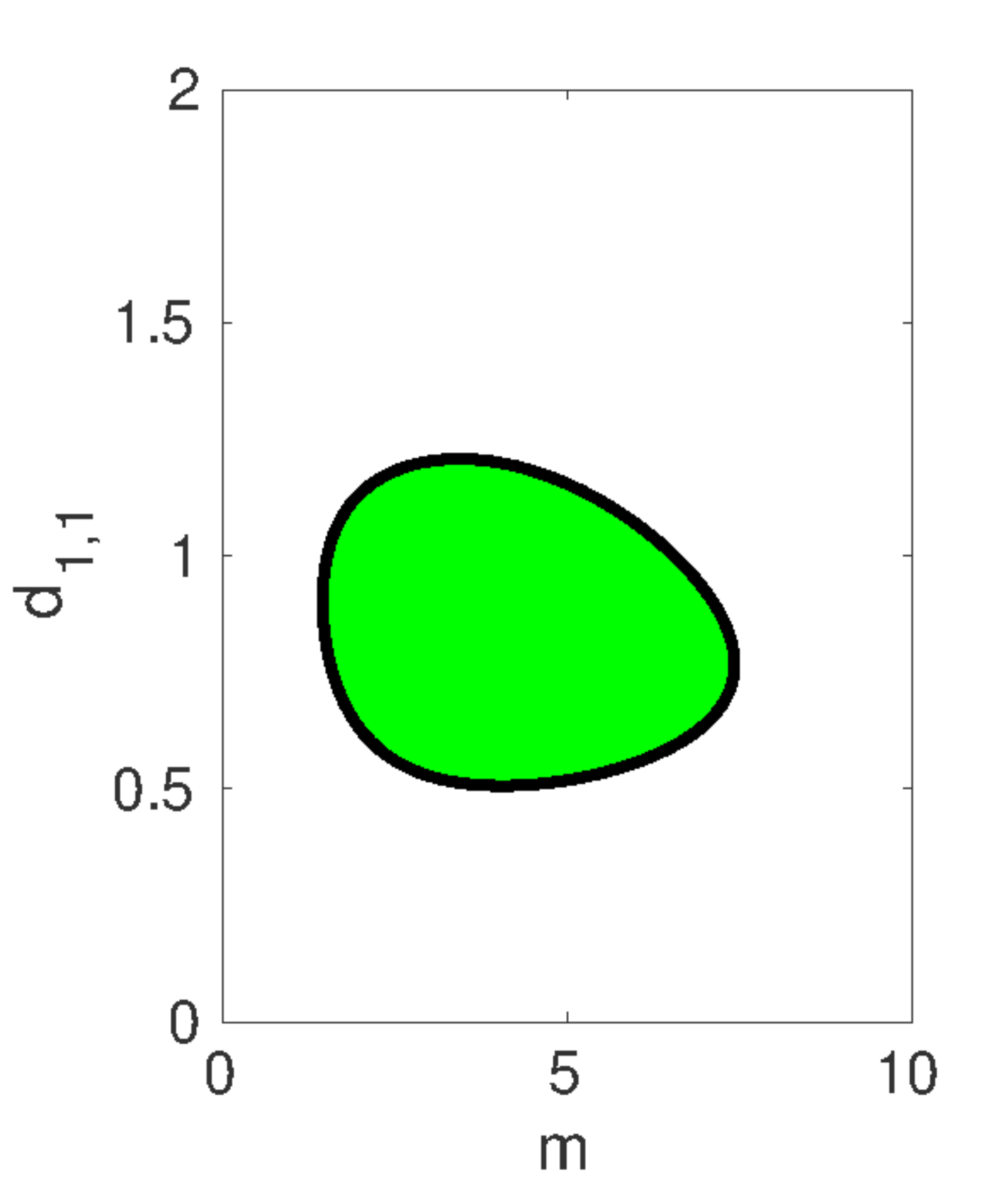}
      \caption{$d_{1,1}$ versus $m$.}
        \label{Def_d1_m}
    \end{subfigure}   
     \quad 
    \begin{subfigure}[t]{0.22\textwidth}
        \includegraphics[width=\textwidth,valign=t]{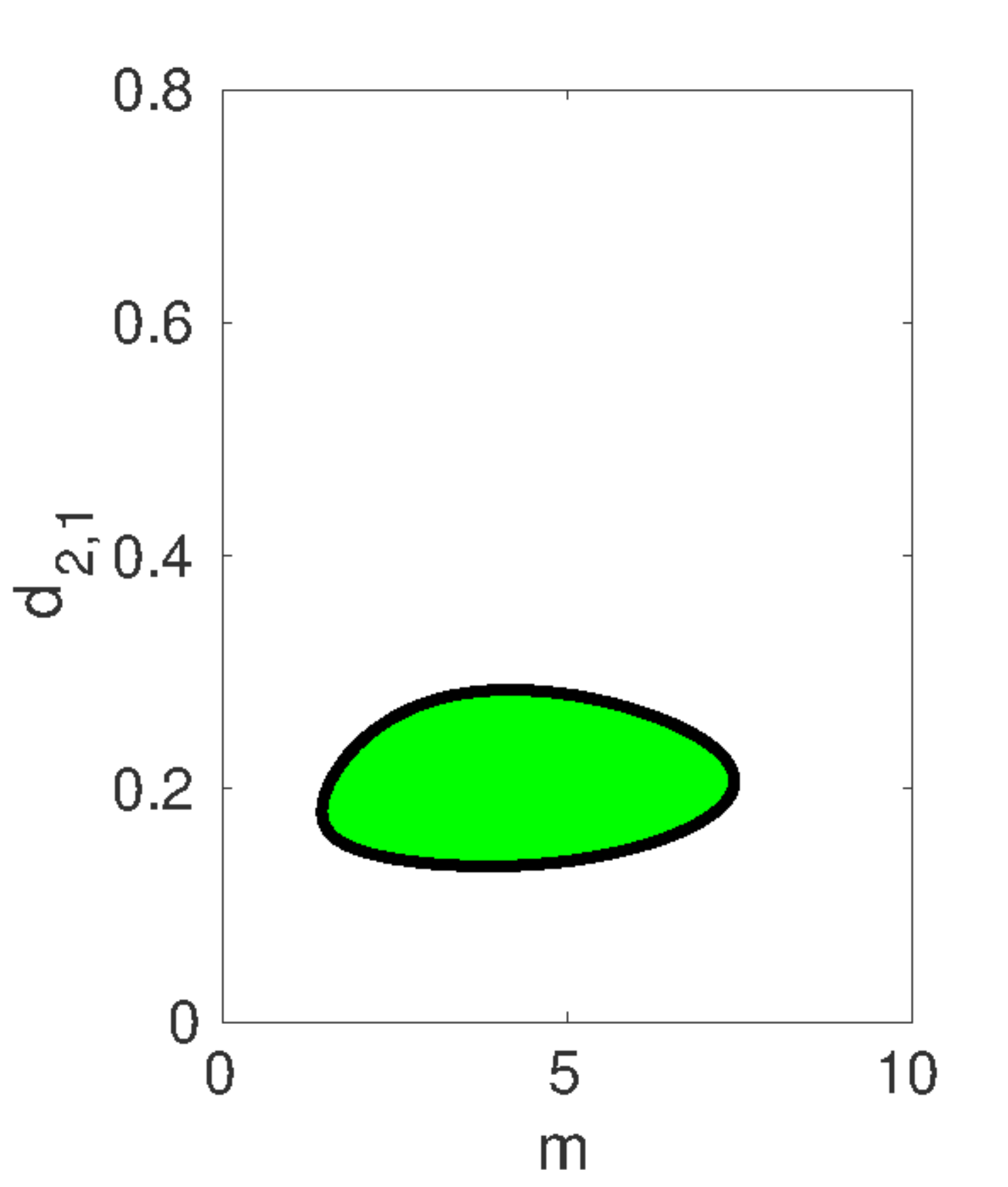}
        \caption{$d_{1,2}$ versus $m$.}
        \label{Def_d2_m}
    \end{subfigure}   
     \quad 
    \begin{subfigure}[t]{0.22\textwidth}
        \includegraphics[width=\textwidth,valign=t]{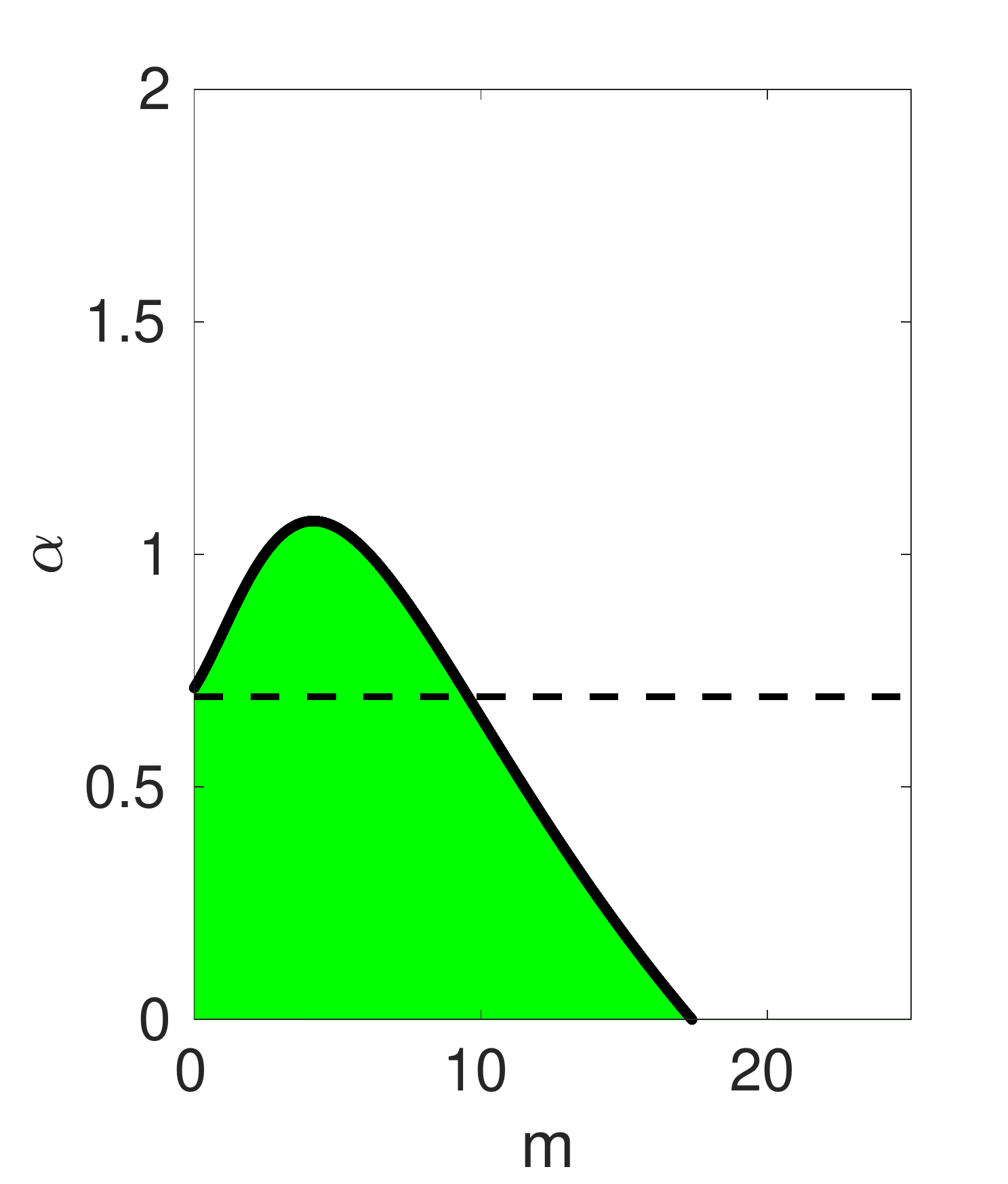}
        \caption{$\alpha$ versus $m$.}
        \label{Def_Alpha_m}
    \end{subfigure}  
     \quad 
    \begin{subfigure}[t]{0.22\textwidth}
        \includegraphics[width=\textwidth,valign=t]{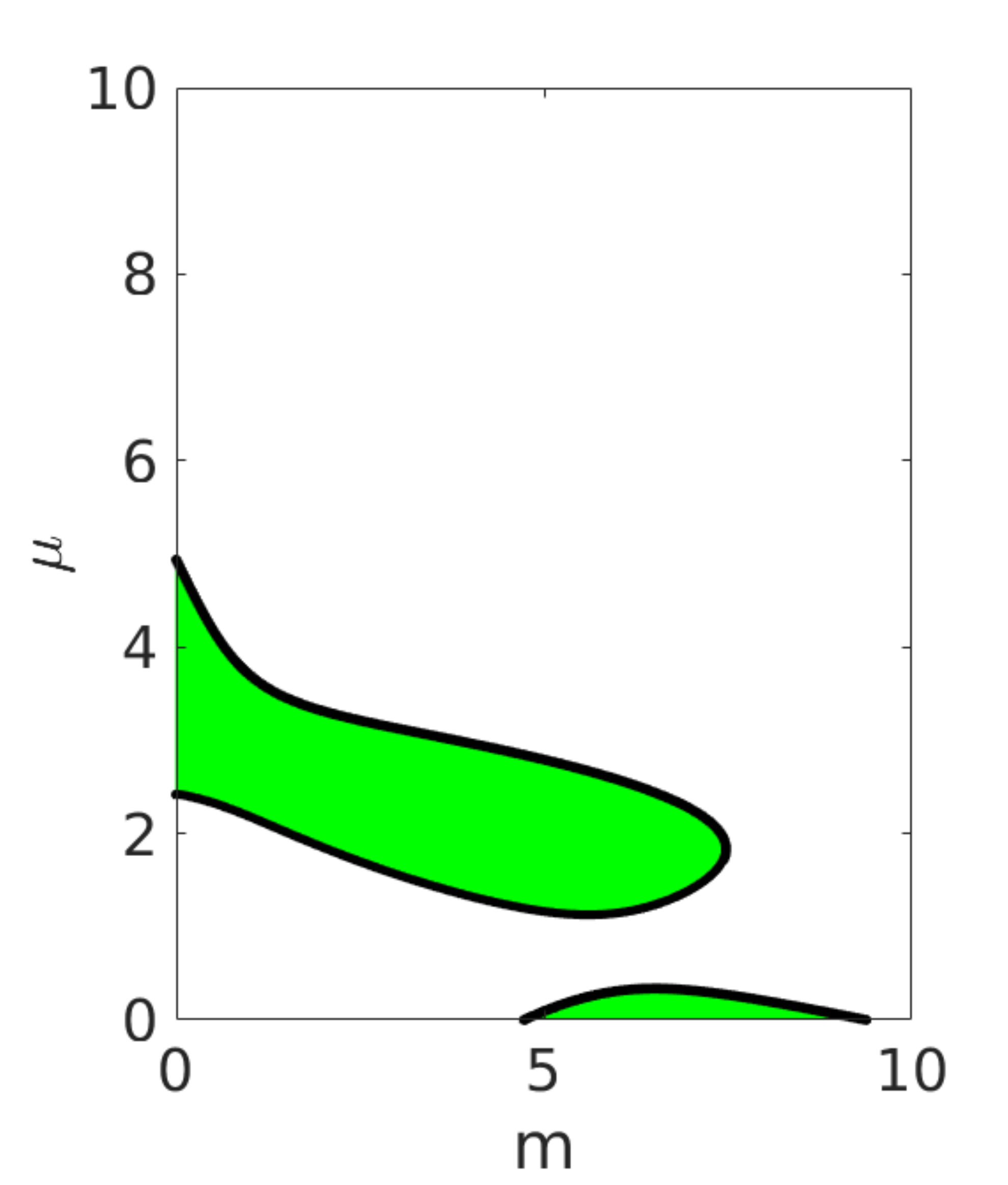}
        \caption{$\mu$ versus $m$.}
        \label{Def_Mu_m}
    \end{subfigure}    	    
    \caption{Two-parameter bifurcation diagrams computed from the ODEs
      \eqref{WM_ODEsys_2D} showing regions of instability with respect
      to certain parameters and the number of cells. Linearly stable
      periodic solutions are predicted in the green-shaded regions
      while the unshaded regions correspond to linearly stable
      steady-state solutions. The parameters are $d_{1,j}=0.8$ and
      $d_{2,j}=0.2$ for $j=1, \dots,m$, with $\mu = 2$, $\alpha = 0.9$
      $, \epsilon_0=0.15$ and $\tau = 0.5$ (except when used as the
      bifurcation parameter). The cells are not all identical, and the
      bifurcation parameters $d_{1,1}$, $d_{2,1}$, $\alpha$ and $\mu$
      are for the defective cell only.  $m$ is the population of
      identical cells. (a) and (b) are for the permeability parameters
      $d_{1,1}$ and $d_{2,1}$, respectively, (c) is for $
      \alpha$. {The black dashed horizontal line at
        $\alpha\approx 0.694$ is the Hopf bifurcation threshold of a
        single isolated cell when uncoupled from the bulk.}  (d) is
      for $\mu$. } \label{Bifur_M_Def}
\end{figure}

{In this subsection, we determine how parameter regimes for
  linearly stable periodic solutions change as the number of cells
  increases. Our findings are presented as two-parameter Hopf
  bifurcation diagrams, where the vertical axis is one of our main
  bifurcation parameters and the horizontal axis is the number of
  cells, the latter of which is treated for convenience as a
  continuous rather than a discrete variable.} In Figure \ref{Bifur_M}
and Figure \ref{Bifur_M_Def}, linearly stable periodic solutions are
predicted in the green-shaded regions, while linearly stable solutions
are predicted in the unshaded regions. The parameter values are in the
caption of these figures. Since synchronous oscillations occur as the
cell population increases, these results can also be used to study
quorum sensing behavior for a fixed value of each parameter. {Starting
  with the case of identical cells, Figure \ref{ID_d1_m} shows the
  result for the permeability parameter $d_{1}$, which agrees with the
  result in Figure $18$ of \cite{jia2016}. It predicts that linearly
  stable periodic solutions can be triggered in the system for
  $m \leq 35 $. For each population size $m$, and for fixed $d_2$,
  there is a range of $d_1$ for which synchronous oscillations exist,
  and this range widens as the number of cells decreases. This trend
  is qualitatively different than that shown in Figure~\ref{ID_d2_m}
  with regards to $d_2$ at a fixed value of $d_1$. For this
  permeability parameter, linearly stable periodic solutions exist for
  all values of $m$ in some finite interval of $d_2$, and this range
  widens as the number of cells increases.  Since $d_2$ is the efflux
  rate out of a cell, the observation that the Hopf bifurcation
  threshold for $d_2$ increases as $m$ increases, indicates that more
  signaling chemical is needed to synchronize the intraceulluar
  dynamics of a larger number of cells.  Figure~\ref{ID_alpha_m} shows
  the corresponding result for the Sel'kov parameter $\alpha > 0$. The
  black dashed horizontal line in this figure at $\alpha\approx 0.694$
  is the Hopf bifurcation threshold of a single isolated cell when
  uncoupled from the bulk. We observe from this figure that linearly
  stable periodic solutions can occur for any number of cells, even
  when each cell has a linearly stable steady-state when uncoupled
  from the bulk. However, as seen in Figure \ref{ID_Mu_m}, this is not
  the case for $\mu> 0$, where linearly stable synchronous solutions
  are predicted only for $m \leq 8$ on some range of $\mu$ that
  decreases as the number of cells increases. No oscillation is
  predicted for any value of $\mu$ when $m \geq 9$.}

{Similar results, but predicting qualitatively different
  outcomes, are presented in Figure~\ref{Bifur_M_Def} for the case of
  a single defective cell coupled to a group of identical cells. In
  these results, the bifurcation parameters $d_1$, $d_2$, $\alpha$ and
  $\mu$ are for the defective cell only, and $m$ is the population of
  identical cells. From Figure~\ref{Def_d1_m}, for $2\leq m\leq 7$,
  there is a finite range of permeabilities $d_{1,1}$ for which
  intracellular oscillations can occur. On this range of $m$,
  variations in $d_{1,1}$ can either trigger or extinguish
  intracellular oscillations. No oscillations can occur for
  $m \geq 8$.  From Figure \ref{Def_d2_m} similar conclusions hold for
  the second permeability parameter $d_{2,1}$.  We observe from
  Figures~\ref{Def_d1_m} and \ref{Def_d2_m} that variations in the 
  cell population $m$ also play a dual role of triggering and
  quenching intracellular oscillations.}

{In Figure~\ref{Def_Alpha_m} we show that when the number of
  identical cells with a common Sel'kov parameter $\alpha=0.9$
  satisfies $m\leq 18$, intracellular oscillations for the group of
  cells can still occur when a defective cell is included, provided
  that the Sel'kov parameter for the defective cell is below some
  threshold. Without the defective cell, if there are more than seven
  identical cells with a common Sel'kov parameter $\alpha=0.9$ there
  are no oscillations, as $m\geq 7$ and $\alpha=0.9$ is not within the
  green-shaded region of Figure \ref{ID_alpha_m}. However, on the
  range $10\leq m\leq 18$, we observe from Figure~\ref{Def_Alpha_m}
  that to maintain intracellular oscillations the defective cell is no
  longer a conditional oscillator, as its required value of $\alpha$
  is below the horizontal black dashed line in
  Figure~\ref{Def_Alpha_m}. With regards to $\mu$,
  Figure~\ref{Def_Mu_m} shows the qualitatively different feature that
  there are two regimes of $\mu$ for which intracellular oscillations
  occur for different ranges of $m$. As $m$ increases, there is
  initially only one regime in $\mu$ for which periodic solutions
  exists. However, as the cell population increases, a new regime
  emerges when $m=5$. These two regimes co-exist until $m=8$, after
  which the first regime vanishes, and the second one continues until
  $m=9$. There are no intracellular oscillations in the system for
  $m>9$.}

\section{Quorum sensing and phase synchronization: The Kuramoto order parameter}\label{Section_QS}

In this section, we use the ODE system \eqref{WM_ODEsys_2D} to study
quorum sensing (QS) behavior, characterized by the emergence of
collective cell dynamics that arises from a quiescent state as the
cell population density exceeds a threshold. Since the ratio of the
area occupied by $m$ circular cells of a common radius $\varepsilon$
to the overall domain area is ${m\pi\varepsilon^2/|\Omega|}$, our
measure of the cell density is taken as $\rho \equiv {m/|\Omega|}$.
With our Sel'kov kinetics, we will define collective dynamics
as the emergence of synchronous oscillations in the intracellular
dynamics as $\rho$ exceeds a critical value. 

In terms of the parameter $\rho \equiv m/|\Omega|$, the ODE system
\eqref{WM_ODEsys_2D} becomes
\begin{equation}\label{QS_ODEsys}
\begin{split}
  \frac{dU_0}{dt} & = - \frac{1}{\tau} U_0  - \frac{\rho}{m}
  \sum_{j=1}^{m}  (k_{1,j}\, U_0 - k_{2,j} u_j^1)\,, \quad
  \frac{d \pmb{u}_j}{dt}  =  \, \pmb{F}_j \left( \pmb{u}_j  \right)  +
  \pmb{e}_1 \, ( k_{1,j} \, U - k_{2,j} \, u_j^1 )\,, \qquad j = 1, \ldots, m\,,
\end{split}
\end{equation}
where we have defined $k_{1,j} \equiv (2\pi d_{1,j})/\tau$ and
$k_{2,j} \equiv (2\pi d_{2,j})/\tau$.  For the case of identical
cells, we will perform a Hopf bifurcation analysis on this system in
order to identify the range of instability of the steady-state as
$\rho$ is varied. For the case of non-identical cells, in \S
\ref{QS_sync} the Kuramoto order parameter (cf.~\cite{Kuramoto},
\cite{Rossler}, \cite{RosslerHeter}) will be used to measure the
degree of phase synchrony of the intracellular oscillations.

\subsection{Hopf bifurcation  analysis for identical cells}\label{hopf:iden}

For the case where the cells are all identical we label
$k_1 \equiv k_{1,j}$ and $k_2 \equiv k_{2,j}$, and introduce the
common local variables $\pmb{u} = \pmb{u}_j$ and
$ \pmb{F}(\pmb{u}) \equiv \pmb{F}_j(\pmb{u}_j)$ for $j=1,\dots,m$. For
this identical cell case, \eqref{QS_ODEsys} reduces to
\begin{equation}\label{QS_Reduce}
\begin{split}
  \frac{dU_0}{dt} & = - \frac{1}{\tau} U_0  - \rho\,
  (k_{1}\, U_0 - k_{2} u^1)\,,\qquad
  \frac{d \pmb{u}}{dt}  =  \, \pmb{F} \left( \pmb{u}  \right)  +
  \pmb{e}_1 \, ( k_{1} \, U - k_{2} \, u^1 ) \,,
\end{split}
\end{equation}
where $U_0 \equiv U_0(t)$ is the leading-order concentration of the
signaling chemical in the bulk region and
$\pmb{u} = (u^{1}, \ldots, u^{n})^T$ represents the $n$ interacting
chemical species in the cells. Upon substituting the Sel'kov reaction
kinetics \eqref{Selkov} into \eqref{QS_Reduce}, for which $n=2$, we readily
calculate that the steady-state solution $ (U_{0e},u_e^1,u_e^2)$ is
\begin{equation}\label{QS_SS}
\begin{split}
  U_{0e}= \frac{\tau \mu \rho k_2}{(1 + k_2+ \tau \rho k_1)}\,, \quad
  u_e^1 = \frac{ \mu(1+ \tau \rho k_1 )}{(1 + k_2+ \tau \rho k_1)}\,,
  \quad u_e^2 = \frac{\mu}{\left(\alpha + (u_e^1)^2\right)}\,, \quad
  \text{where} \quad \pmb{u}_e = (u_e^1, u_e^2)^T\,.
\end{split}
\end{equation}
From \eqref{QS_SS} we observe that both the steady-state bulk
concentration $U_{0e}$ and $u_e^{1}$ are increasing functions of the
cell density parameter $\rho$. Moreover, as expected, $U_{0e}$ is an
increasing function of the cell efflux rate $k_2$, but decreases as
the influx rate $k_1$ into the cells increases. As the influx rate
$k_1$ into the cells from the bulk medium increases, the intracellular
level $u_{e}^{1}$ also increases. In the large cell density limit 
we obtain from \eqref{QS_SS} that
\begin{equation}\label{qs:large_rho}
  u_e^{1}\to \mu \,, \qquad u_{e}^{2} \to \frac{\mu}{\alpha + \mu^2} \,,
  \qquad U_{0e}\to \frac{\mu k_2}{k_1} \,, \qquad \mbox{as} \quad \rho\to
  \infty \,.
\end{equation}
These limiting values for the intracellular species are the same as those
for a single cell that is uncoupled from the bulk medium, as summarized in
\S \ref{GlobalBifur}.

\begin{figure}[htbp]
    \centering
    \begin{subfigure}[t]{0.40\textwidth}
        \includegraphics[width=\textwidth,height=4.5cm]{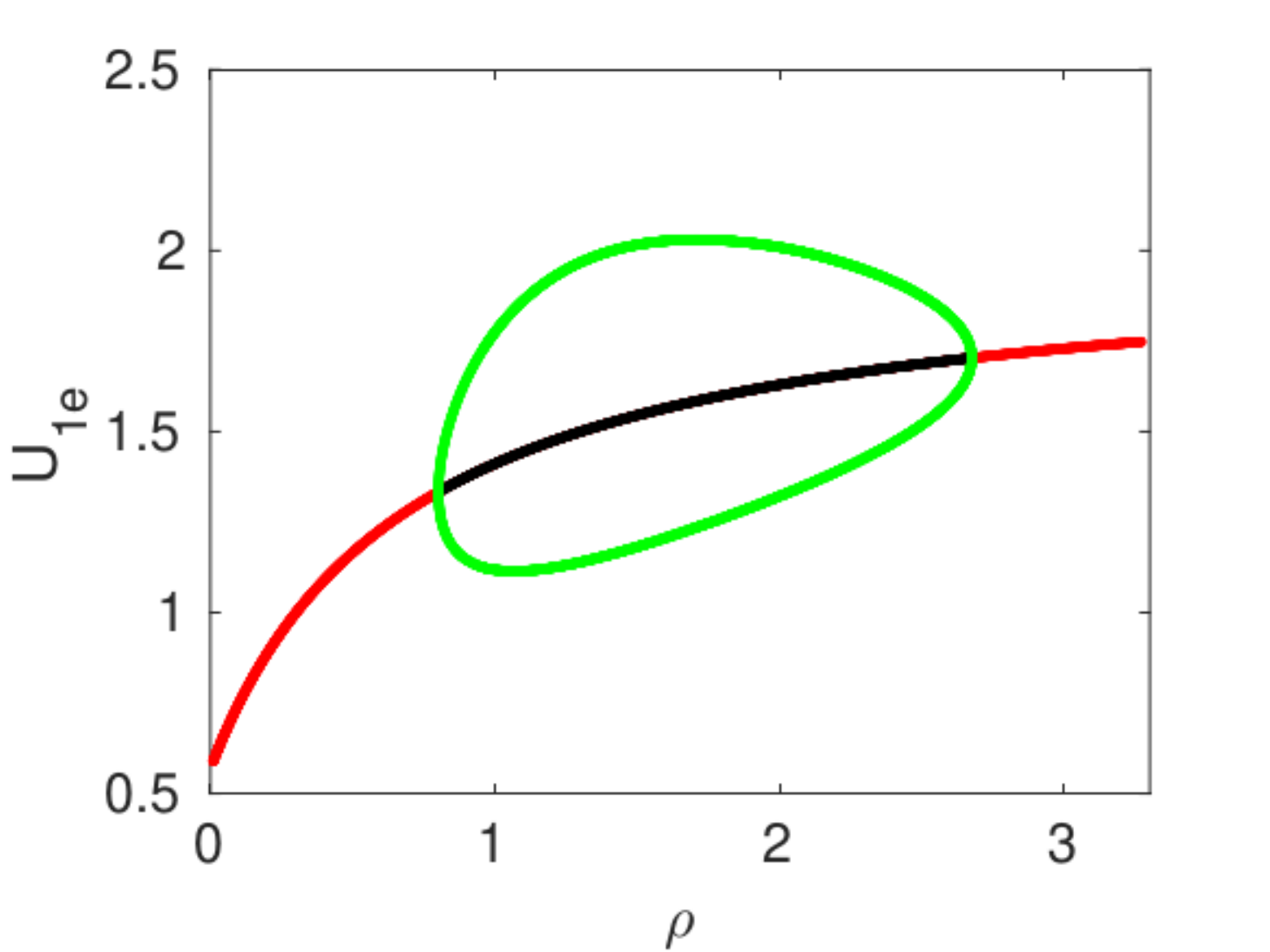}
        \caption{One-parameter Hopf Bifurcation for $u_{1e}$ vs $\rho$ }
        \label{QS_BFa}
    \end{subfigure}
    \qquad
    ~ 
    \begin{subfigure}[t]{0.40\textwidth}
        \includegraphics[width=\textwidth,height=4.5cm]{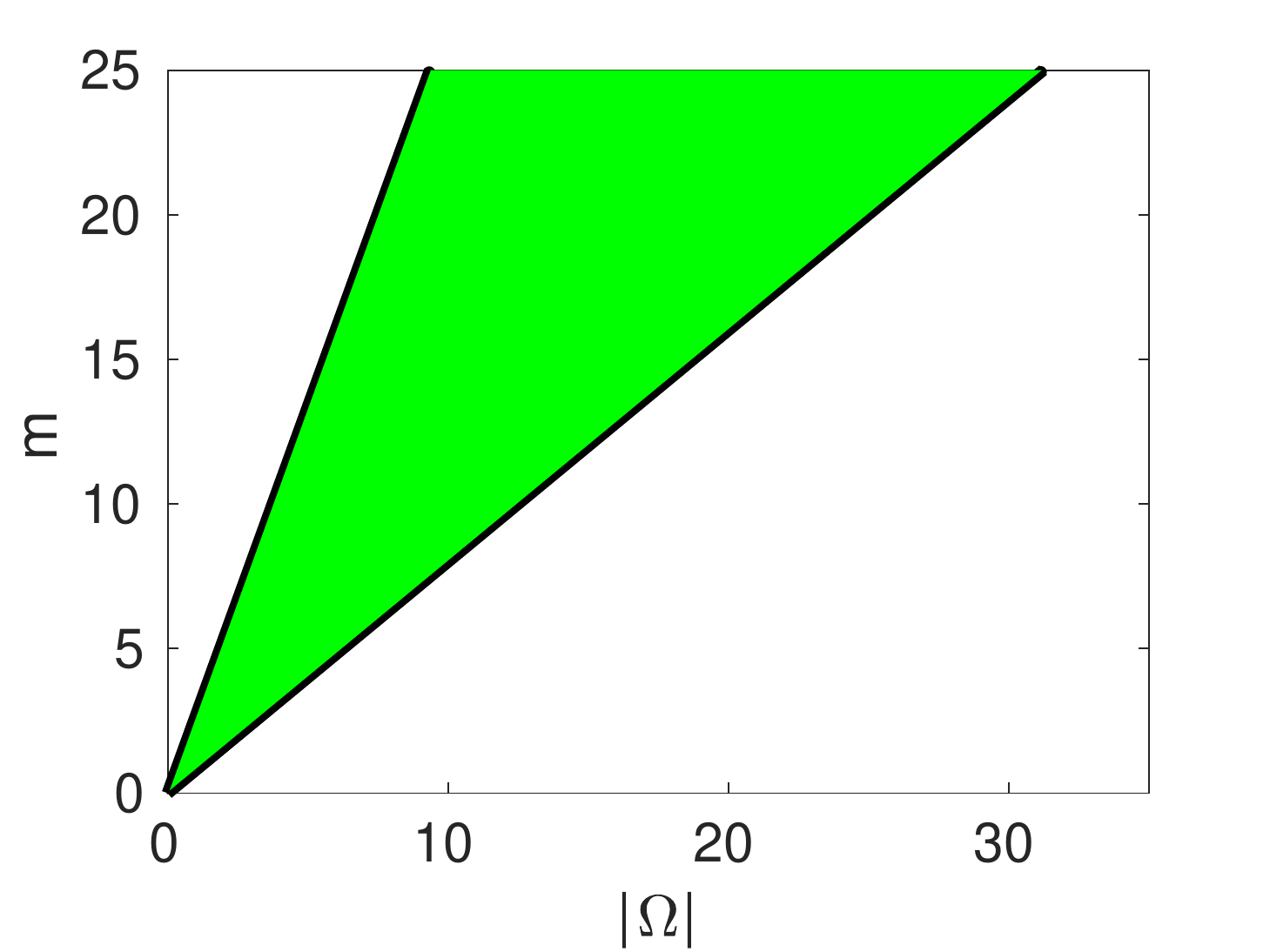}
        \caption{Hopf bifurcation diagrams for $m$ vs $|\Omega|$ }
        \label{QS_BFb}
    \end{subfigure}
    \caption{(a) Global bifurcation diagrams for $u_{1e}$ versus the
      cell population density parameter $\rho \equiv {m/|\Omega|}$, as
      computed from the ODEs \eqref{QS_Reduce}, showing steady-states
      and global branches of periodic solutions for the case of
      identical cells. The red and black lines correspond to linearly
      stable and unstable steady-state solution branches,
      respectively. The green loop indicates a linearly stable branch
      of periodic solutions. The Hopf bifurcation points are at
      $\rho_1 = 0.8010$ and $\rho_2 = 2.6750$. (b) A two-parameter
      bifurcation diagram for the number of cells $m$ and the domain
      area $|\Omega|$. The green wedge-shaped region predicts stable
      periodic solutions while the unshaded regions correspond to
      linearly stable steady-states.  In (a) and (b) the Sel'kov
      parameters are $ \mu=2, \alpha=0.9$ and $ \epsilon_0=0.15$,
      with $\tau=0.5, k_1=10.0531$ and $k_2=2.5133$, which corresponds
      to the permeabilities $d_1=0.8$ and $d_2=0.2$.}  \label{QS_BF}
\end{figure}

To determine the linear stability of this steady-state, we introduce
the perturbation
\begin{equation}\label{QS_SS_pertub}
\begin{split}
U & = U_{0e} + e^{\lambda t}\eta \,,\qquad
 \pmb{u}= \pmb{u}_{e} +  e^{\lambda t} \pmb{\phi} \,,
\end{split}
\end{equation}
where $|\pmb{\phi}|\ll 1$ and $\eta\ll 1$.  Upon substituting
\eqref{QS_SS_pertub} into \eqref{QS_Reduce}, we obtain the linearized
system 
\begin{equation}\label{QS_Linear}
\begin{split}
\lambda \eta &= -\frac{\eta}{\tau} -\rho (k_1\eta - k_2 \phi_1)\,,\qquad	
\lambda \pmb{\phi} = J_e \pmb{\phi} + \pmb{e}_1 (k_1\eta - k_2 \phi_1)\,,
\end{split}
\end{equation}
where $J_e$ is the Jacobian matrix of the local kinetics
$\pmb{F}(\pmb{u}) = (F(u^1, u^2),G(u^1, u^2))^T$ evaluated at the
steady-state solution $\pmb{u}_e = (u_e^1, u_e^2)^T$, and
$\pmb{\phi} = (\phi_1,\phi_2)^T $. For convenience of notation, we let
$(v,w) \equiv (u_e^1, u_e^2)$, and write the linearized system
\eqref{QS_Linear} in matrix form as
\begin{equation}\label{QS_Matrix}
\begin{split}
\mathcal{H}(\lambda) \pmb{\Psi}  = \pmb{0} \,,
\end{split}
\end{equation}
where $\pmb{\Psi} = (\eta, \phi_1, \phi_2)^T$ and
$\mathcal{H}(\lambda)$ is the $3 \times 3$ matrix defined by
\begin{equation}\label{QS_MatrixH}
\begin{split}
\mathcal{H}(\lambda) = \begin{pmatrix}
-\left(\frac{1}{\tau} + \rho k_1 \right)  -\lambda & \rho k_2 & 0 \\
\\
k_1   & (F_v^e - k_2 -\lambda)   & F_w^e \\
\\
0  & G_v^e  & (G_w^e - \lambda)
\end{pmatrix}\,.
\end{split}
\end{equation}
Here $F_v^e, F_w^e, G_v^e$, and $G_w^e$ denote the partial derivatives of
$F$ or $G$ with respect to $v$ or $w$ evaluated at the steady-state
$ \pmb{u}_e$.  A simple calculation shows that there is a
nontrivial solution to \eqref{QS_Matrix}  if only if $\lambda$ is a root
of the cubic
\begin{subequations}\label{QS_cubic}
\begin{equation}\label{QS_Poly}
\begin{split}
\lambda^3 + p_1 \lambda^2 + p_2 \lambda + p_3 = 0\,,
\end{split}
\end{equation}
whose coefficients are given by
\begin{equation}\label{QS_CoePoly}
\begin{split}
  p_1 \equiv \left(k_2 + \frac{1}{\tau} + \rho k_1\right)& - \text{tr}(J_e)\,,
  \qquad p_2 \equiv (1+k_2) \det (J_e) - \left( \frac{1}{\tau} + \rho k_1\right)
  \text{tr}(J_e) + \frac{k_2}{\tau}\,,\\
  p_3& \equiv \left( \frac{1}{\tau} + \rho k_1 + \frac{k_2}{\tau}\right)
  \det (J_e)  \,.
\end{split}
\end{equation}
\end{subequations}
Here $\det (J_e) = F_v^e G^e_w - F_w^e G^e_v>0$ and
$\text{tr}(J_e) = F_v^e + G_w^e$ are the determinant and trace of
$J_e$ for Sel'kov reaction kinetics, for which we readily calculate
that $\det(J_e)=-G_w^{e}=\epsilon_0(\alpha + v_e^2)$.  By the
Routh-Hurwitz criterion for cubic functions, the three roots of the
polynomial \eqref{QS_Poly} satisfy $\mbox{Re}(\lambda) < 0$ if and
only if the following conditions hold
\begin{equation}\label{QS_RouthHurtwiz}
\begin{split}
p_1 > 0\,, \qquad p_3 > 0\,, \qquad \text{and} \quad p_1p_2 > p_3\,.
\end{split}
\end{equation}
From \eqref{QS_CoePoly} we observe that $p_3>0$ always holds. Since we are
interested in finding Hopf bifurcation points, we need to consider a
special cubic polynomial whose roots satisfy $\lambda_1 = a < 0$,
$ \lambda_{2,3} = \pm i \omega$, for which
\begin{equation}\label{QS_RT_Hopf}
\begin{split}
(\lambda - a)(\lambda - i\omega)(\lambda + i\omega) = \lambda^3 - a\lambda^2 + \omega^2 \lambda - a\omega^2 = 0 \,.
\end{split}
\end{equation}
Upon comparing the polynomials \eqref{QS_Poly} and \eqref{QS_RT_Hopf}, we
conclude that a Hopf bifurcation threshold must satisfy
\begin{equation}\label{QS_Hopf}
\begin{split}
p_1 > 0\,, \quad p_3 > 0\,, \quad \text{and} \quad p_1p_2 = p_3 \,.
\end{split}
\end{equation}
For the parameter values in the caption of Figure \ref{QS_BF}, we use
\eqref{QS_Hopf} to numerically compute two Hopf bifurcation points at
$\rho_1 = 0.8010$ and $\rho_2 = 2.6750$. These results agree with
those computed in Figure \ref{QS_BF} using XPPAUT \cite{xpp2002}.

Figure \ref{QS_BFa} shows a one-parameter Hopf bifurcation diagram for
the ODE system \eqref{QS_Reduce} with respect to the cell population
density $\rho$ parameter. Observe from this figure that the range of
$\rho$ that predicts stable periodic solutions is bounded. This
indicates that the cell population density plays a dual role of both
triggering and then quenching intracellular oscillations, which agrees
with our earlier observation in Figure~\ref{Bifur_M_Def}.  Our results
in Figure \ref{QS_BFa} show that there are no synchronous oscillations
for $\rho < 0.8010$ and for $\rho > 2.6750$, and that linearly stable
periodic solutions exist only on the range
$0.8010 \leq \rho \leq 2.6750$.  As $\rho$ increases above
$\rho = 0.8010$, synchronous oscillations are triggered through
quorum sensing behavior. These two Hopf bifurcation values for $\rho$
generate the green wedge-shaped region shown in Figure \ref{QS_BFb} in
the $m$ versus $|\Omega|$ parameter plane where synchronous
intracellular oscillations occur. For any vertical slice through the
bifurcation diagram in Figure \ref{QS_BFb}, corresponding to a domain
of fixed area $|\Omega|$, there is a quorum sensing threshold for the
number of cells required to trigger synchronous
oscillations. Moreover, there is an additional threshold on the number
of cells for which the oscillations are quenched.  This clearly shows
that the cells adjust their intracellular dynamics in accordance with
the number of cells in the population. When the domain area is small,
fewer cells are required to trigger oscillations, while as the domain
area increases the number of cells required for quorum sensing also
increases. On the other hand, the quenching of oscillations predicted
by Figure \ref{QS_BFb} results from having too many cells in a given
domain area. As the domain area increases, the number of cells
required to quench oscillations also increases, as expected
intuitively. The qualitative mechanism underlying the quenching of
oscillations is that when $\rho\gg 1$ the collection of cells behaves
like a single ``effective'' cell, which has no intracellular
oscillations and a unique linearly stable steady-state as result of the
Sel'kov parameters chosen. Mathematically, as $\rho\to \infty$, we
have $v_e\to \mu$ and $w_e\sim {\mu/(\alpha+\mu^2)}$ from
\eqref{qs:large_rho}, and so from \eqref{selkov:trace} and the fact
that the Sel'kov kinetic parameters were chosen so that a single
isolated cell has a linearly steady-state, we must have
$\text{tr}(J_e)<0$ for $\rho\gg 1$. Therefore, from \eqref{QS_CoePoly}
we obtain that $p_1>0$ for $\rho$ sufficiently large. Moreover, since
from \eqref{QS_CoePoly} we have
\begin{equation}\label{quench:estimate}
  p_1p_2\sim \rho^2k_1^2|\text{tr}(J_e)|={\mathcal O}(\rho^2) \,, \qquad
  p_3\sim \rho k_1\left(\det{J_e}\right)={\mathcal O}(\rho) \,, \qquad
  \mbox{as} \quad \rho\to \infty \,,
\end{equation}
we conclude that $p_1p_2>p_3$ for $\rho$ sufficiently
large. Therefore, when $\rho$ is sufficiently large the Routh-Hurwitz
conditions in \eqref{QS_RouthHurtwiz} are satisfied. As a result,
there must be a critical value $\rho_m$ of the cell density parameter
$\rho$, with $\rho_m$ sufficiently large, for which the steady-state
of the ODE system \eqref{QS_Reduce} is linearly stable for all
$\rho>\rho_m$.

We emphasize that the two bifurcation diagrams in Figure \ref{QS_BF}
are related. A vertical slice of the one-parameter Hopf bifurcation
diagram in Figure \ref{QS_BFa} at a population density of
$\rho = \rho_s$ corresponds to a straight line with slope $\rho_s$
(since $\rho = m/|\Omega|$) passing through the origin in the
two-parameter bifurcation diagram in Figure \ref{QS_BFb}. The Hopf
bifurcation points $\rho_1 = 0.8010$ and $\rho_2 = 2.6750$ in Figure
\ref{QS_BFa} are the slopes of the Hopf bifurcation boundaries in
Figure \ref{QS_BFb}. Any straight line passing through the origin and
contained between these boundaries corresponds to a vertical slice of
the one-parameter bifurcation diagram in Figure \ref{QS_BFa} at a
$\rho$ value between $\rho_1$ and $\rho_2$.

\subsection{Phase synchronization and the order parameter}\label{QS_sync}

Next, we study quorum sensing using the ODE system \eqref{QS_ODEsys}
together with the Kuramoto order parameter. The Kuramoto order
parameter was originally used to measure the degree of phase synchrony
of a system of coupled oscillators \cite{kuramoto1975self}. We will
use the version of the order parameter presented in
\cite{RosslerHeter} and \cite{Rossler}, given by
\begin{align}\label{QS_Kura_Par}
  R = \left\langle \left| N^{-1} \sum_{j=1}^N \,\, \exp[i \theta_j(t)] -
  \left\langle  N^{-1} \sum_{j=1}^N \,\,
  \exp[i \theta_j(t)] \right\rangle \right| \right\rangle \,,
\end{align}
which was used in \cite{RosslerHeter} and \cite{Rossler} to measure
the degree of phase synchrony of indirectly coupled chaotic
oscillators. Here $N$ is the number of oscillators, $\theta_j(t)$ is
the instantaneous phase of the $j^{\text{th}}$ oscillator, and
$\langle \dots \rangle$ represents average over time. The value of $R$
indicates the level of phase synchronization of the oscillators, and
it varies between 0 and 1, inclusive. When $R = 1$, the oscillators
are perfectly in phase, while the value $R=0$ indicates that they are
perfectly out of phase. Two oscillators are said to be in phase if
their frequency and phase angles are the same at each cycle as they
oscillate. Our goal is to use the system of ODEs \eqref{QS_ODEsys}
together with the order parameter \eqref{QS_Kura_Par} to measure the
degree of phase synchronization of the intracellular dynamics within
the signaling compartments as the population density $\rho$ is varied.
In our numerical examples below, the cell population will be taken to
be a mixture of identical and defective cells, where the heterogeneous
cells have one different Sel'kov parameter.

To measure the degree of synchronization of the cells, we first solved
the coupled system of ODEs \eqref{QS_ODEsys} numerically using
\textit{ODE45} in MATLAB \cite{shampine1997matlab}. The solution over
the transient period was discarded, after which we extracted the time
evolution of the same chemical species in all the cells. Only one of
the intracellular chemical species was used for this analysis because
both of them started oscillating and synchronizing at the same time. A
single mode Fourier series expansion was fitted to each of the
selected solutions, and their instantaneous phase $\theta(t)$ was
computed from the coefficients of the series. At each time point, the
phase of each oscillator was mapped to the complex number
$e^{i \theta}$ (where $i = \sqrt{-1}$) on the unit circle, and an
average of these points was computed using
$z = N^{-1} \sum_{j=1}^N e^{i \theta_j}$. For better intuition of the
meaning of this average, take each point on the unit circle to be a
unit vector. Then $z$ is a vector whose direction is the average
direction of all the vectors on the unit circle.  The length of this
average vector is bounded between 0 and 1, which indicates the degree
of phase synchrony of the oscillators at that time point. For
instance, if two oscillators are perfectly synchronized at a specific
time point, their phases at that point will be the same and they will
both be mapped to the same point on the unit circle. The average of
these two points/vectors gives another unit vector in the same
direction. The fact that their average is a unit vector verifies that
the two oscillators are in perfect synchrony at this point. On the
other hand, if the two oscillators are completely out of phase, their
phases will be mapped to opposite ends of the unit circle, and their
average will be the zero vector, so that $|z|=0$. The more
synchronized oscillators are at a time point, the closer the
corresponding mapping of their phases are on the unit circle, and as a
result the closer the value of $|z|$ is to 1. Thus, $|z|$ indicates
the level of synchrony of the oscillators at each time point. Upon
computing the instantaneous averages at each time point, we computed
their average over a specified time interval (after the system has
reached a quasi steady-state). Lastly, the modulus of the difference
between the instantaneous averages $z$ and the time-average
$\langle z \rangle$ was computed for each time point, and then the
corresponding result was averaged over time to calculate the order
parameter $R$, as given in \eqref{QS_Kura_Par}
(cf.~\cite{Rossler}, \cite{Kuramoto}).  In
our computations below, we set $R=0$ when the cells are in a quiescent
state and whenr oscillations have an amplitude less than
$1 \times 10^{-4}$.

\begin{figure}[htbp]
    \centering
    \begin{subfigure}[t]{0.31\textwidth}
        \includegraphics[width=\textwidth]{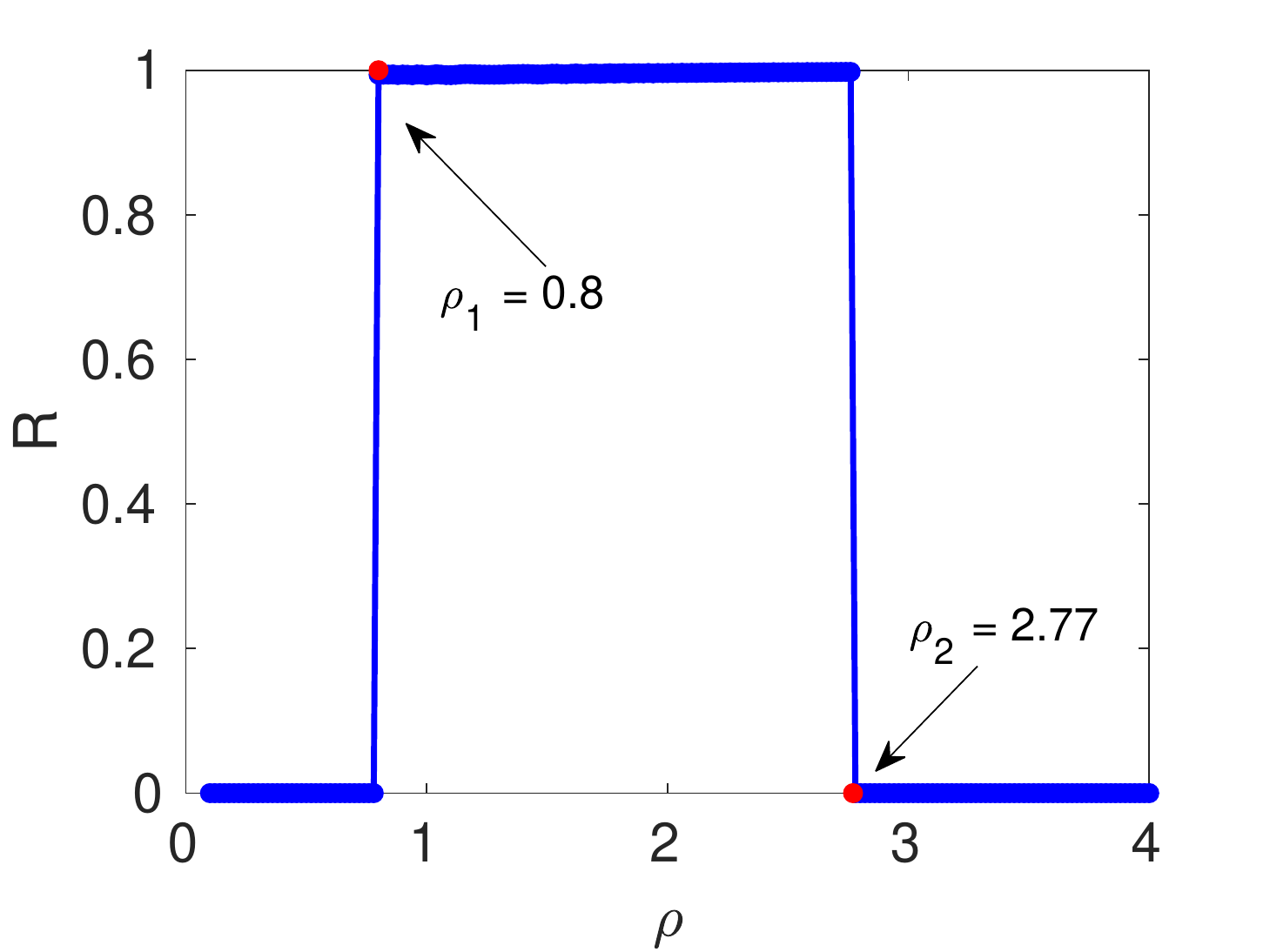}
        \caption{$1000$ identical cells}
        \label{QS_R_Ampa}
    \end{subfigure}
  \quad
    ~ 
    \begin{subfigure}[t]{0.31\textwidth}
        \includegraphics[width=\textwidth]{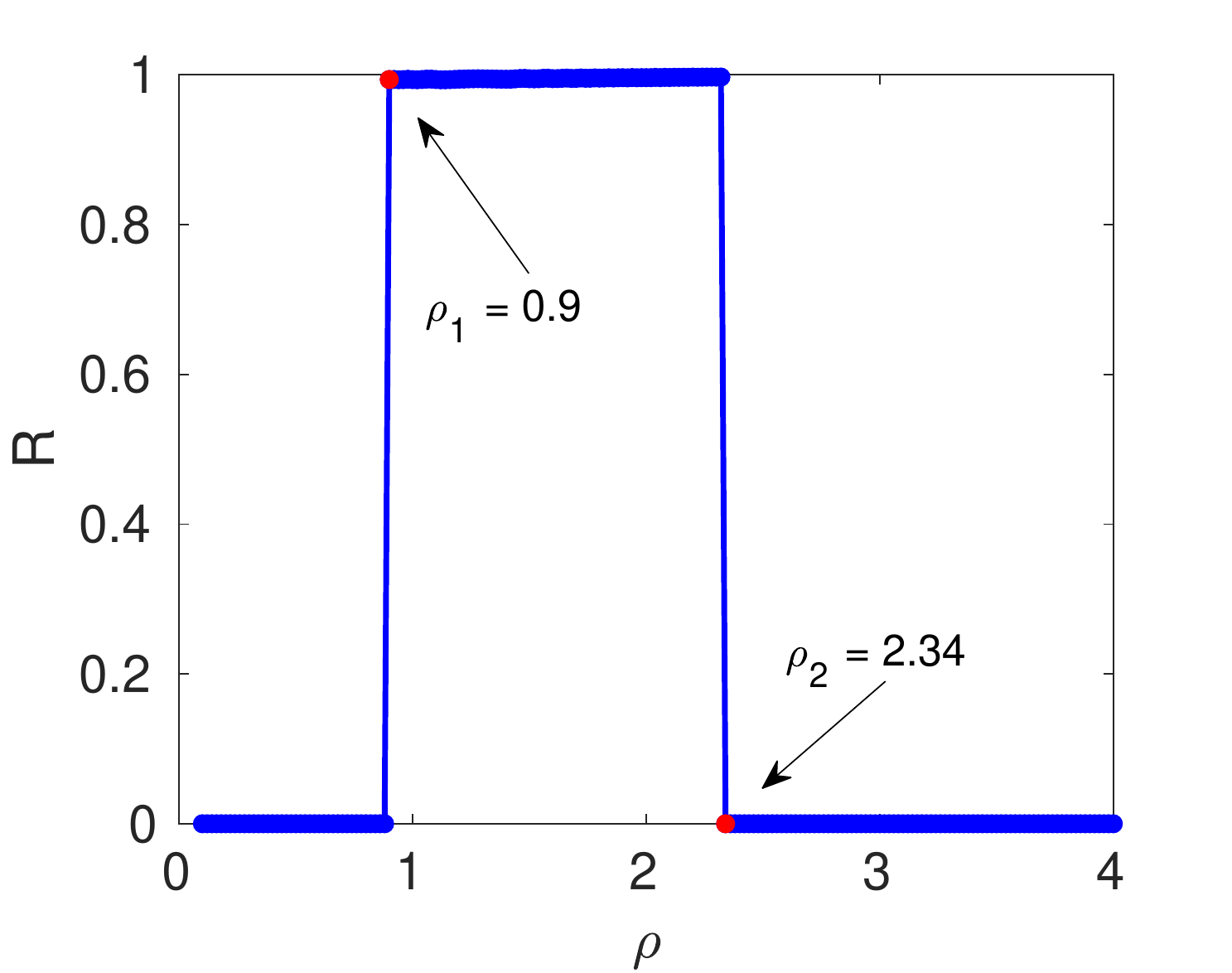}
        \caption{$600$ identical and $400$ defective cells}
        \label{QS_R_Ampb}
    \end{subfigure} 
    ~ 
    \begin{subfigure}[t]{0.31\textwidth}
        \includegraphics[width=\textwidth]{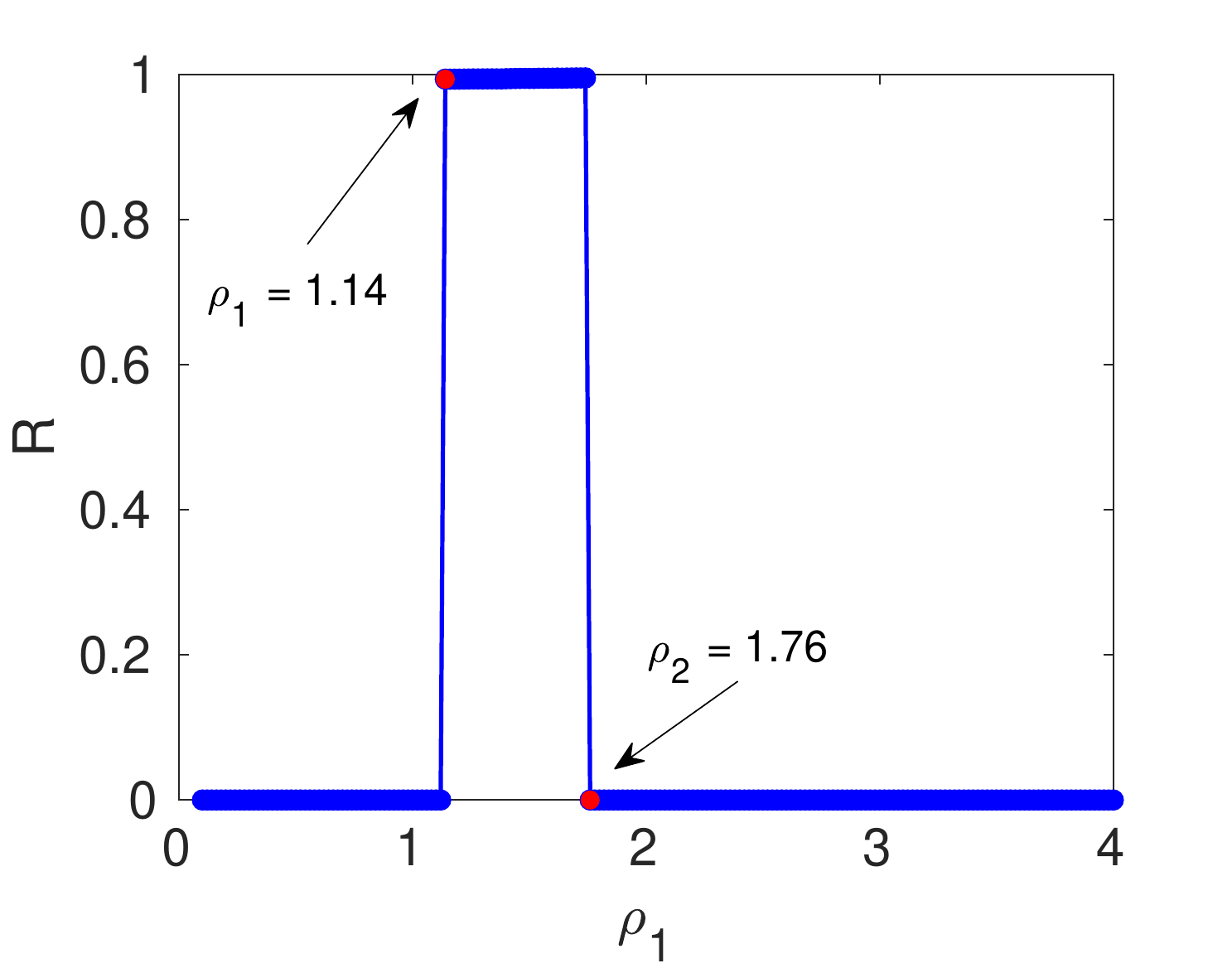}
        \caption{$100$ identical and $900$ defective cells}
        \label{QS_R_Ampc}
    \end{subfigure}\\
    \begin{subfigure}[t]{0.31\textwidth}
        \includegraphics[width=\textwidth]{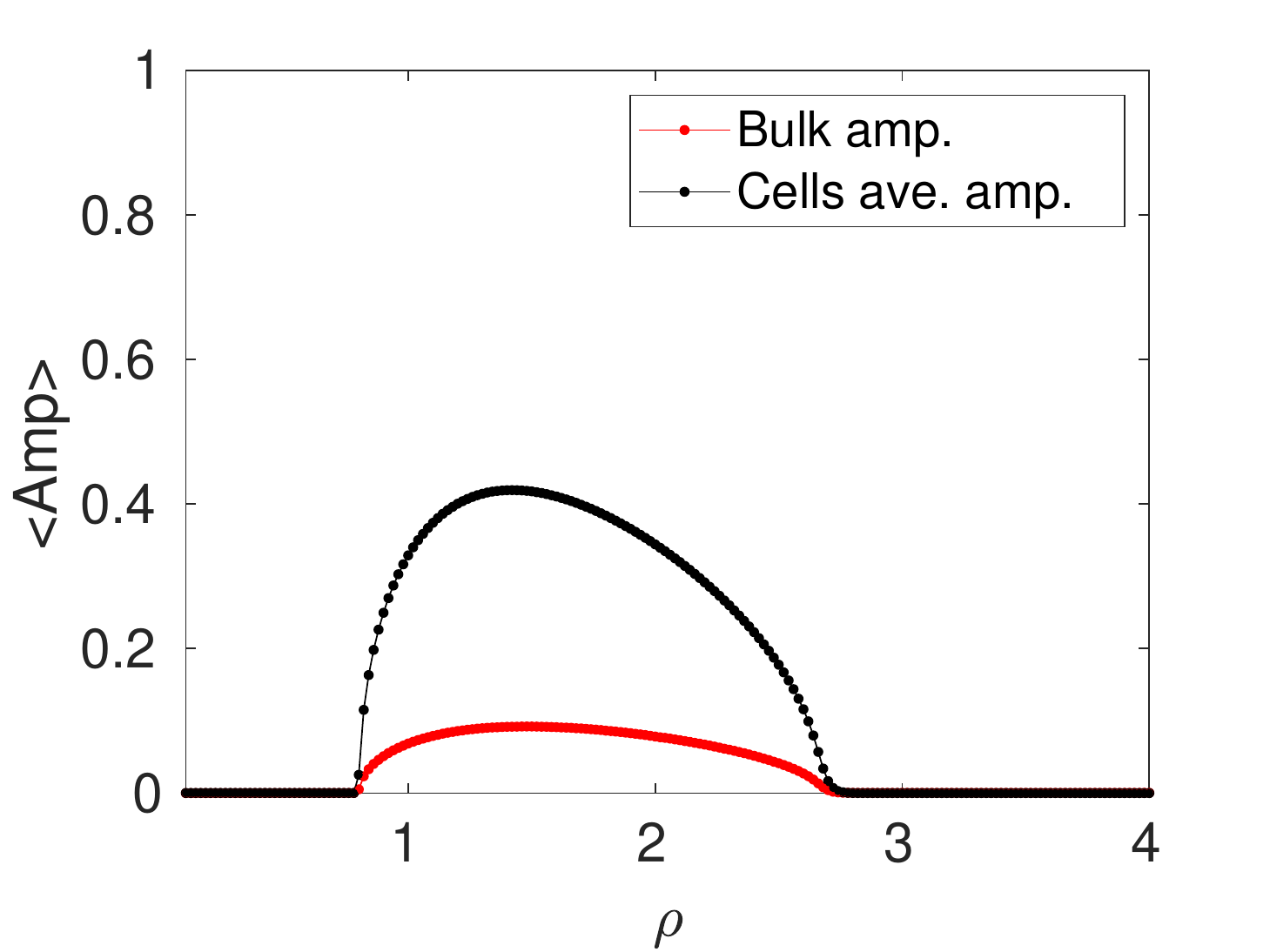}
        \caption{$1000$ identical cells}
        \label{QS_R_Ampd}
    \end{subfigure}
    \quad
    ~ 
    \begin{subfigure}[t]{0.31\textwidth}
        \includegraphics[width=\textwidth]{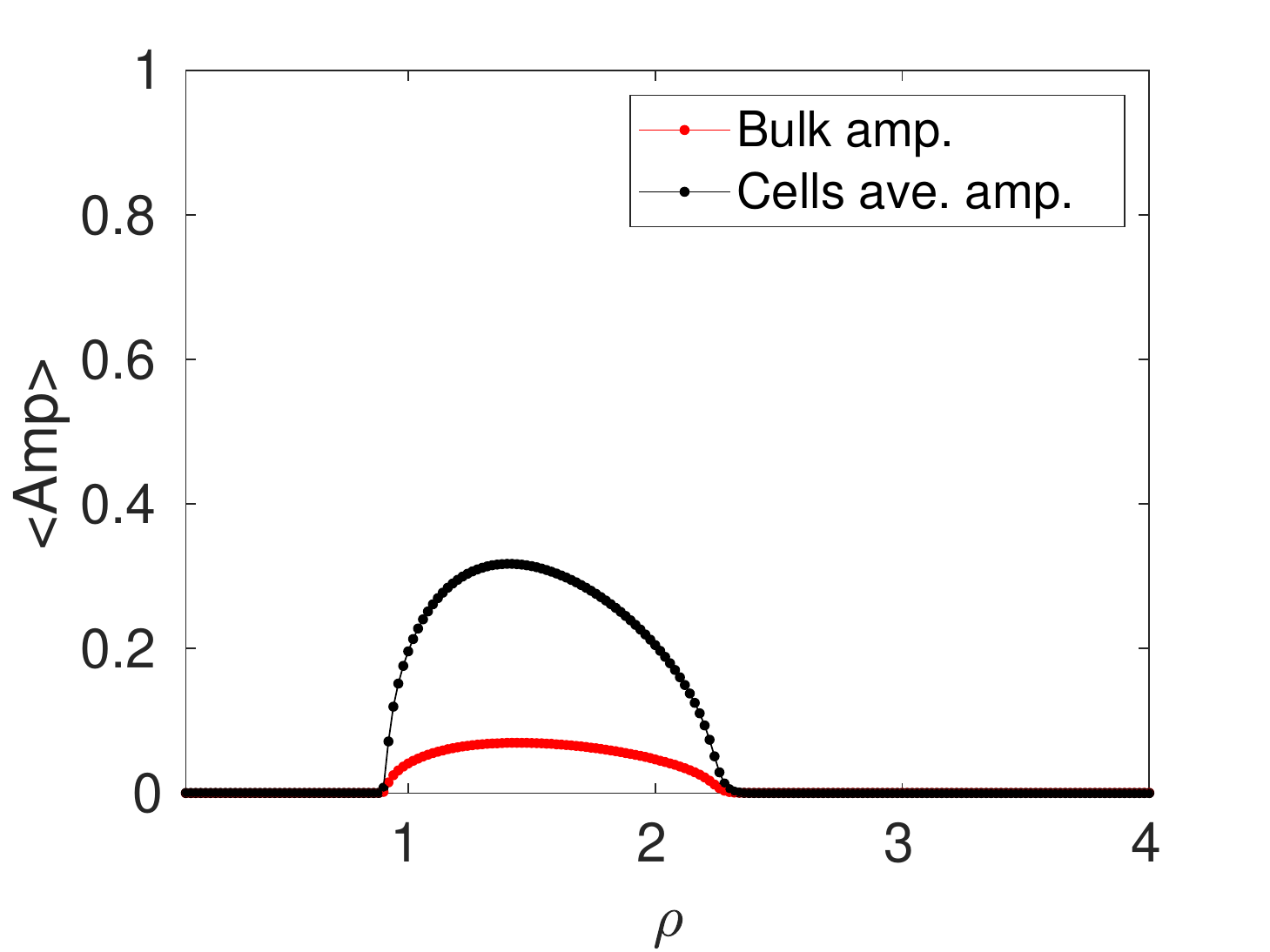}
        \caption{$600$ identical and $400$ defective cells}
        \label{QS_R_Ampe}
    \end{subfigure}
    ~ 
    \begin{subfigure}[t]{0.31\textwidth}
        \includegraphics[width=\textwidth]{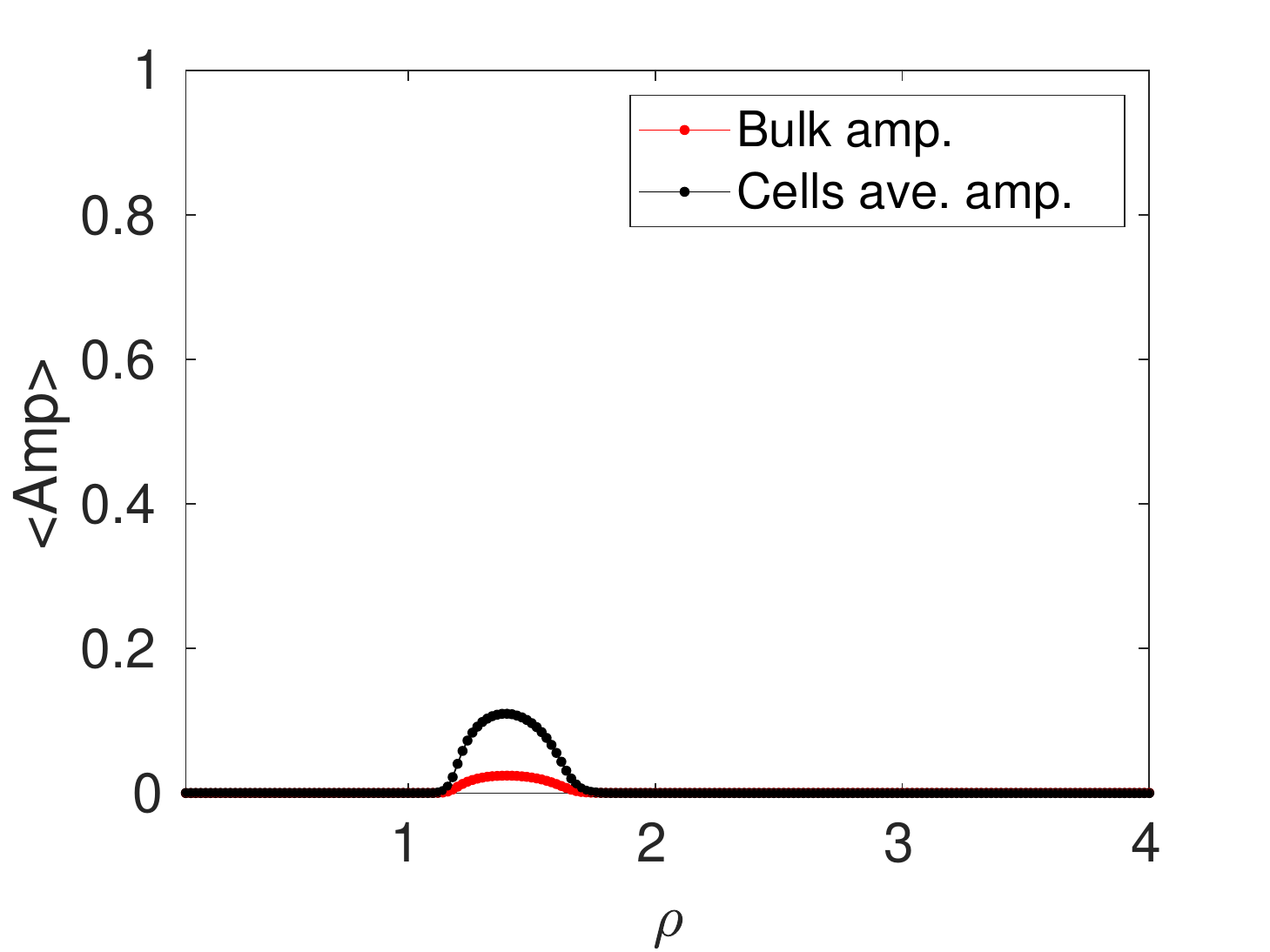}
        \caption{$100$ identical and $900$ defective cells}
        \label{QS_R_Ampf}
    \end{subfigure}
    \caption{The degree of phase synchrony and amplitudes of the
      oscillations predicted in $1000$ coupled cellular compartments
      measured with the order parameter $R$ of \eqref{QS_Kura_Par}, as
      computed from the ODE system \eqref{QS_ODEsys}. (a), (b), and
      (c) show the computed values of $R$ with respect to $\rho$, for
      $1000$ identical cells, $600$ identical and $400$
      \textit{defective} cells, and $100$ identical and $900$
      \textit{defective} cells, respectively.  (d), (e), and (f) show
      the corresponding average amplitudes of the oscillations in the
      cells (black curve) and in the bulk region (red curve). The
      Sel'kov parameter $ \alpha = 0.9$ is used for the identical
      cells, while for the defective cells it is chosen uniformly from
      the range $ 0.9220 \leq \alpha \leq 0.9520$. The other
      parameters are 
      $ \mu=2, \epsilon_0=0.15, \tau=0.5, k_1=10.0531$ and
      $k_2=2.5133$. }\label{QS_R_Amp}
\end{figure}

In terms of the cell population density $\rho$, in Figure
\ref{QS_R_Amp} we plot the numerically computed order parameter $R$
and the corresponding average amplitudes of predicted oscillations, as
computed from the ODEs \eqref{QS_ODEsys}, for $1000$ cells.  Since the
cell population is fixed at $m=1000$, increasing the population
density $\rho$ corresponds to decreasing the area $|\Omega|$ of the
2-D bounded domain $\Omega$. The results in Figure \ref{QS_R_Amp} are
for the Sel'kov kinetic parameters $\mu=2$, $\epsilon_0=0.15$, the
reaction-time parameter $\tau=0.5$, and the permeabilities
$k_1=10.0531$ and $k_2=2.5133$. The remaining Sel'kov parameter value
$\alpha$ was fixed at $\alpha =0.9$ for the identical cells, and was
chosen uniformly from the interval $ 0.9220 \leq \alpha \leq 0.9520$
for the defective cells. As seen from Figure \ref{fig:selkov}, each
cell has Sel'kov kinetic parameters that are chosen so that
intracellular steady-state is linearly stable when there is no
coupling to the bulk medium. From Figure \ref{QS_R_Amp} we observe
that there are unique critical values of $\rho$ at which the order
parameter $R$ switches from $0$ to $1$ (perfectly out of phase to
perfect synchrony) and from $1$ to $0$ (perfect synchrony to perfectly
out of phase). These transitions occur in a switch-like manner, in the
sense that when intracellular oscillations emerge they all synchronize
simultaneously. When $R$ switches from $0$ to $1$, the synchronous
intracellular oscillations that are triggered can be interpreted as a
quorum sensing transition in the sense that it occurs for $1000$ cells
only when the domain area $|\Omega|$ is small enough.  Similarly, the
cells desynchronize simultaneously as $\rho$ increases past a further
threshold, leading to a linearly stable steady state. For identical
cells, this linearly stable steady-state when $\rho$ exceeds a
threshold was predicted theoretically in \S \ref{hopf:iden}. Having
two critical values of $\rho$, labeled by $\rho_1$ and $\rho_2$, at
which the dynamics of the cells change state agrees with our earlier
observation in Figure \ref{QS_BF} that cell population density plays a
dual role of initiating and then quenching intracelluar
oscillations. When the cells are all identical, the critical
population densities are $\rho_1 = 0.8$ and $\rho_2=2.77$, when 600 of
the cells are identical (remaining 400 are defective) we have
$\rho_1=0.9$ and $\rho_2=2.34$ , and when only $100$ of the cells are
identical (remaining 900 are defective) we have $\rho_1=1.14$ and
$\rho_2=1.76$. For any $\rho$ in $\rho_1 \leq \rho < \rho_2$, we have
$R=1$ and perfect synchrony in the intracellular dynamics.  On the
other hand, when $\rho < \rho_1$, we have $R=0$ so that the cells are
either perfectly out of phase or there are no oscillations.  This
occurs when there are too few cells relative to the area of the
domain, as there is insufficient secreted chemical from the cells to
trigger synchronous oscillations. Similarly, $R=0$ when
$\rho > \rho_2$, since in this case the domain area is sufficiently
small so much that the cells behave like a single \textit{effective}
cell, which has a linearly stable steady state.

In Figure \ref{QS_R_Amp} we observe that an increase in the relative
proportion of defective cells in the population significantly alters
the critical values $\rho_1$ and $\rho_2$ for which synchronous
oscillations are triggered and quenched, respectively. As the number
of defective cells increases, the range of $\rho$ that predicts
perfect synchrony shrinks (see Figures \ref{QS_R_Ampa},
\ref{QS_R_Ampb}, and \ref{QS_R_Ampc}), and so does the amplitudes of
the predicted oscillations (see Figures \ref{QS_R_Ampd},
\ref{QS_R_Ampe}, and \ref{QS_R_Ampf}). This shows that synchronous
intracellular dynamics, resulting from effective communication between
the cells, becomes more difficult to sustain as the relative number of
defective cells increases.

\section{Discussion}\label{Sec:Discussion}

{We have studied the emergence of oscillatory intracellular
  dynamics for a collection of $m$ small dynamically active
  disk-shaped signaling compartments, referred to as ``cells'', that
  are coupled together through a scalar passive bulk diffusion field
  in a bounded 2-D domain. In our theoretical model, originating from
  \cite{muller2006} (see also \cite{muller2013} and
  \cite{Mueller2014uecke}) and later extended and formalized in
  \cite{jia2016}, each cell secretes across its boundary only one
  signaling chemical into the common bulk medium, while it receives
  feedback across its boundary from the entire collection of
  cells. This boundary efflux and influx of signaling material is
  regulated by permeability parameters. The resulting PDE-ODE model
  \eqref{Dim_bulk}, given in dimensionless form in
  \eqref{DimLess_bulk}, was studied in the limit of a small common
  cell radius $\varepsilon\ll 1$ and in the limit where the bulk
  diffusion coefficient $D$ is asymptotically large, on the range
  $D \gg \mathcal{O} (\nu^{-1}) \gg \mathcal{O}(1)$ where
  $\nu\equiv {-1/\log\varepsilon}\ll 1$. In this limit, where the bulk
  region becomes well-mixed and the bulk concentration becomes
  spatially homogeneous, a matched asymptotic analysis was used to
  reduce the dimensionless PDE-ODE model into the system of ODEs
  \eqref{WM_ODEsys_2D} with global coupling. A similar reduction, but
  for identical cells, was done in   \cite{jia2016}.

  For the specific case of Sel'kov reaction kinetics, we have used
  this ODE system to provide a detailed study of the emergence of
  intracellular oscillations in a collection of cells that is due to
  the cell-cell coupling through the bulk medium, and which otherwise
  would not occur for isolated cells. As such we have assumed that the
  Sel'kov kinetic parameters are chosen so that each cell is a {\em
    conditional oscillator}, and so its dynamics has only a stable
  steady-state when uncoupled from the bulk. For identical cells, the
  emergence of intracellular oscillations is found to arise via a Hopf
  bifurcation of the globally coupled system as parameters are varied.

  One key focus has been to provide a detailed bifurcation study of
  the role of cell membrane permeabilites and the Sel'kov kinetic
  parameters on the emergence of intracellular oscillations. Our
  numerical bifurcation study using XPPAUT \cite{xpp2002} has
  considered two main scenarios: one where all the cells are
  identical, and the other where there is a \textit{defector} cell
  that is coupled to a group identical cells. A cell is considered
  defective if its Sel'kov kinetic parameters or permeabilities are
  different from those of the group of identical cells. Our results
  were presented in the form of one- and two-parameter global
  bifurcation diagrams of steady-states and periodic solution branches
  of the ODEs \eqref{WM_ODEsys_2D}. Some highlights of the
  effect of a defector cell on a small group of identical cells
  include the following: Small changes in the influx rate of a single
  defective cell, relative to the common influx rate of the group, can
  act as a switch that extinguishes intracellular oscillations for the
  {\em entire collection} of cells (see Figures~\ref{Bifur_d1_3cells}
  and \ref{Bifur_d1_8cells}).  Intracellular oscillations of an entire
  group of cells can be extinguished by a single defective cell that
  reduces its efflux rate (see Figures~\ref{Bifur_d2_3cells} and
  \ref{Bifur_d2_8cells}). Intracellular oscillations of an entire
  group of identical cells can be extinguished (see Figure
  \ref{Bf_alpha}) or triggered (see Figure \ref{nBf_alpha}) through
  increases or decreases, respectively, in the Sel'kov kinetic
  parameter $\alpha$ of {\em one defective cell.} A more detailed
  discussion of the effect of a defector cell on a small group of
  identical cells is given in \S \ref{GlobalBifur}.

  Our second key focus was to use the ODE system \eqref{WM_ODEsys_2D}
  to study quorum sensing behavior associated with a large collection
  of either identical cells or a mixture of identical and defective
  cells. In terms of a cell density parameter, defined by
  $\rho\equiv {m/|\Omega|}$, we showed that quorum sensing is
  characterized by the switch-like emergence of synchronous
  intracellular oscillations as $\rho$ crosses through a critical
  threshold $\rho_1$ (see Figures \ref{QS_BF} and
  \ref{QS_R_Amp}). Moreover, we showed that further increases in the
  cell density parameter eventually leads to a switch-like quenching
  of the oscillations when a second threshold density $\rho_2$ is
  reached.  For the case of identical cells, we showed that these two
  thresholds $\rho_1$ and $\rho_2$ are Hopf bifurcation points
  associated with a cubic polynomial (see \eqref{QS_cubic}), while the
  existence of the second threshold follows from
  \eqref{quench:estimate}. The qualitative explanation underlying the
  quenching threshold is that for large $\rho$ the collection of cells
  behaves like a single ``effective'' cell, which has no intracellular
  oscillations since it is a conditional Sel'kov oscillator. For two
  different mixtures of identical and defective cells, for which the
  Sel'kov kinetic parameter $\alpha$ was uniformly distributed in some
  interval, the Kuramoto order parameter in \eqref{QS_Kura_Par} was
  computed numerically to establish intervals in the cell density
  parameter $\rho$ where synchronous intracellular oscillations occur.

  There are several possible extensions of our modeling framework and
  analysis that should be considered. One main direction would be to
  analyze \eqref{DimLess_bulk} for the finite bulk diffusion regime
  $D={\mathcal O}(1)$ for an arbitrary collection of non-identical
  small cells in a bounded 2-D domain. In particular, it would be
  interesting to investigate the effect of various spatial factors,
  such as diffusion sensing and the role of cell-clustering
  (cf.~\cite{gao2016crucial}), on the emergence of intracellular
  oscillations for a collection of cells whose reaction kinetics are
  conditional Sel'kov oscillators. In this context, the spatial
  diffusion gradient and the overall spatial configuration of cells
  should determine which cells can effectively communicate and
  synchronize their dynamics. In particular, is a chimera-type state
  involving distinct intracellular oscillations for different
  cell-clusters within the entire group possible in this parameter
  regime?

  Our PDE-ODE system \eqref{DimLess_bulk} with Sel'kov reaction
  kinetics provides a simple conceptual model that predicts the
  switch-like emergence of intracellular oscillations with increasing
  cell density, known as quorum sensing, and this feature is
  qualitatively similar to that seen in many biological and chemical
  settings (cf.~\cite{de2007dynamical}, \cite{gregor2010},
  \cite{review-yeast}, \cite{mina}, \cite{leaman2018}, \cite{taylor1},
  \cite{taylor2}, \cite{tinsley1}, \cite{tinsley2}).  As a second main
  direction, it would be worthwhile to investigate triggered
  intracellular oscillations and quorum sensing behavior when the
  Sel'kov kinetics in our theoretical PDE-ODE system
  \eqref{DimLess_bulk} are replaced with specific
  biologically-motivated reaction kinetics, such as the detailed model
  of \cite{wolf} and \cite{henson} for the glycolytic pathway in yeast
  cells and the model of \cite{mina} for bacterial communication of
  {\em Escherichia coli} cells. In other settings, such as in the
  modeling of bioluminescence emitted by the marine bacterium {\em
    Vibrio fischeri}, the quorum sensing behavior of microbial cells
  due to an autoinducer bulk field does not involve the triggering of
  intracellular oscillations, but instead is believed to involve the
  sudden transition to a new stable steady-state (the bioluminescent
  state) as the cell population increases past a threshold
  (cf.~\cite{qs_jump}). It would be interesting to incorporate the
  intracellular signaling pathways of \cite{qs_jump} into our PDE-ODE
  system \eqref{DimLess_bulk} to show that quorum sensing in this
  context results from a saddle-node structure of equilibria and the
  sudden transition between stable states as the number of cells
  increases. Other modeling approaches to analyze quorum sensing
  behavior associated with sudden transitions between stable
  steady-states as the cell density increases past a threshold are
  given in \cite{king2001} and \cite{dock}.

  Finally, other cell-cell communication problems mediated by a bulk
  diffusion field involve two-bulk diffusing species that mediate
  interactions between small signalling compartments. In this
  direction, it would be interesting to reframe the class of
  agent-based models considered in \cite{rauch} for Turing pattern
  formation into an extended version of our PDE-ODE system
  \eqref{DimLess_bulk} where there are now two-bulk diffusing species,
  consisting of an activator and an inhibitor field. For this
  scenario, it would be interesting to investigate, by using the
  permeabilities as bifurcation parameters, whether a Turing
  bifurcation can occur even when the activator and inhibitor bulk
  diffusion fields have very similar diffusion coefficients, as is
  typical in real-world biological systems.}

\section*{Acknowledgements}\label{sec:ak}
Michael Ward gratefully acknowledges the financial support from the
NSERC Discovery grant program.

\begin{appendices}
\section{ Non-dimensionalization of the coupled PDE-ODE model} \label{Append_A}

In this appendix, the dimensional coupled PDE-ODE model
\eqref{Dim_bulk} is non-dimensionalized. With $[z]$ denoting the
dimension of $z$, the dimensional units of the variables in this model
are 
\begin{subequations}\label{App_A_Dim}
\begin{align} 
  [\mathcal{U}]& = \frac{\text{moles}}{(\text{length})^2}\,, \qquad
       [\mu_c] =  \text{moles}\,, \qquad [D_B] =
    \frac{(\text{length})^2}{\text{time}} \,,  \qquad [k_R]=[k_B]=
     \frac{1}{\text{time}}\,,\\
  \quad [\pmb{\mu}_j] & = \text{moles}\,, \qquad [\beta_{1,j}] =
    \frac{\text{length}}{\text{time}}\,,  \qquad [ \beta_{2,j} ] =
     \frac{1}{\text{length} \times \text{time} } \quad \text{for}
      \quad j=1,\ldots,m\,.
\end{align}
\end{subequations}

We assume that the cells are circular with a common radius $R_0$, which is
small relative to the length-scale of the domain $L$, and so we define a
small parameter $\varepsilon = {R_0/L} \ll 1$. We then introduce 
dimensionless variables defined by
\begin{align} \label{App_A_dimvar}
  U = \frac{L^2}{\mu_c}\, \mathcal{U}\,, \qquad
  \pmb{x} = \frac{\pmb{X}}{L}\,, \qquad  t= k_R \, T\,, \qquad
  \pmb{u}_j = \frac{\pmb{\mu}_j }{\mu_c}\,.
\end{align}
Observe that the dimensionless time-scale that is chosen is the
time-scale for the reaction kinetics of the cells. By introducing
\eqref{App_A_dimvar} into \eqref{Dim_bulk}, we obtain that the
dimensionless bulk concentration satisfies
\begin{subequations}\label{App_A1_dimlessB}
\begin{align} 
  \frac{k_R}{k_B} \, U_t = D \Delta U - \, U\,, \qquad & \pmb{x}
         \in \Omega \setminus \cup_{j=1}^{m} \Omega_{\varepsilon_j}\,; \\
  \varepsilon \,  D \, \partial_n U = \frac{\varepsilon \,\beta_{1,j}}{L \, k_B}
  \,\, U - \frac{\varepsilon \, \beta_{2,j}L}{k_B} \,\, u_j^1& \qquad
      \text{on} \quad \partial \Omega_{\varepsilon_j}\,; \qquad
        \partial_n U = 0  \quad \text{on} \quad \partial \Omega\,,
\end{align}
where $D \equiv D_B/(L^2\, k_B)$ is the dimensionless bulk
diffusivity, $\Omega_{\varepsilon_j}$ for $j=1,\dots,m$ are disks of a common
radius $\varepsilon$, and $\Delta$ is the Laplace operator with respect to
$\pmb{x}$. This PDE for the dimensionless bulk concentration is
coupled to the intracellular dynamics of the $j^{\text{th}}$ cell
using the dimensionless integro-differential equation given by
\begin{align}\label{App_A1_dimlessI}
  \frac{\text{d} \pmb{u}_j}{\text{d} t}  = \pmb{F}_j( \pmb{u}_j) +
  \frac{ \pmb{e}_1}{\varepsilon}\, \frac{k_B}{k_R} \,
  \int_{\partial \Omega_{\varepsilon_j}} \left(\frac{\varepsilon \,\beta_{1,j}}
  {L \, k_B}\, U - \frac{\varepsilon \, \beta_{2,j}L}{k_B}\, u_j^1 \right)\,\,
  \text{d} S_x \,,
\end{align}
\end{subequations}
where we used $\text{d}S_{\pmb{X}} = L \, \text{d} S_{\pmb{x}}$. In terms of the
dimensionless variables
\begin{align} \label{App_A_newvar}
  \tau \equiv  \frac{k_R}{k_B}\,, \qquad D \equiv \frac{D_B}{L^2\, k_B}\,,
\qquad d_{1,j} \equiv
  \frac{\varepsilon \,\beta_{1,j}}{L \, k_B}\,, \quad \text{and} \quad
  d_{2,j} \equiv \frac{\varepsilon \, \beta_{2,j}L}{k_B}\,,
\end{align}
the dimensionless form of the coupled PDE-ODE model \eqref{Dim_bulk}
is given by
\begin{subequations}\label{App_A_dimlessB}
\begin{align} 
  \tau \, U_t = D \Delta U - \, U\,, \qquad & \pmb{x} \in
        \Omega \setminus \cup_{j=1}^{m} \Omega_{\varepsilon_j}\,; \\
  \varepsilon \,  D \, \partial_n U = d_{1,j}\,\, U - d_{2,j} \,\, u_j^1& \qquad
      \text{on} \quad \partial \Omega_{\varepsilon_j} \,; \qquad                           \partial_n U = 0  \quad \text{on} \quad \partial \Omega \,,
\end{align}
in the bulk region. This PDE is coupled to the intracellular dynamics of the
$j^{\text{th}}$ cell through 
\begin{align}\label{App_A_dimlessI}
  \frac{\text{d} \pmb{u}_j}{\text{d} t}  = \pmb{F}_j( \pmb{u}_j) +
  \frac{ \pmb{e}_1}{\varepsilon \, \tau}\,  \,
  \int_{\partial \Omega_{\varepsilon_j}} \left(d_{1,j}\, U - d_{2,j}\, u_j^1 \right)
  \,\, \text{d} S_x\,.
\end{align}
\end{subequations}

\section{ Gallery of Numerical Simulations } \label{Append_B}

Here, we present some numerical simulations of the reduced system of
ODEs \eqref{WM_ODEsys_2D}.

\begin{figure}[htbp]
    \centering
    \begin{subfigure}[b]{0.2 \textwidth}
        \includegraphics[width=\textwidth,height=3.7cm]{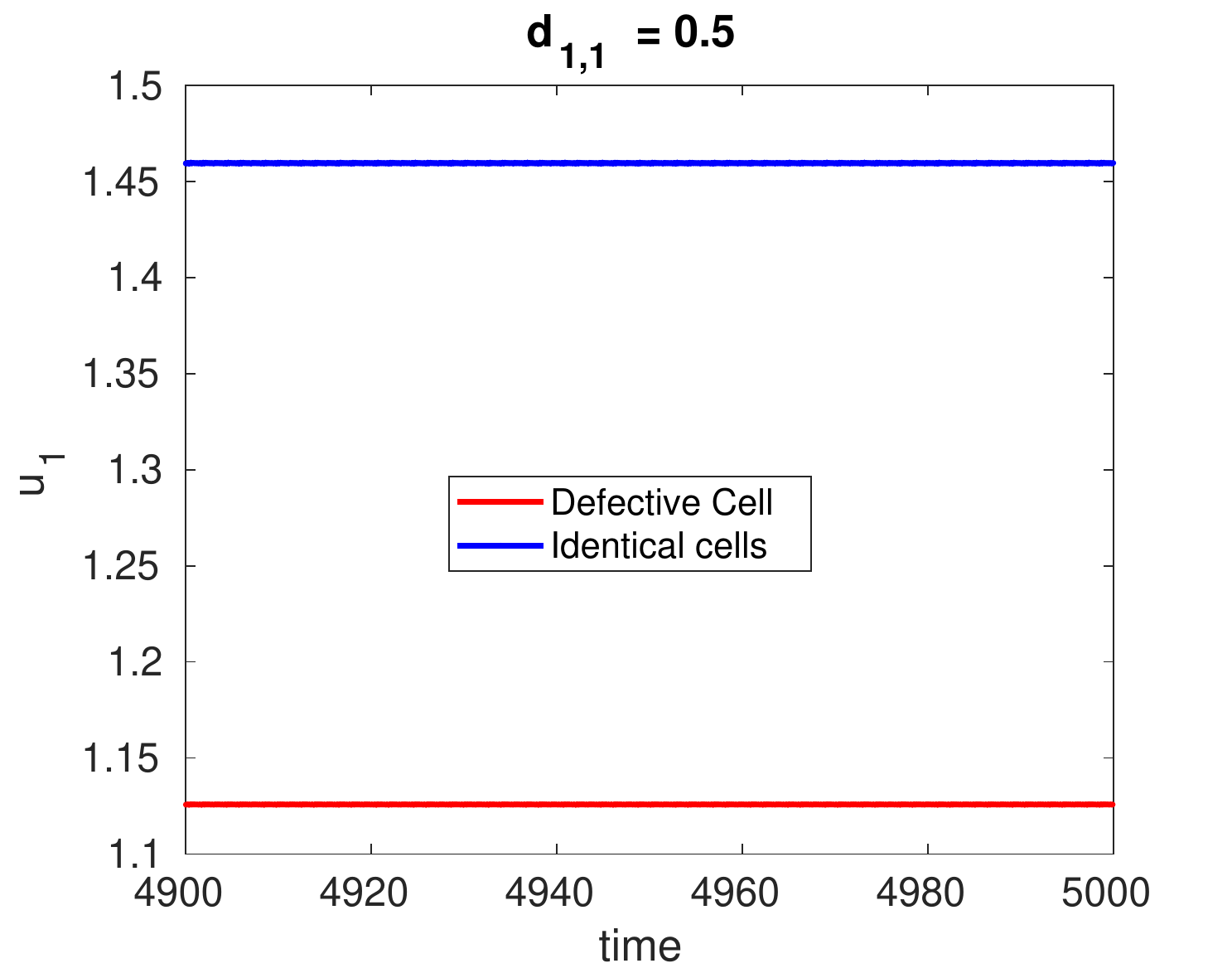}
        \caption{2 identical,1 defective}
        \label{GL_1a}
    \end{subfigure}
    \quad
    ~ 
    \begin{subfigure}[b]{0.2 \textwidth}
        \includegraphics[width=\textwidth,height=3.7cm]{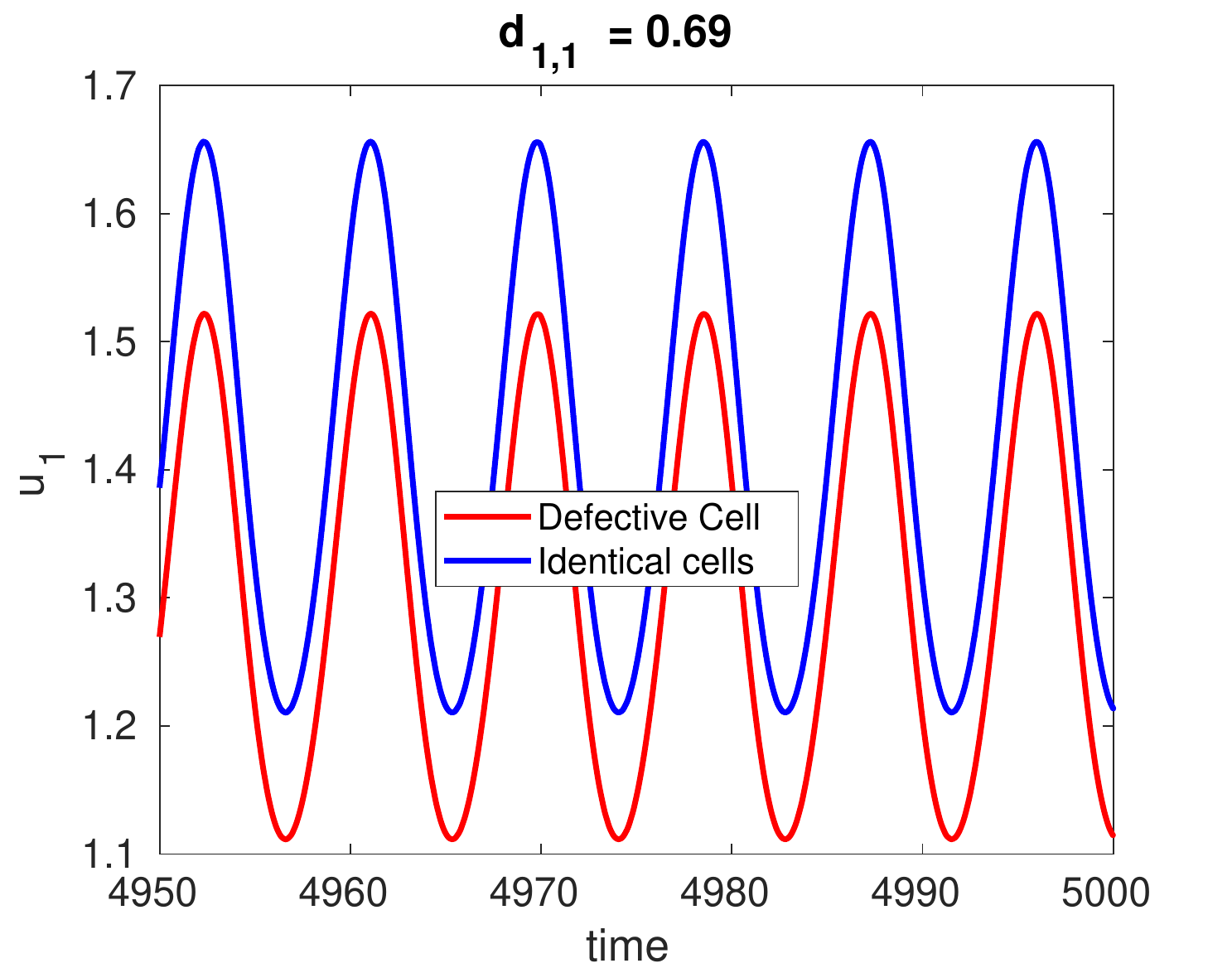}
        \caption{2 identical,1 defective}
        \label{GL_1b}
    \end{subfigure}
     \quad
    ~ 
    \begin{subfigure}[b]{0.2 \textwidth}
        \includegraphics[width=\textwidth,height=3.7cm]{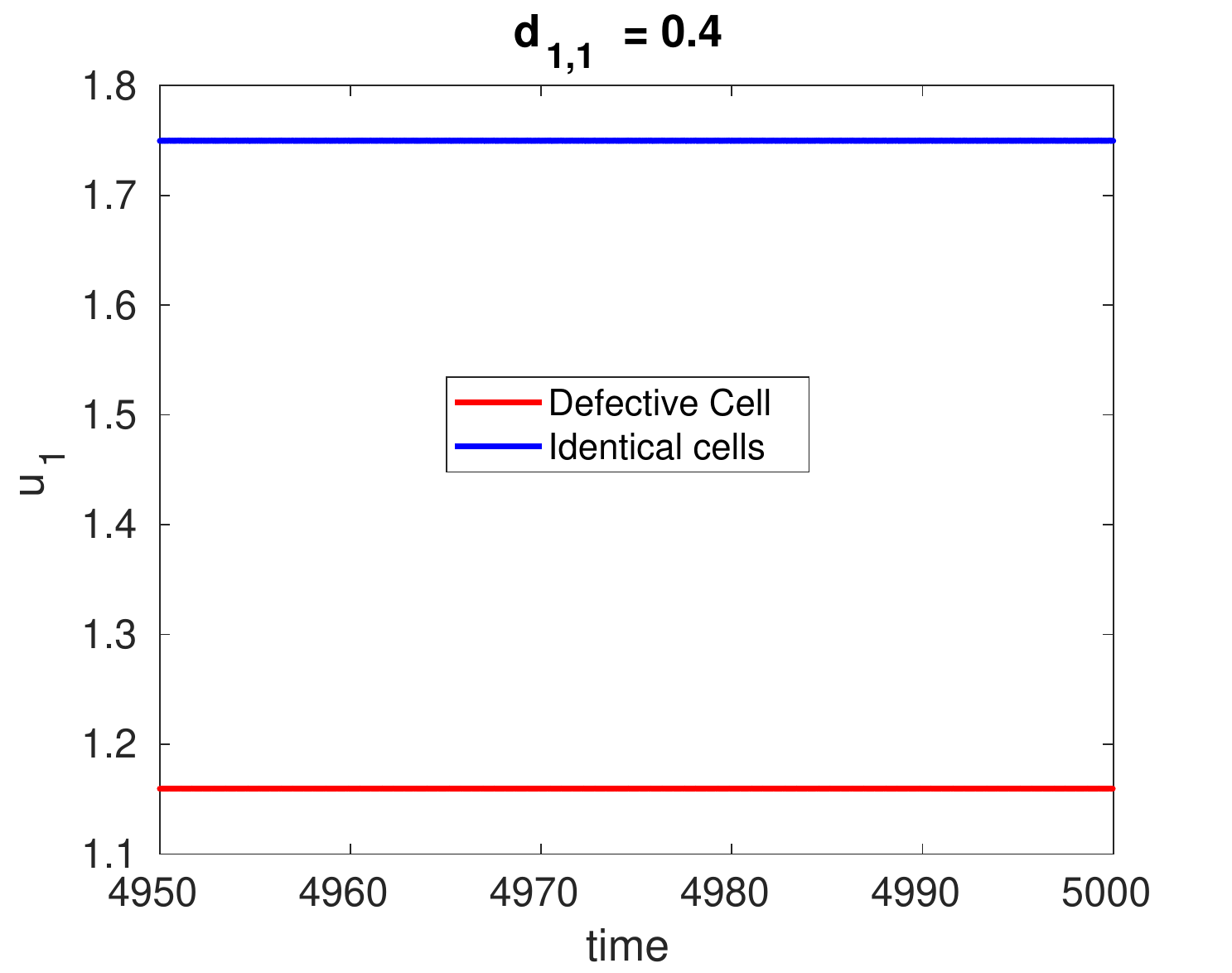}
        \caption{7 identical,1 defective}
        \label{GL_1c}
    \end{subfigure}
     \quad
    ~ 
    \begin{subfigure}[b]{0.2 \textwidth}
        \includegraphics[width=\textwidth,height=3.7cm]{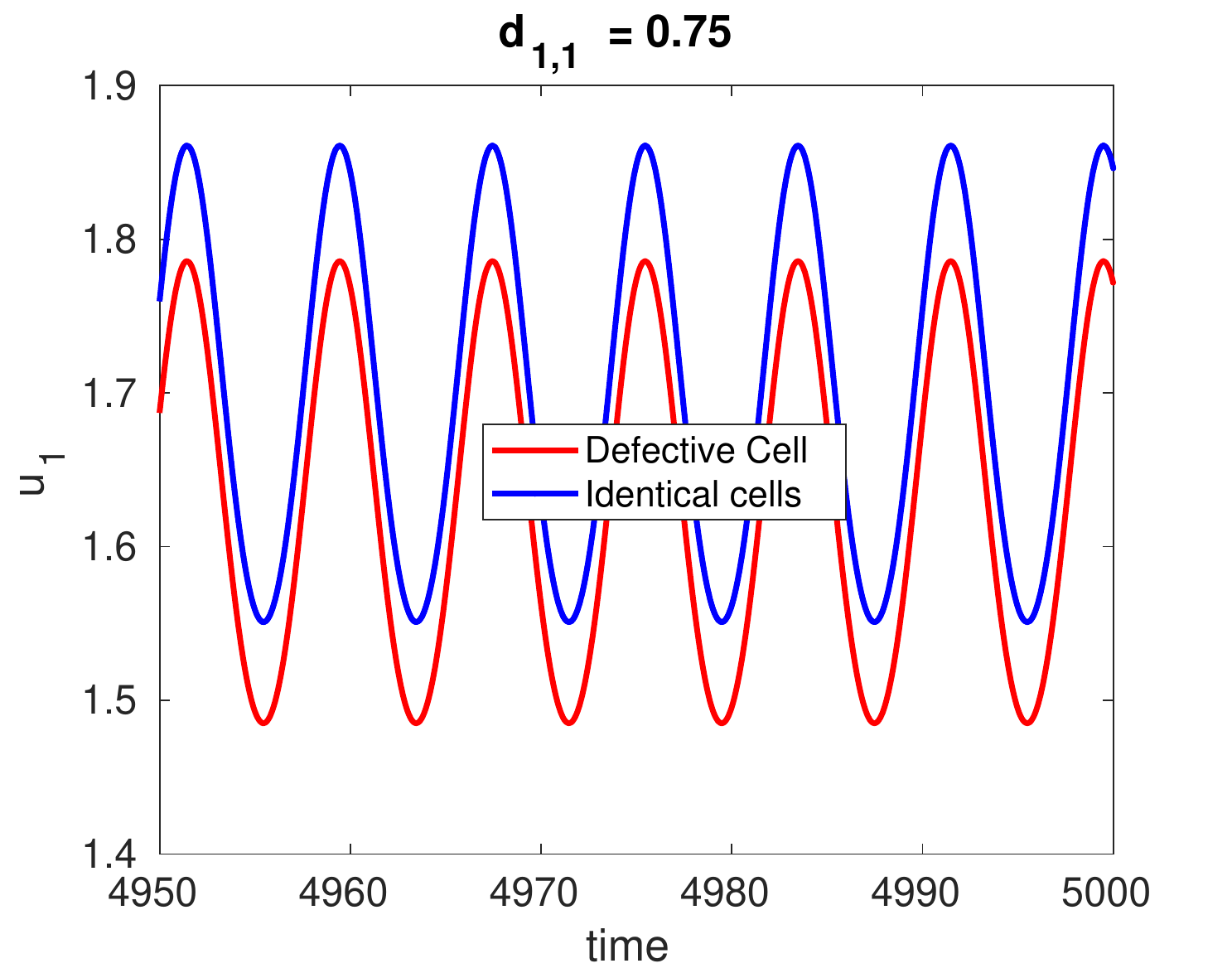}
        \caption{7 identical,1 defective}
        \label{GL_1d}
    \end{subfigure}\\
    ~ 
    \begin{subfigure}[t]{0.2 \textwidth}
        \includegraphics[width=\textwidth,height=3.7cm]{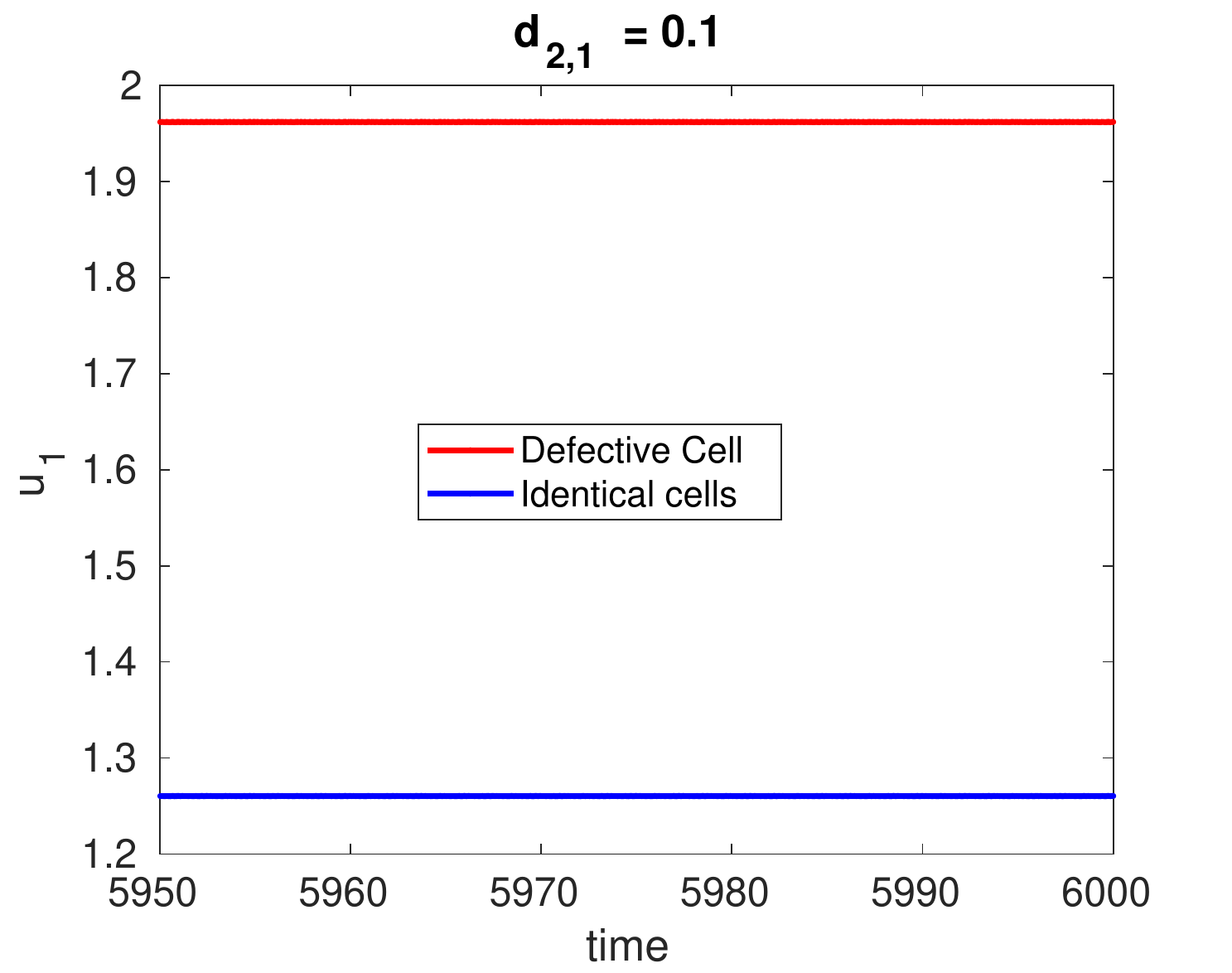}
        \caption{2 identical,1 defective}
        \label{GL_1e}
    \end{subfigure}
     \quad
    ~ 
    \begin{subfigure}[t]{0.2 \textwidth}
        \includegraphics[width=\textwidth,height=3.7cm]{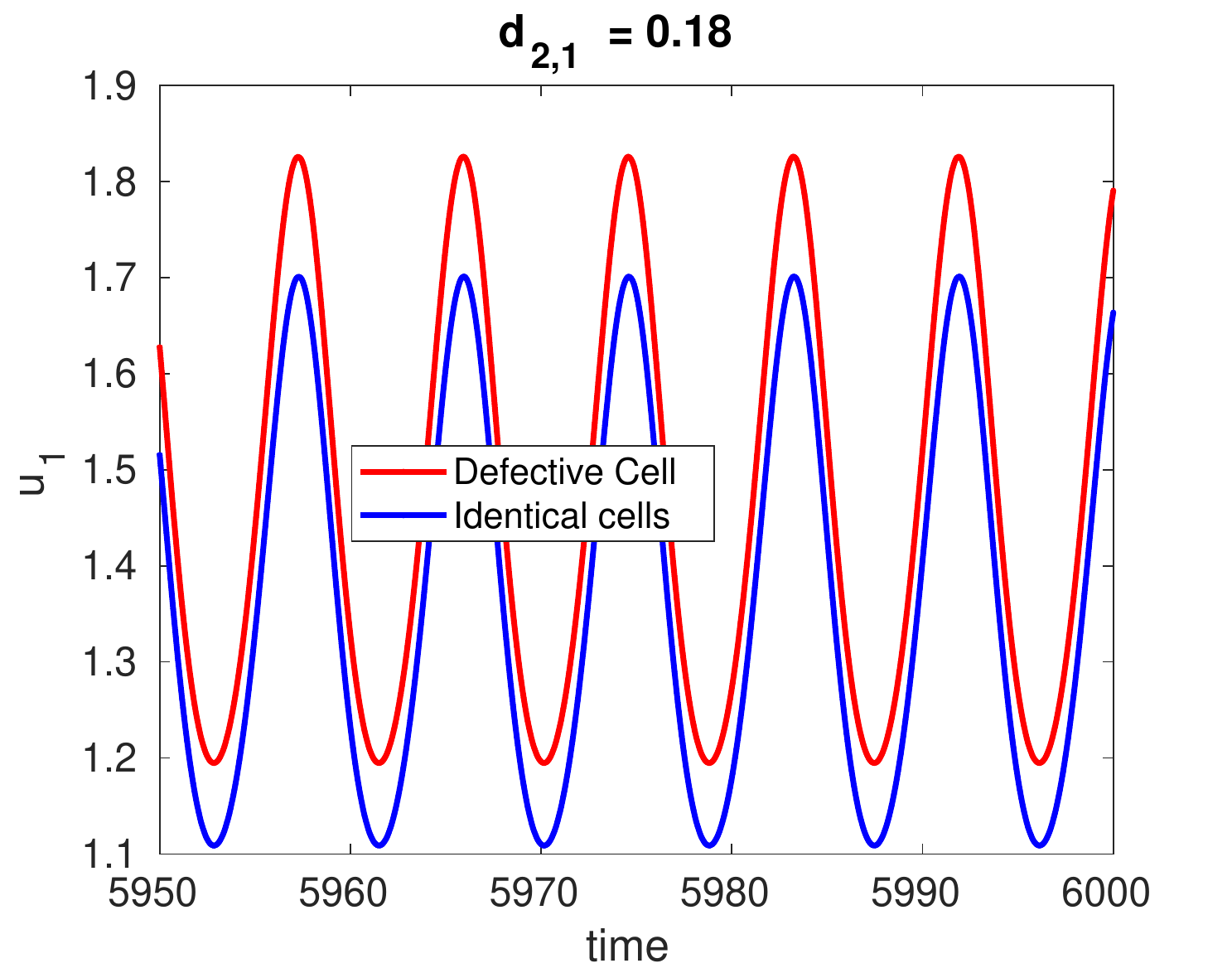}
        \caption{2 identical,1 defective }
        \label{GL_1f}
    \end{subfigure}
     \quad
    ~ 
    \begin{subfigure}[t]{0.2 \textwidth}
        \includegraphics[width=\textwidth,height=3.7cm]{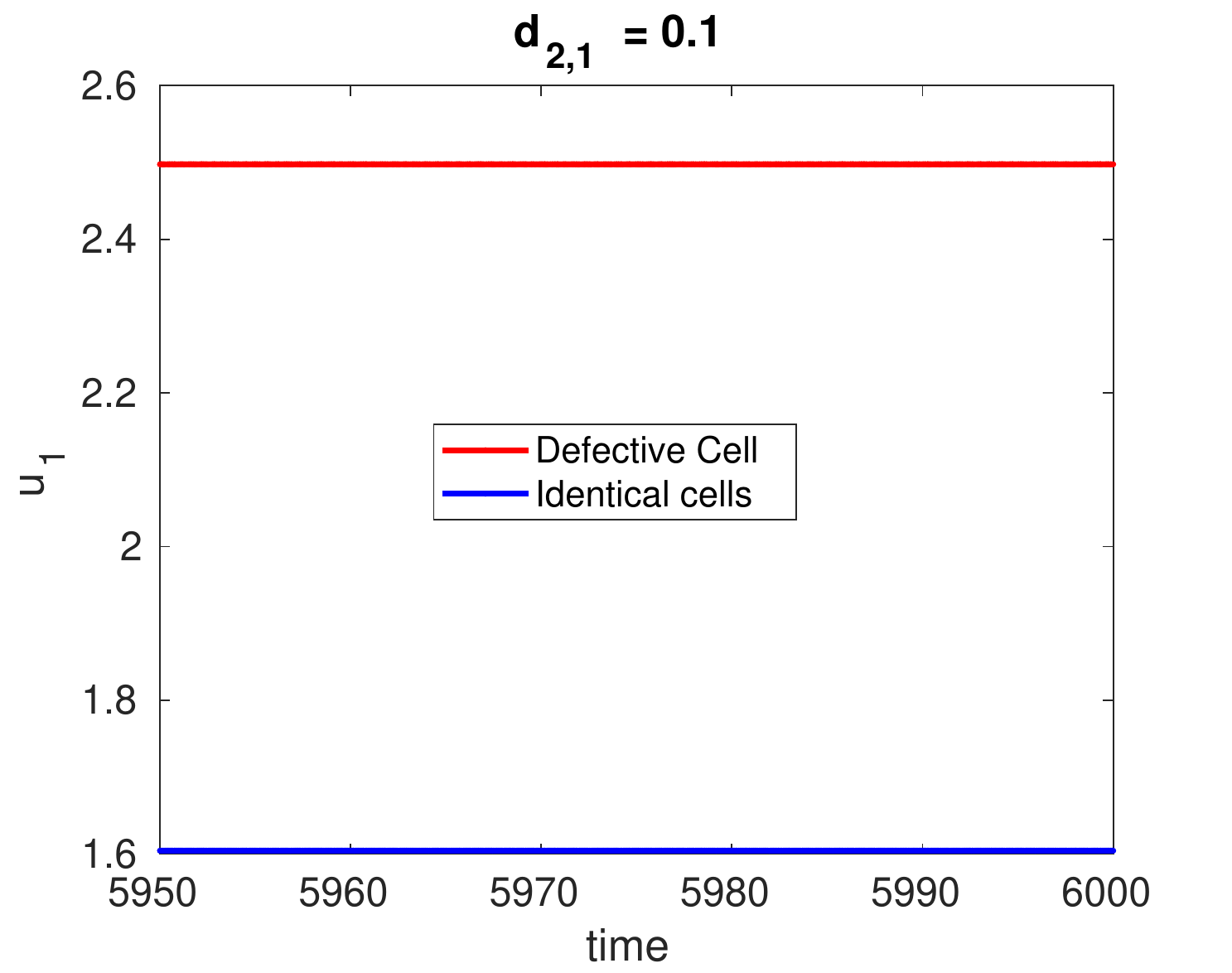}
        \caption{7 identical,1 defective}
        \label{GL_1g}
    \end{subfigure}
    ~\quad
    \begin{subfigure}[t]{0.2 \textwidth}
        \includegraphics[width=\textwidth,height=3.7cm]{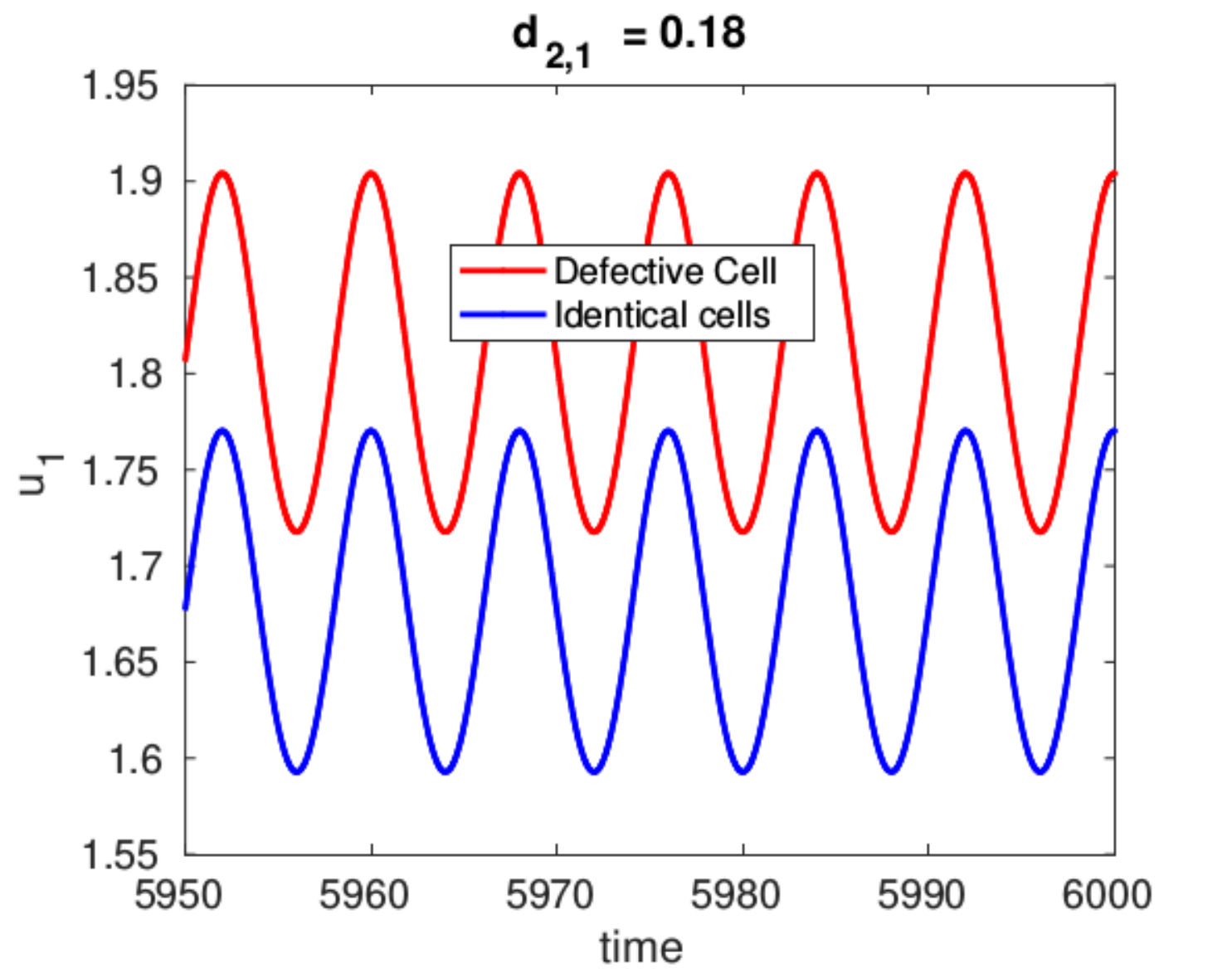}
        \caption{7 identical,1 defective}
        \label{GL_1h}
    \end{subfigure}\\~ 
    \begin{subfigure}[b]{0.2 \textwidth}
        \includegraphics[width=\textwidth,height=3.7cm]{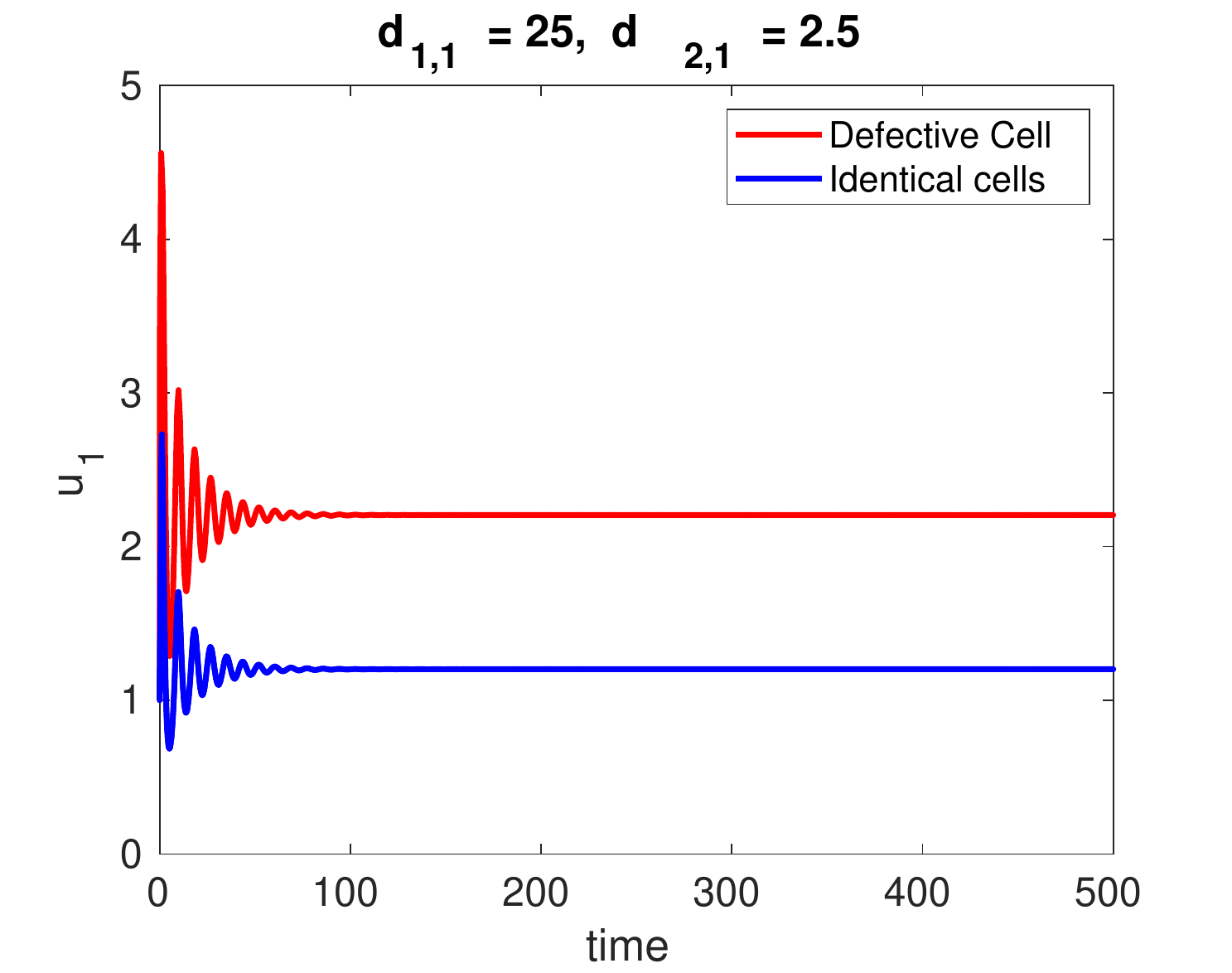}
        \caption{2 identical,1 defective}
        \label{GL_1i}
    \end{subfigure}
     \quad
    ~ 
    \begin{subfigure}[b]{0.2 \textwidth}
        \includegraphics[width=\textwidth,height=3.7cm]{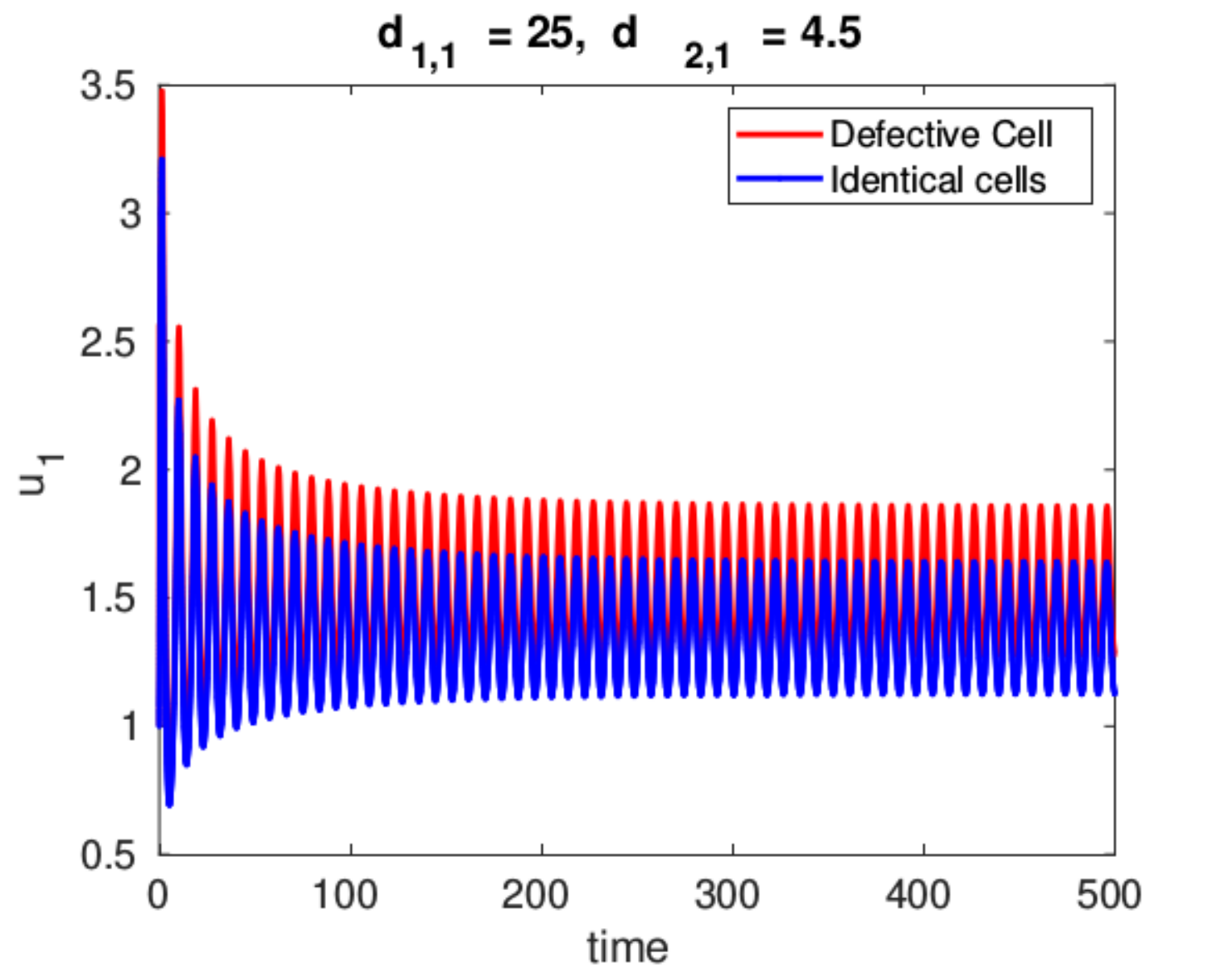}
        \caption{2 identical,1 defective }
        \label{GL_1j}
    \end{subfigure}
     \quad
    ~ 
    \begin{subfigure}[b]{0.2 \textwidth}
        \includegraphics[width=\textwidth,height=3.7cm]{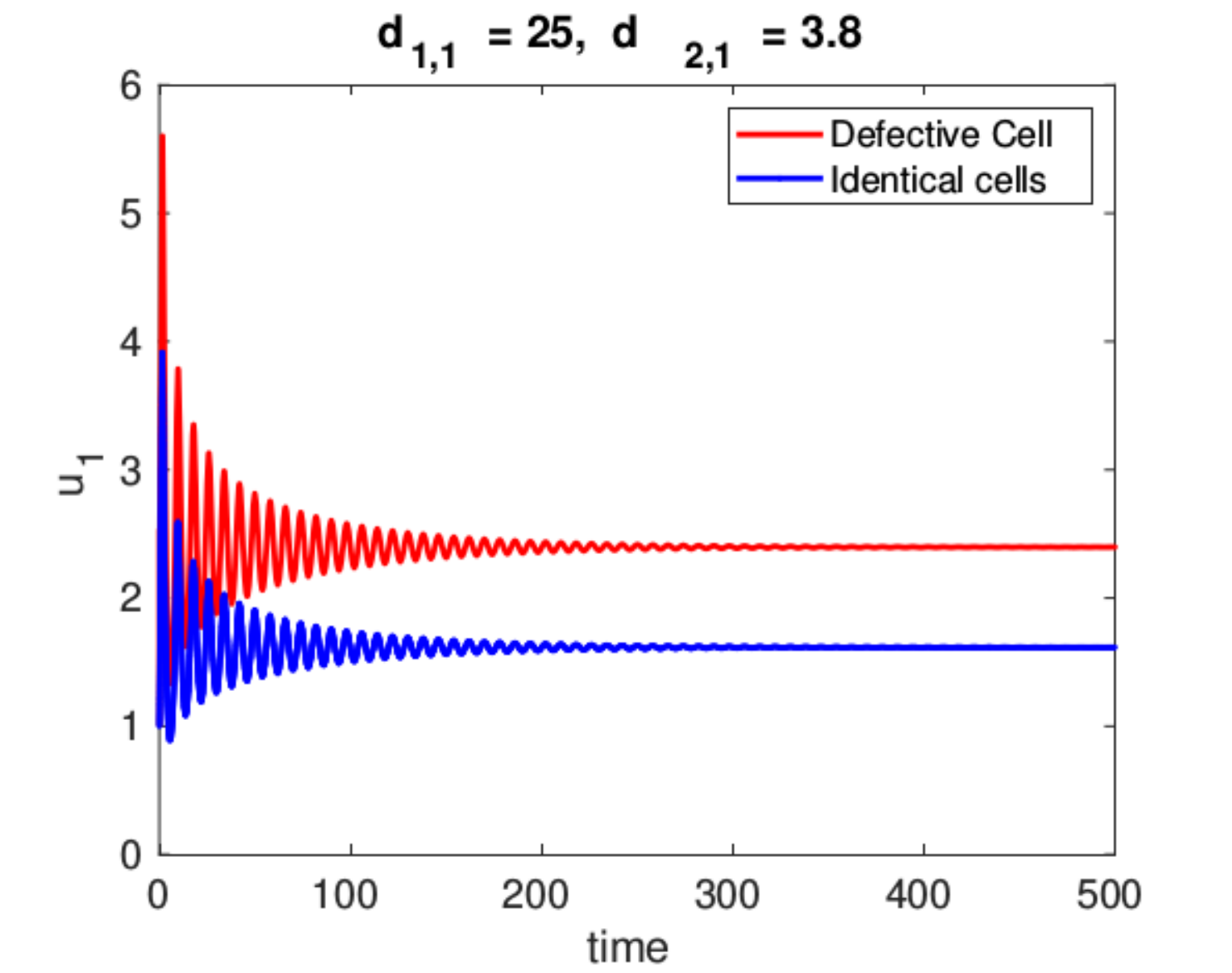}
        \caption{7 identical,1 defective}
        \label{GL_1k}
    \end{subfigure}
    ~\quad
    \begin{subfigure}[b]{0.2 \textwidth}
        \includegraphics[width=\textwidth,height=3.7cm]{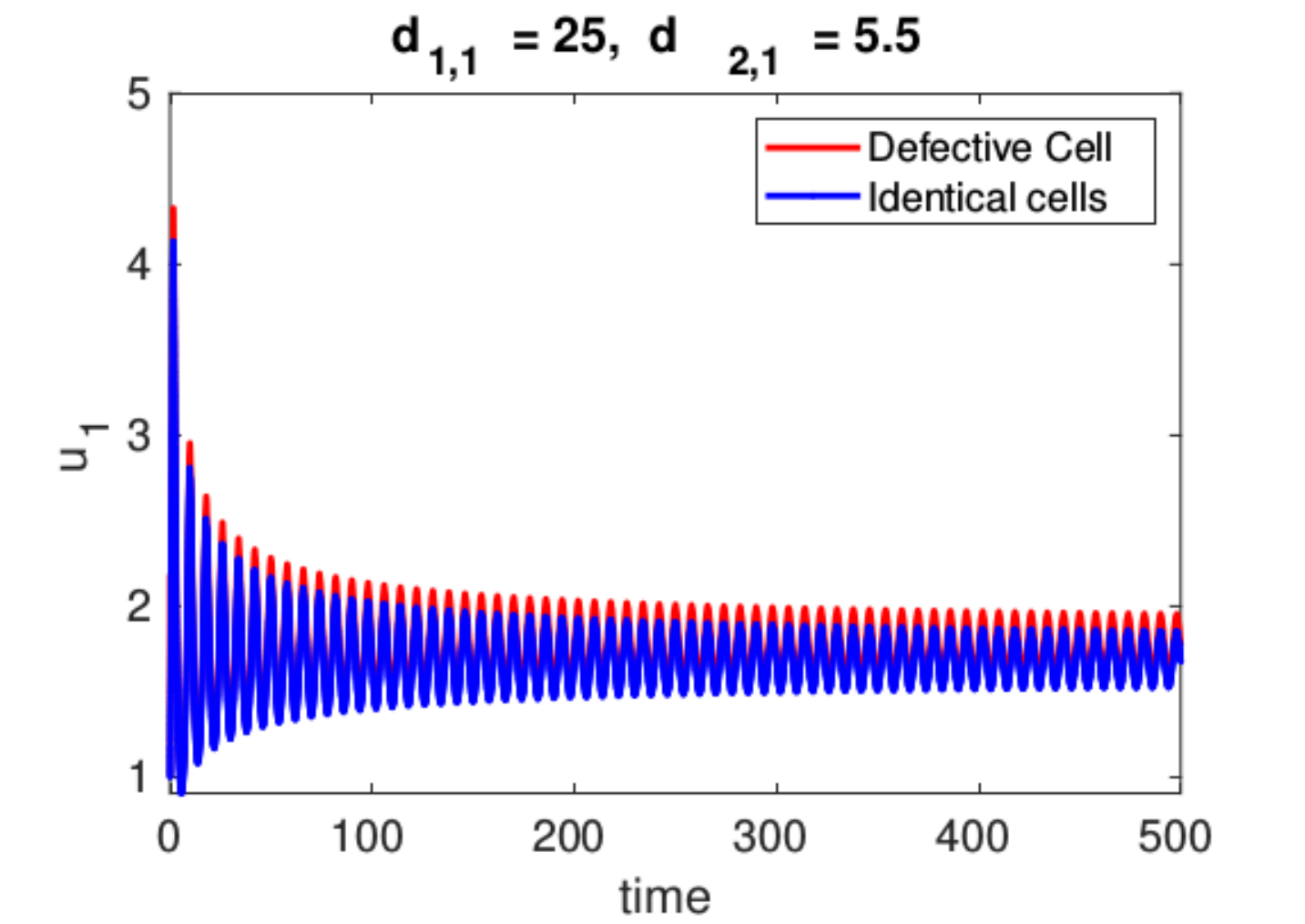}
        \caption{7 identical,1 defective}
        \label{GL_1l}
    \end{subfigure}
    \caption{\textit{Numerical simulations of the system of ODEs
        \eqref{WM_ODEsys_2D} for different values of the permeability
        parameters $d_1$ and $d_2$ sampled from the bifurcation
        diagrams in Figure \ref{Bifur_d1} (a-d), Figure \ref{Bifur_d2}
        (e-f), and Figure \ref{Bifur_2ParA} (i-l).}  }\label{GL_1}
\end{figure}

\begin{figure}[htbp]
    \centering
    \begin{subfigure}[b]{0.2 \textwidth}
        \includegraphics[width=\textwidth,height=3.7cm]{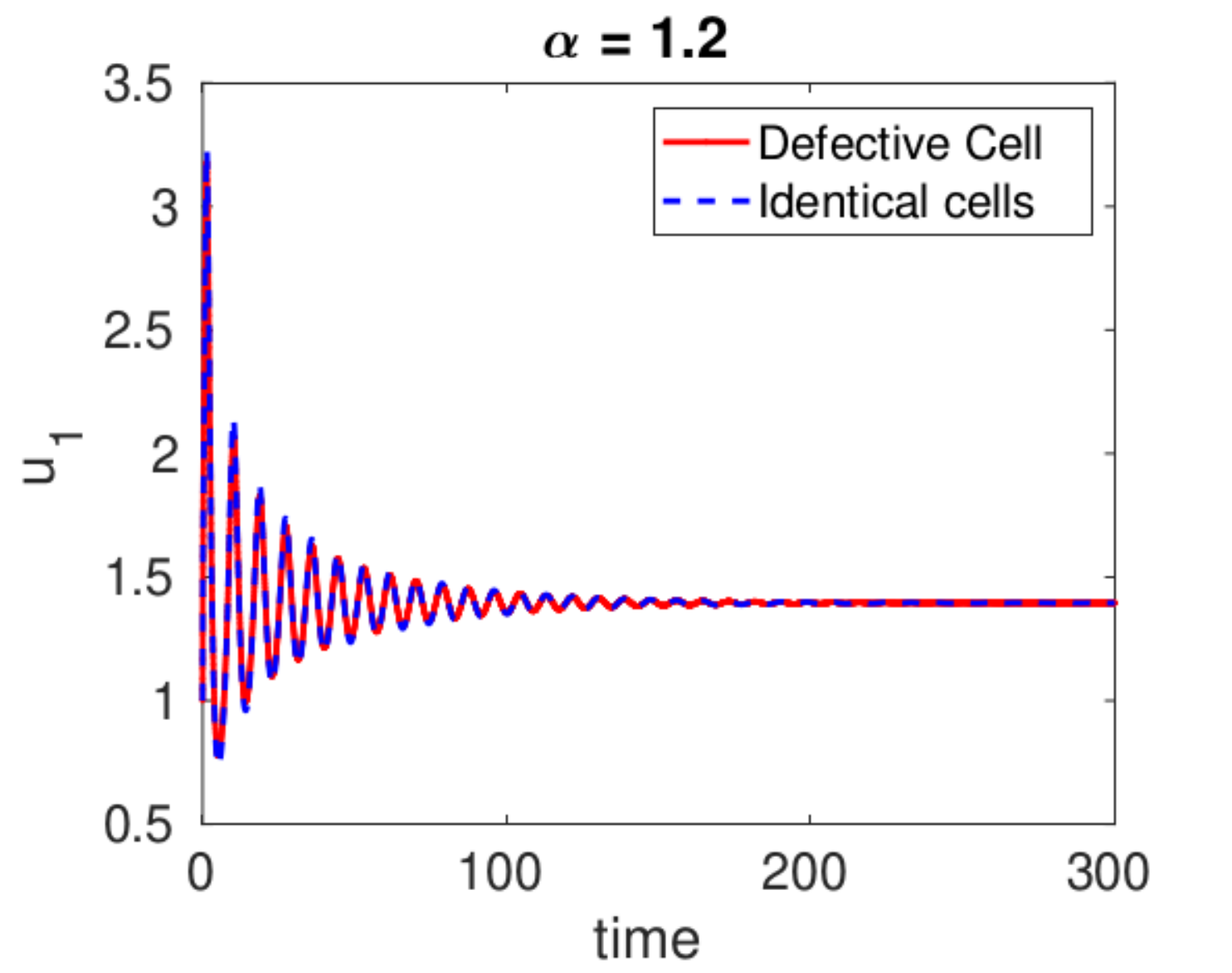}
        \caption{2 identical,1 defective}
        \label{GL1_1a}
    \end{subfigure}
    \quad
    ~ 
    \begin{subfigure}[b]{0.2 \textwidth}
        \includegraphics[width=\textwidth,height=3.7cm]{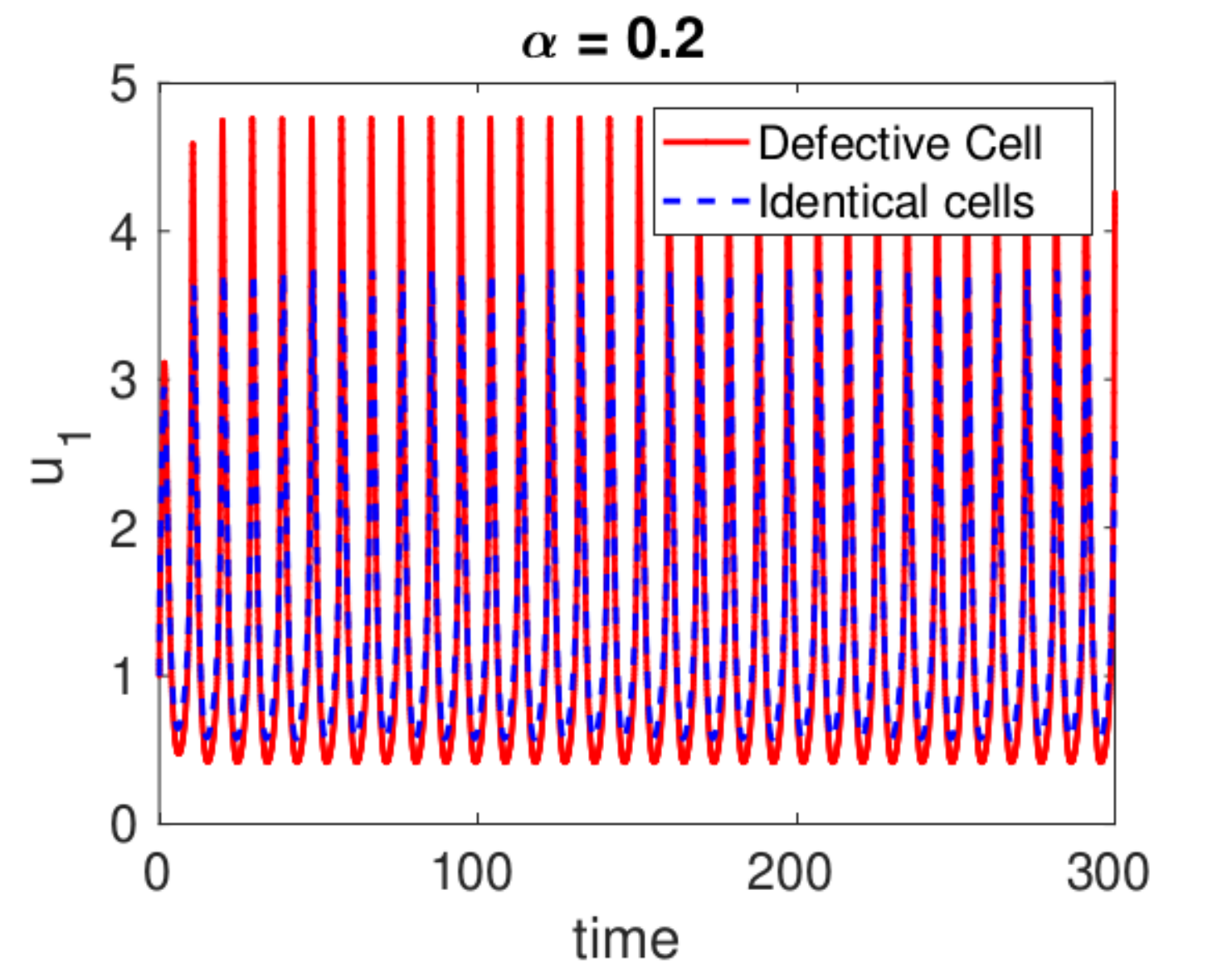}
        \caption{2 identical,1 defective}
        \label{GL1_1b}
    \end{subfigure}
     \quad
    ~ 
    \begin{subfigure}[b]{0.2 \textwidth}
        \includegraphics[width=\textwidth,height=3.7cm]{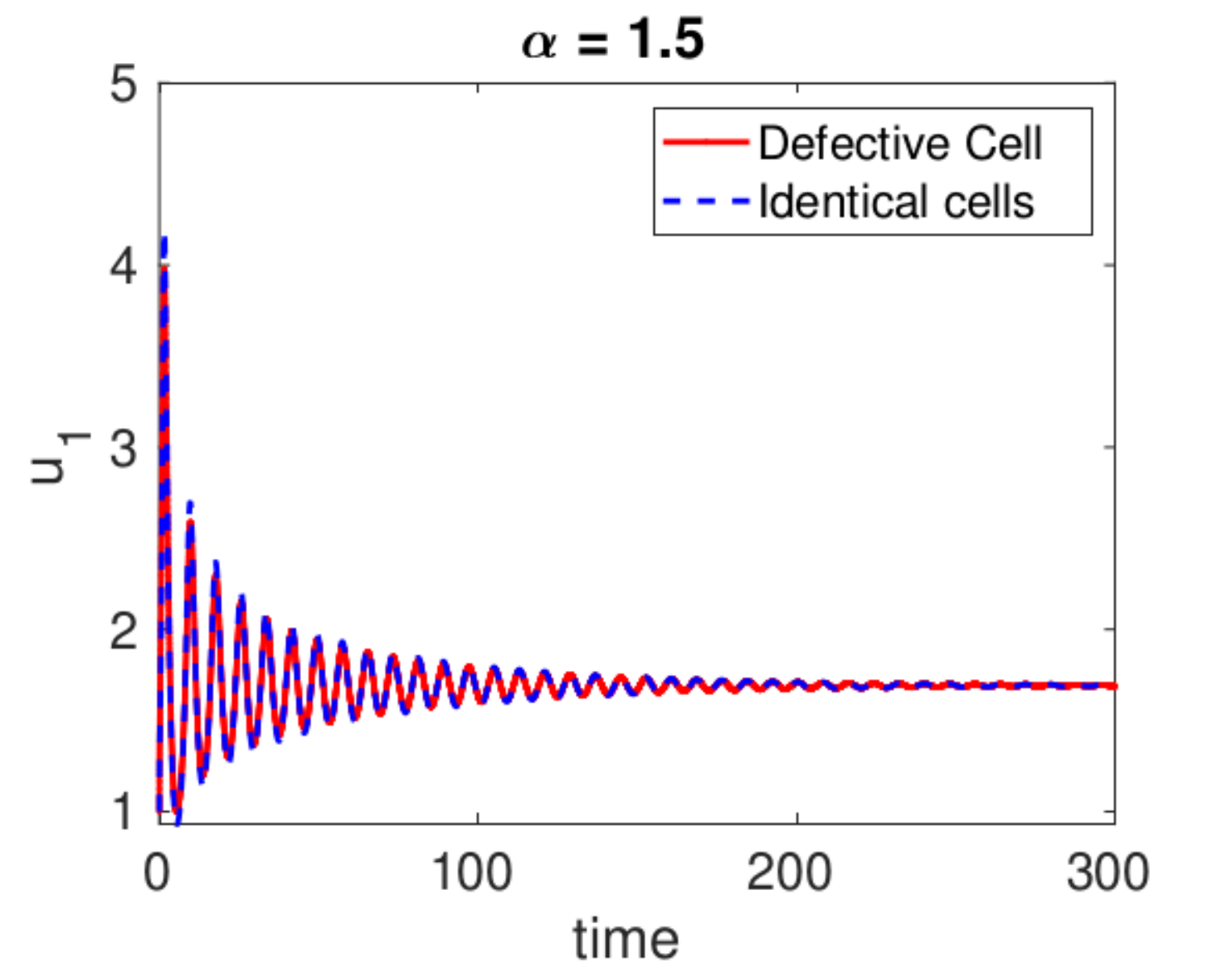}
        \caption{7 identical,1 defective}
        \label{GL1_1c}
    \end{subfigure}
     \quad
    ~ 
    \begin{subfigure}[b]{0.2 \textwidth}
        \includegraphics[width=\textwidth,height=3.7cm]{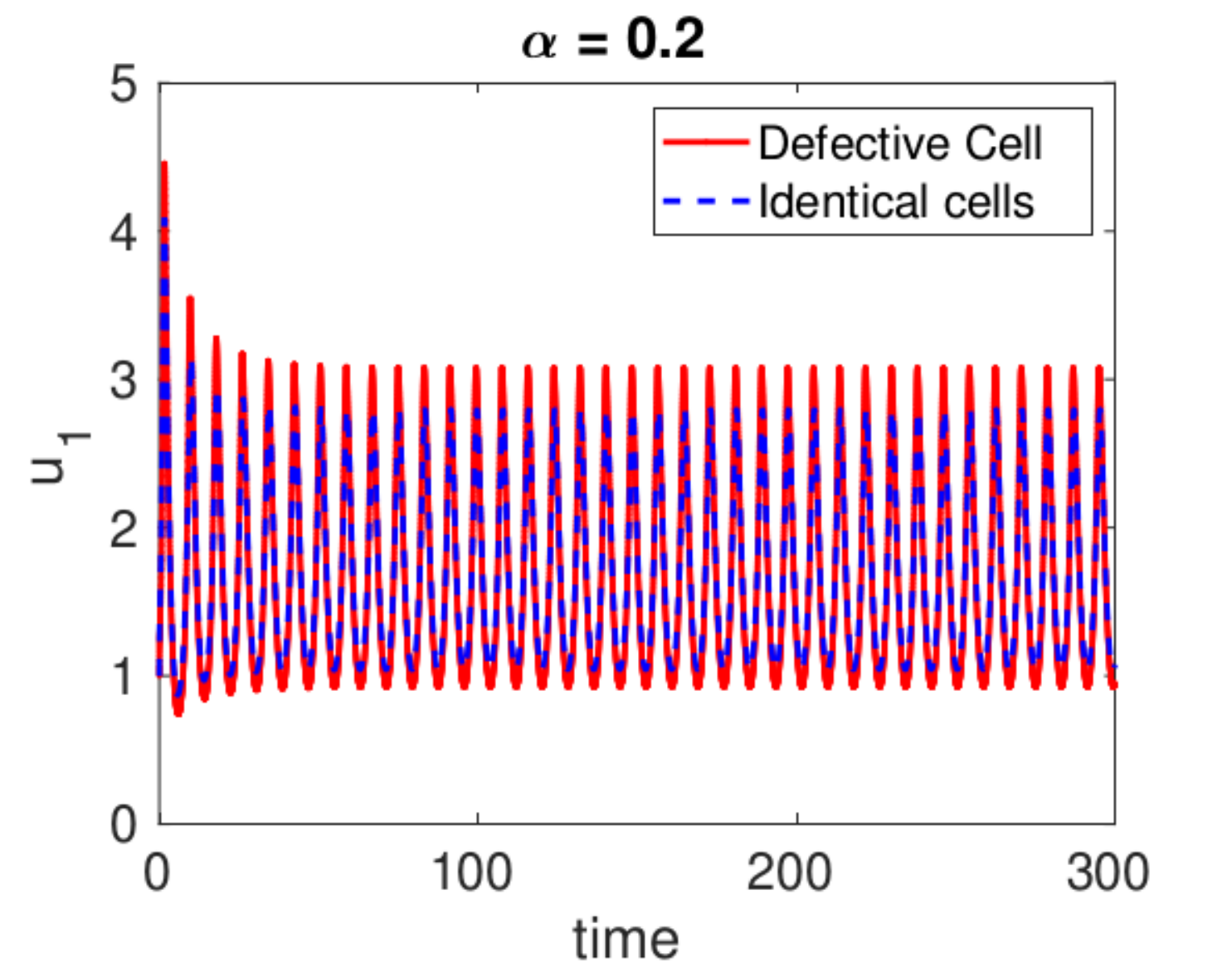}
        \caption{7 identical,1 defective}
        \label{GL1_1d}
    \end{subfigure}\\
    ~ 
    \begin{subfigure}[b]{0.2 \textwidth}
        \includegraphics[width=\textwidth,height=3.7cm]{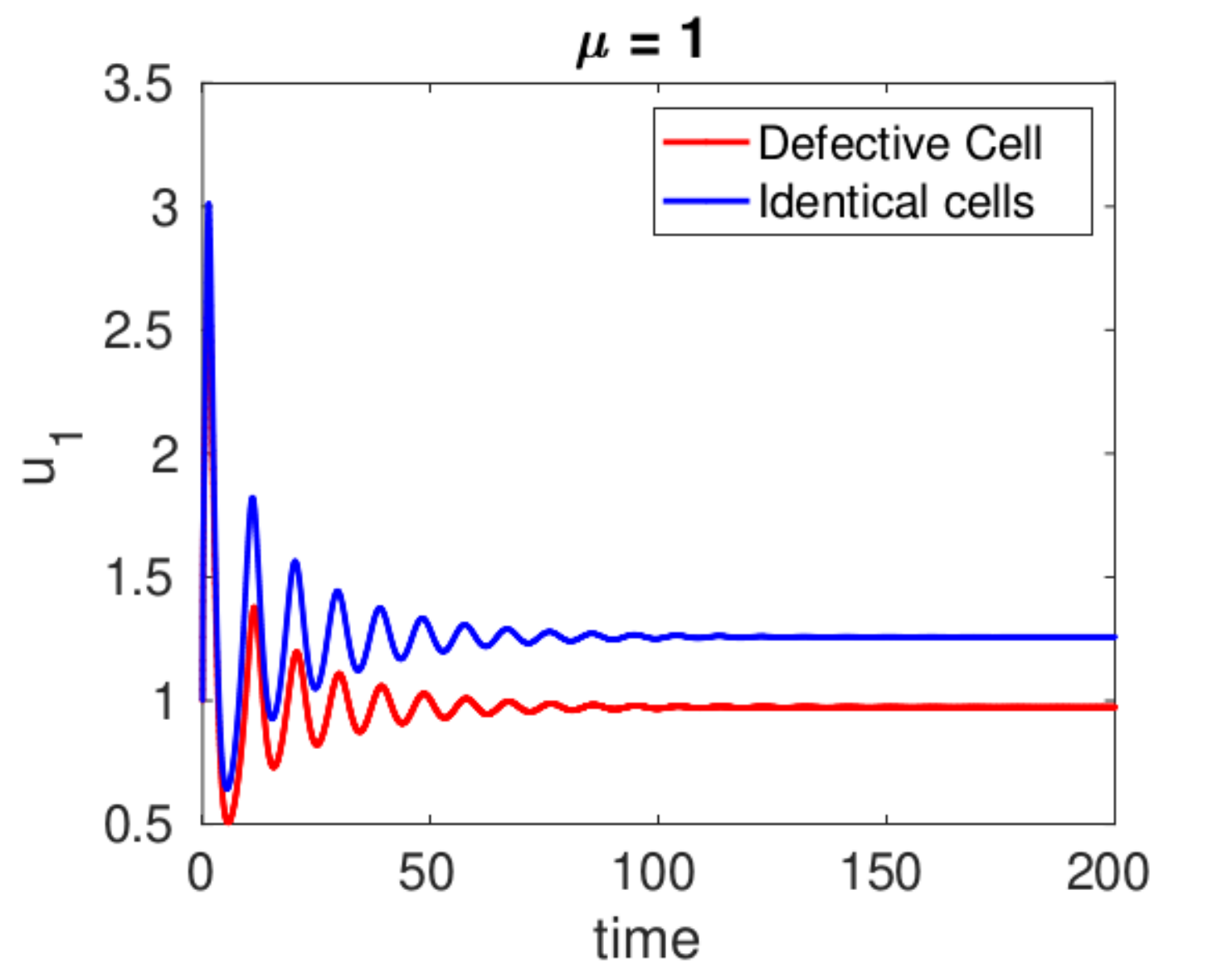}
        \caption{2 identical,1 defective}
        \label{GL1_1e}
    \end{subfigure}
     \quad
    ~ 
    \begin{subfigure}[b]{0.2 \textwidth}
        \includegraphics[width=\textwidth,height=3.7cm]{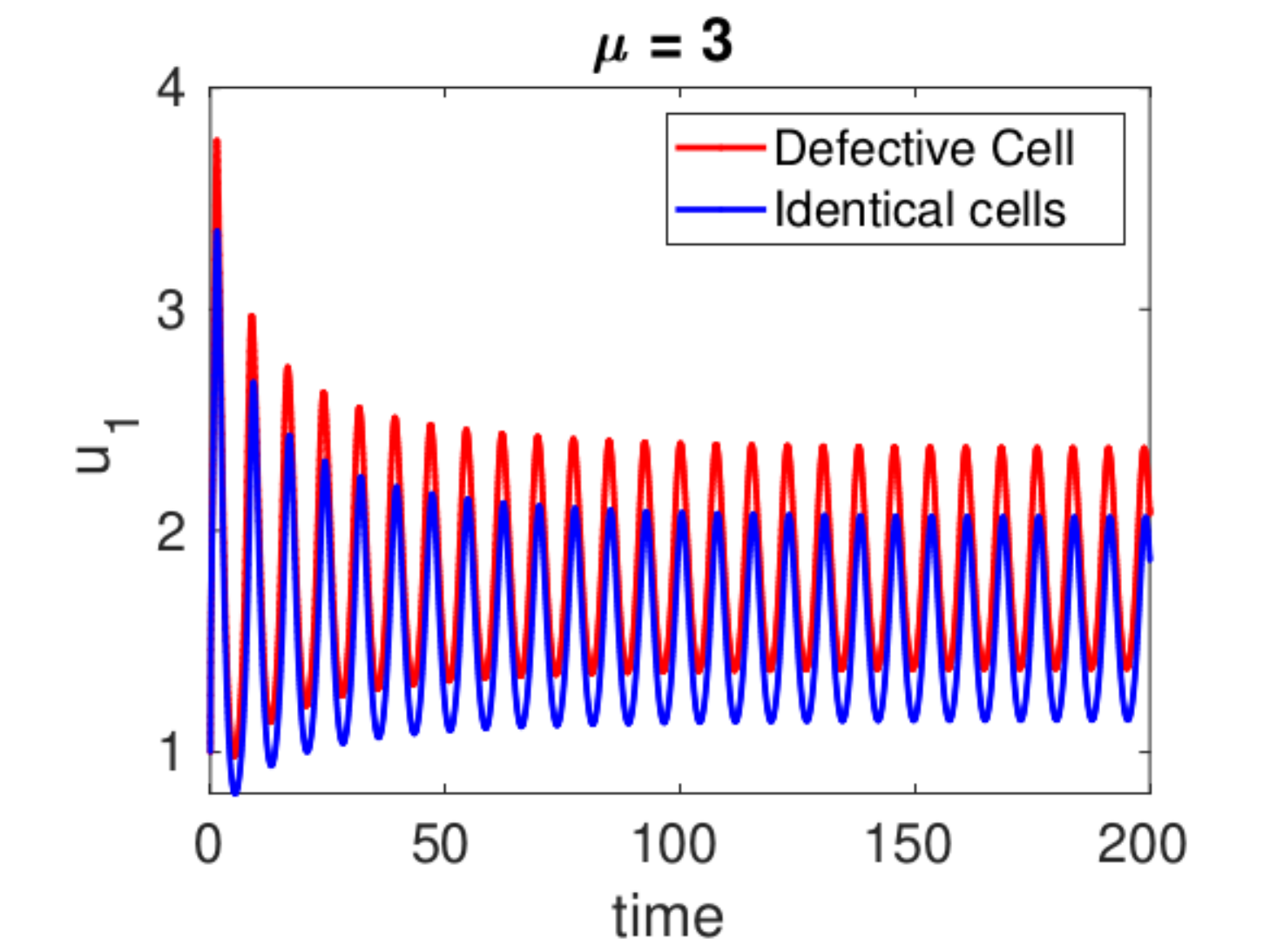}
        \caption{2 identical,1 defective }
        \label{GL1_1f}
    \end{subfigure}
     \quad
    ~ 
    \begin{subfigure}[b]{0.2 \textwidth}
        \includegraphics[width=\textwidth,height=3.7cm]{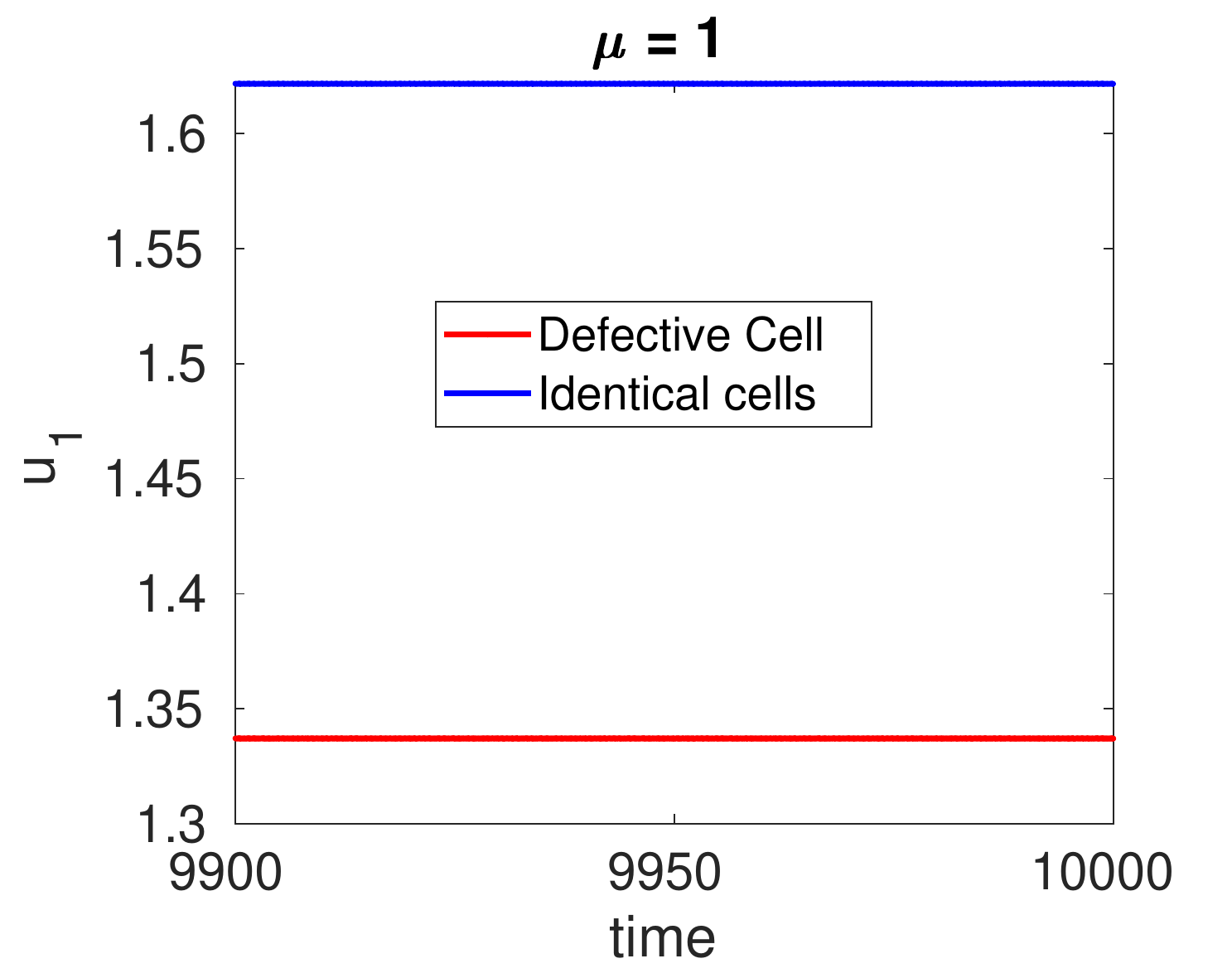}
        \caption{7 identical,1 defective}
        \label{GL1_1g}
    \end{subfigure}
    ~\quad
    \begin{subfigure}[b]{0.2 \textwidth}
        \includegraphics[width=\textwidth,height=3.7cm]{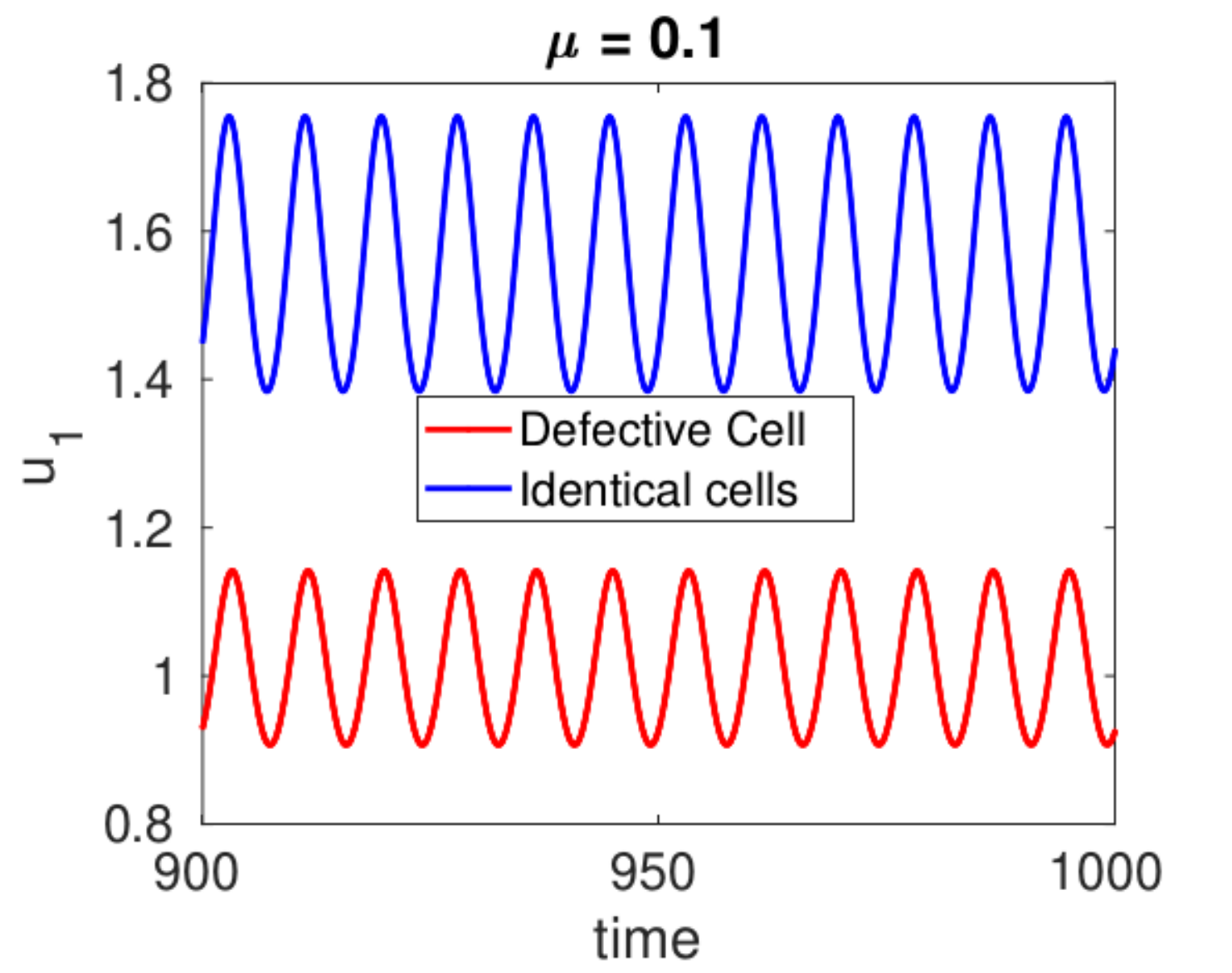}
        \caption{7 identical,1 defective}
        \label{GL1_1h}
    \end{subfigure}\\~ 
    \begin{subfigure}[b]{0.2 \textwidth}
        \includegraphics[width=\textwidth,height=3.7cm]{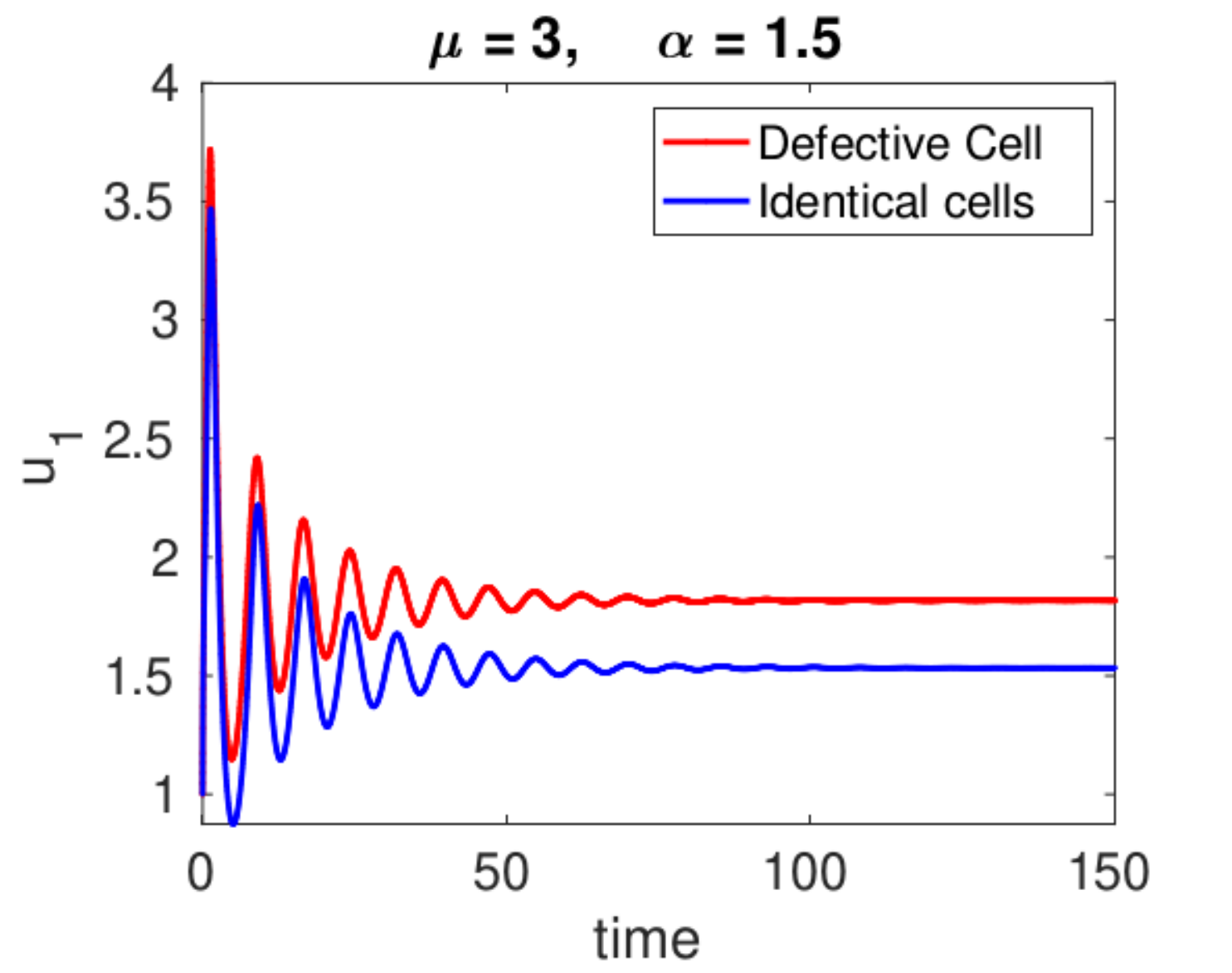}
        \caption{2 identical,1 defective}
        \label{GL1_1i}
    \end{subfigure}
     \quad
    ~ 
    \begin{subfigure}[b]{0.2 \textwidth}
        \includegraphics[width=\textwidth,height=3.7cm]{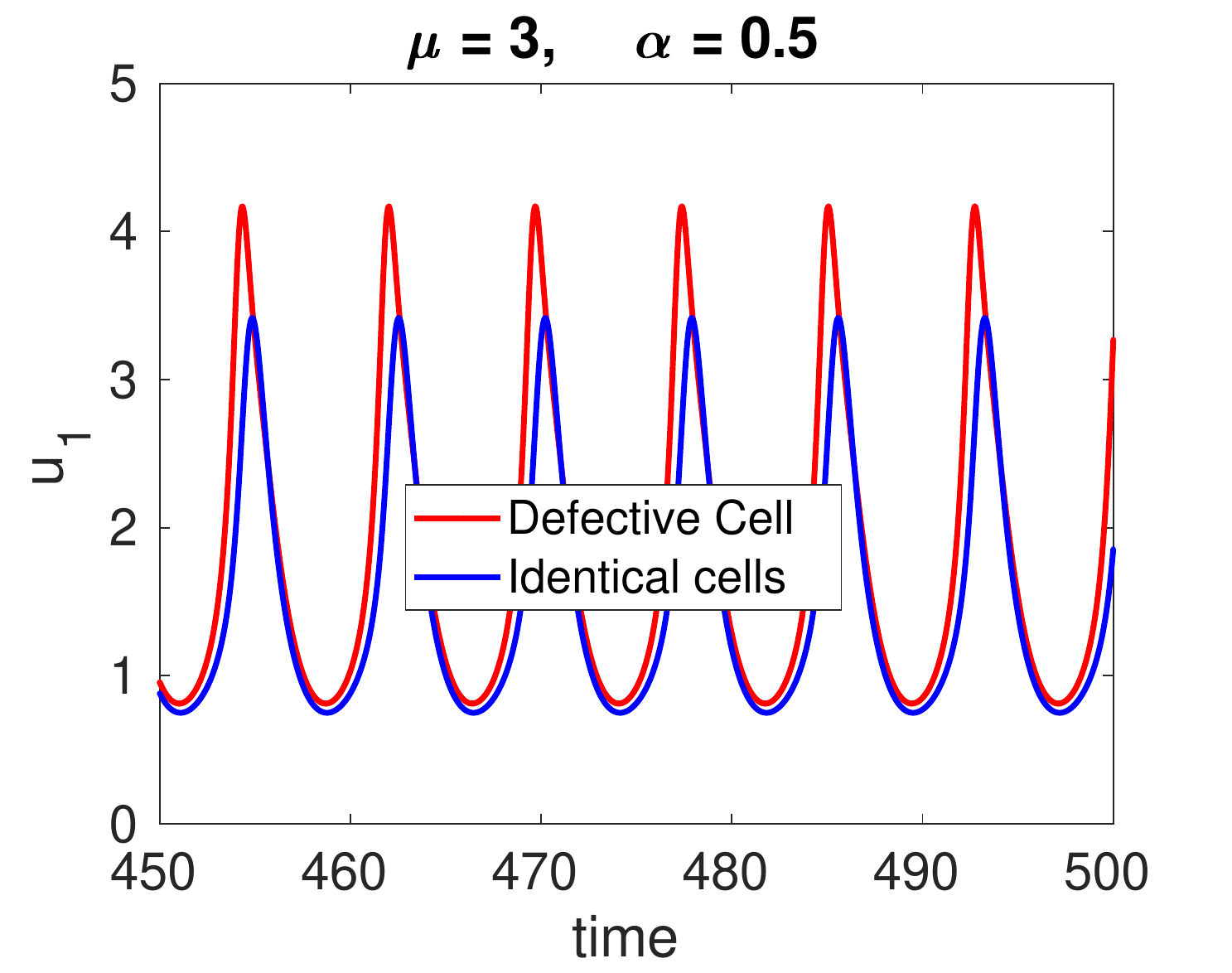}
        \caption{2 identical,1 defective}
        \label{GL1_1j}
    \end{subfigure}
     \quad
    ~ 
    \begin{subfigure}[b]{0.2 \textwidth}
        \includegraphics[width=\textwidth,height=3.7cm]{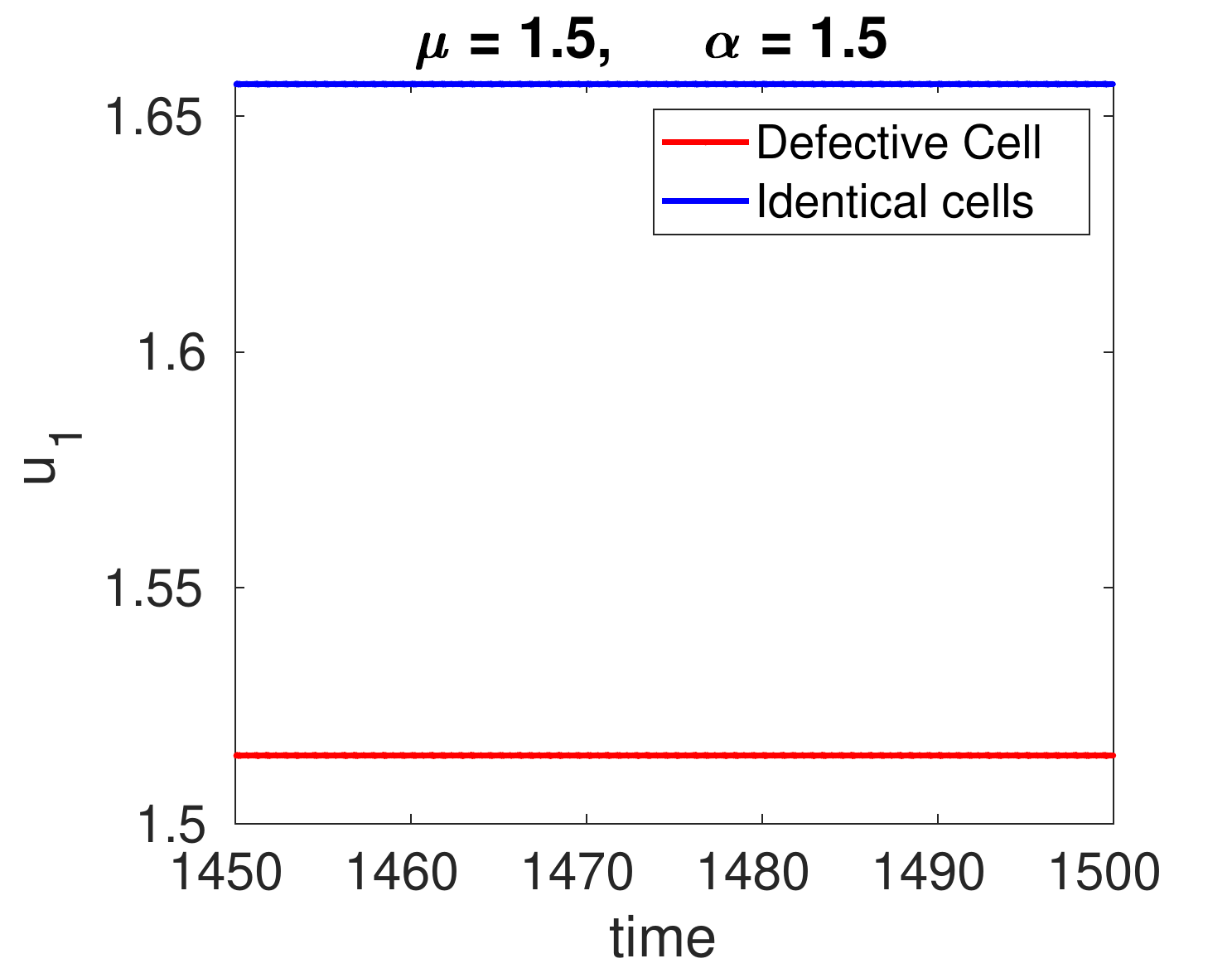}
        \caption{7 identical,1 defective}
        \label{GL1_1k}
    \end{subfigure}
    ~\quad
    \begin{subfigure}[b]{0.2 \textwidth}
        \includegraphics[width=\textwidth,height=3.7cm]{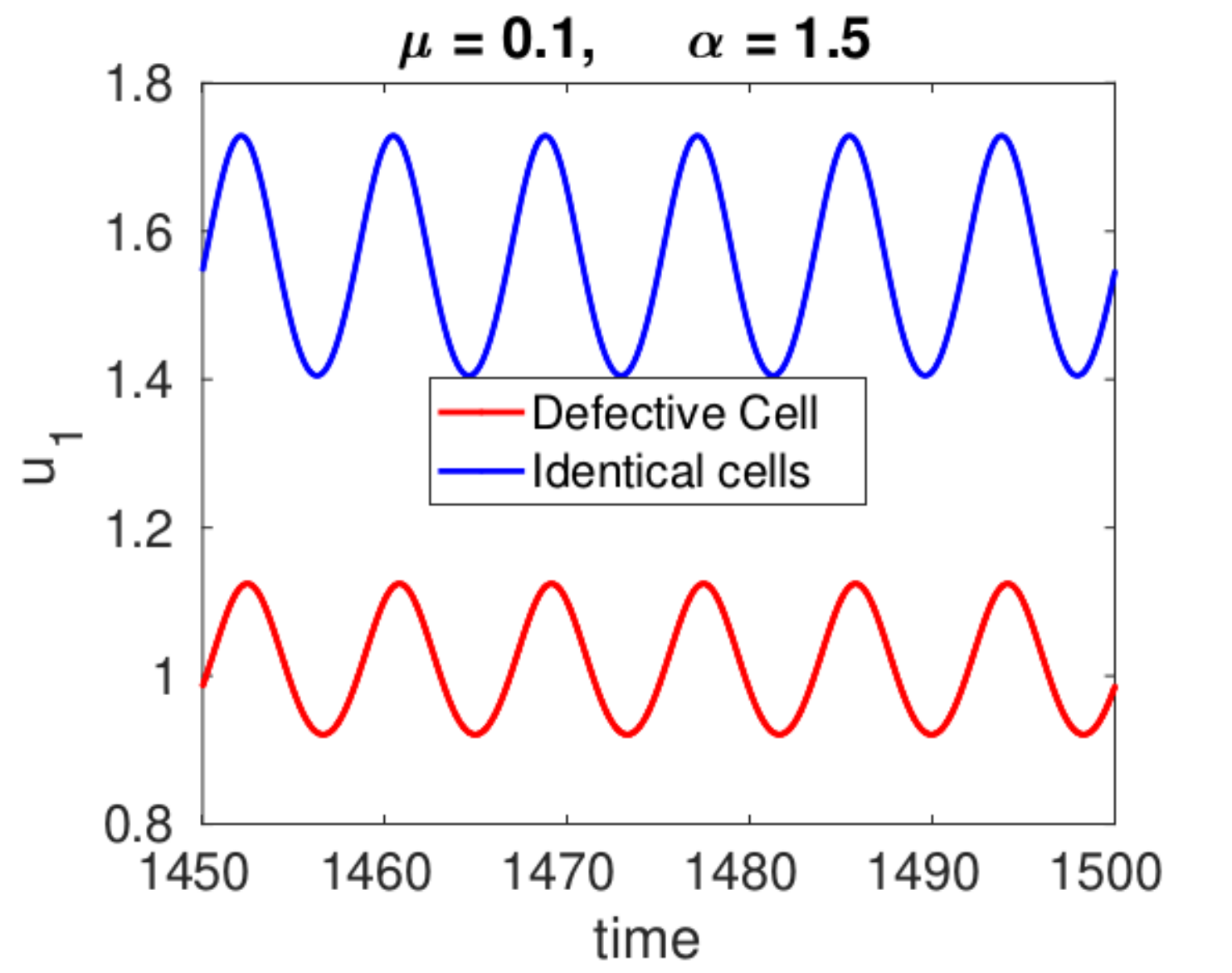}
        \caption{7 identical,1 defective}
        \label{GL1_1l}
    \end{subfigure}
    \caption{\textit{Numerical simulations of the system of ODEs
        \eqref{WM_ODEsys_2D} for different values of the Sel'kov
        parameters $\alpha$ and $\mu$ sampled from the bifurcation
        diagrams in Figure \ref{Bf_alpha} (a-d), Figure \ref{Bf_mu}
        (e-f), and Figure \ref{Bifur_2Par} (i-l).  }  }\label{GL1_1}
\end{figure}

\end{appendices}

\newpage
\bibliographystyle{plain}
\bibliography{ReferenceProposal.bib}

\end{document}